\documentclass{aa}  

\usepackage{natbib}
\usepackage{graphicx}
\usepackage{txfonts}
\usepackage{hyperref}
\usepackage{subfig}
\usepackage{lineno}
\usepackage[export]{adjustbox}
\makeatletter
\renewcommand*\aa@pageof{, page \thepage{} of \pageref*{LastPage}}
\makeatother

\newcommand{\hrieuv}{HRI\textsubscript{EUV}\xspace}
\newcommand{\hrilya}{HRI\textsubscript{Lya}\xspace}

\usepackage[normalem]{ulem}

\begin{document} 

   \title{EUV fine structure and variability associated with coronal rain revealed by Solar Orbiter/EUI \hrieuv {and SPICE}}
   \author{
        P. Antolin\inst{\ref{i:unn}}\fnmsep\thanks{Corresponding author: Patrick Antolin \email{patrick.antolin@northumbria.ac.uk}}
        \and
        A. Dolliou\inst{\ref{i:ias}}
        \and
        F. Auch\`ere\inst{\ref{i:ias}}
        \and
        L. P. Chitta\inst{\ref{i:mps}}
        \and
        S. Parenti\inst{\ref{i:ias}, \ref{i:rob}}
        \and
        D. Berghmans\inst{\ref{i:rob}}
        \and
        R. Aznar Cuadrado\inst{\ref{i:mps}}
        \and
        K. Barczynski\inst{\ref{i:pmod},\ref{i:eth}}
        \and
        S. Gissot\inst{\ref{i:rob}}
       \and
        L. Harra\inst{\ref{i:pmod},\ref{i:eth}}
        \and
        Z. Huang\inst{\ref{i:mps}}
        \and
        M. Janvier\inst{\ref{i:esa}, \ref{i:ias}}
        \and
        E. Kraaikamp\inst{\ref{i:rob}}
        \and
        D.~M. Long\inst{\ref{i:mssl}, \ref{i:qub}}
        \and
        S. Mandal\inst{\ref{i:mps}}
       \and
        H. Peter\inst{\ref{i:mps}}
        \and
         L. Rodriguez\inst{\ref{i:rob}}
        \and
        U. Sch\"uhle\inst{\ref{i:mps}}
        \and
        P.~J. Smith\inst{\ref{i:mssl}}
        \and
        S.~K.~Solanki\inst{\ref{i:mps}}
        \and
        K. Stegen\inst{\ref{i:rob}}
        \and 
        L. Teriaca\inst{\ref{i:mps}}
        \and
        C. Verbeeck\inst{\ref{i:rob}}
        \and
        M.~J. West\inst{\ref{i:swri}}
        \and
        A.~N.~Zhukov\inst{\ref{i:rob},\ref{i:sinp}}
        \and
        T. Appourchaux\inst{\ref{i:ias}}
        \and
        G. Aulanier\inst{\ref{i:ObservatoireDeParis},\ref{i:RoCS}}
        \and
        E. Buchlin\inst{\ref{i:ias}}
        \and
        F. Delmotte\inst{\ref{i:iogs}}
        \and
        J.~M. Gilles\inst{\ref{i:csl}}
        \and
        M. Haberreiter\inst{\ref{i:pmod}}
        \and
        J.-P. Halain\inst{\ref{i:csl},\ref{i:esa}}
        \and
        K. Heerlein\inst{\ref{i:mps}}
        \and 
        J.-F. Hochedez\inst{\ref{i:aester},\ref{i:latmos}}
        \and
        M. Gyo\inst{\ref{i:pmod}}
        \and
        S. Poedts\inst{\ref{i:kuleuven},\ref{i:umcs}}
        \and
        P. Rochus\inst{\ref{i:csl}}
       }
        \institute{
            Department of Mathematics, Physics and Electrical Engineering, Northumbria University, Newcastle upon Tyne, NE1 8ST, UK\label{i:unn}
            \and
            Université Paris-Saclay, CNRS, Institut d'Astrophysique Spatiale, 91405, Orsay, France\label{i:ias}
            \and
             Max Planck Institute for Solar System Research, Justus-von-Liebig-Weg 3, 37077 G\"ottingen, Germany\label{i:mps}
            \and
            Solar-Terrestrial Centre of Excellence -- SIDC, Royal Observatory of Belgium, Ringlaan -3- Av. Circulaire, 1180 Brussels, Belgium\label{i:rob}
            \and
            Physikalisch-Meteorologisches Observatorium Davos, World Radiation Center, 7260, Davos Dorf, Switzerland\label{i:pmod}
            \and 
            ETH Z\"urich, Institute for Particle Physics and Astrophysics , Wolfgang-Pauli-Strasse 27, 8093 Z\"urich \label{i:eth}
            \and
            European Space Agency, ESTEC, Keplerlaan 1, PO Box 299, NL-2200 AG Noordwijk, The Netherlands\label{i:esa}
            \and
            UCL-Mullard Space Science Laboratory, Holmbury St.\ Mary, Dorking, Surrey, RH5 6NT, UK\label{i:mssl}
            \and
            Southwest Research Institute, 1050 Walnut Street, Suite 300, Boulder, CO 80302, USA\label{i:swri}
            \and
            Skobeltsyn Institute of Nuclear Physics, Moscow State University, 119992 Moscow, Russia\label{i:sinp}
            \and
            Sorbonne Université, Observatoire de Paris - PSL, École Polytechnique, Institut Polytechnique de Paris, CNRS, Laboratoire de physique des plasmas (LPP), 4 place Jussieu, F-75005 Paris, France\label{i:ObservatoireDeParis}
            \and
            Rosseland Centre for Solar Physics, University of Oslo, P.O. Box 1029, Blindern, NO-0315 Oslo, Norway\label{i:RoCS}
            \and
            Laboratoire Charles Fabry, Institut d'Optique Graduate School, Universit\'e Paris-Saclay, 91127 Palaiseau Cedex, France\label{i:iogs}
            \and
            Centre Spatial de Li\`ege, Universit\'e de Li\`ege, Av. du Pr\'e-Aily B29, 4031 Angleur, Belgium\label{i:csl}
            \and 
            AESTER INCOGNITO, 75008 Paris, France\label{i:aester}
            \and
            LATMOS, CNRS - UVSQ - Sorbonne Université, 78280, Guyancourt, France\label{i:latmos}
            \and
            Centre for mathematical Plasma Astrophysics, KU Leuven, 3001 Leuven, Belgium\label{i:kuleuven}
            \and 
            Institute of Physics, University of Maria Curie-Sk{\l}odowska, Pl.\ M.\ Curie-Sk{\l}odowskiej 5, 20-031 Lublin, Poland\label{i:umcs}
            \and
            Astrophysics Research Centre, School of Mathematics and Physics, Queen’s University Belfast, University Road, Belfast, BT7 1NN, Northern Ireland, UK\label{i:qub}
             }
    \titlerunning{EUV structure and variability associated with coronal rain revealed by SolO/\hrieuv and SPICE}            

   \date{Received ; accepted }

 
  \abstract
   {Coronal rain is the most dramatic cooling phenomenon of the solar corona. Recent observations in the visible and UV spectrum have shown that coronal rain is a pervasive phenomenon in active regions. Its strong link with coronal heating through the Thermal Non-Equilibrium (TNE) - Thermal Instability (TI) scenario, makes it an essential diagnostic tool for the heating properties. Another puzzling feature of the solar corona, besides the heating, is its filamentary structure and variability, particularly in the EUV. }
   {We aim to identify observable features of the TNE-TI scenario underlying coronal rain at small and large spatial scales, to understand the role it plays in the solar corona. }
   {We use EUV datasets at unprecedented spatial resolution of $\approx240~$km from the High Resolution Imager (HRI) in the EUV (\hrieuv) of the Extreme Ultraviolet Imager (EUI) {and SPICE} on board Solar Orbiter from the spring 2022 perihelion.}
   {EUV absorption features produced by coronal rain are detected at scales as small as 260~km. As the rain falls, heating and compression is produced immediately downstream, leading to a small EUV brightening accompanying the fall and producing a {`fireball'} phenomenon in the solar corona. Just prior to impact, a flash-like EUV brightening downstream of the rain, lasting a few minutes is observed for the fastest events. For the first time, we detect the {atmospheric response} to the rain's impact on the chromosphere and {consists of upward propagating rebound shocks and} flows {partly reheating the loop}. The observed widths of the rain clumps are $500\pm200~$km. They exhibit a broad velocity distribution of $10-150~$km~s$^{-1}$, peaking below 50~km~s$^{-1}$. Coronal strands of similar widths are observed along the same loops {co-spatial with cool filamentary structure seen with SPICE,} which {we interpret as} the Condensation Corona Transition Region. {Matching with the expected cooling, prior to the rain appearance sequential loop brightenings are detected in gradually cooler lines from corona to chromospheric temperatures.} Despite the large rain showers, most cannot be detected in AIA~171 in quadrature, indicating that line-of-sight effects play a major role in coronal rain visibility. Still, AIA~304 and {SPICE} observations reveal that only a small fraction of the rain can be captured by \hrieuv.}
   {Coronal rain generates EUV structure and variability over a wide range of scales, from coronal loop to the smallest resolvable scales. This establishes the major role that TNE-TI plays in the observed EUV morphology and variability of the corona.}
\maketitle


\section{Introduction}
\label{sec:intro}
The solar corona is mysteriously heated to millions of degrees by yet unclear mechanisms of magnetic energy conversion and transport in plasmas. Coronal loops, the building blocks of the inner solar corona, constitute the prime target for the coronal heating investigation, due to their visibility above the diffuse background. Many of their properties remain a puzzle, such as their lifetimes (EUV variation), and morphologies (in particular their sub-structure) \citep{Reale_2010LRSP....7....5R, Klimchuk_2015RSPTA.37340256K, Viall_2021}. For instance, a long-standing puzzle of the corona is the observed filamentary / strand-like structure of loops (as opposed to being diffused) and their high variability, particularly in the upper transition region (TR) spectral lines \citep{Kjeldseth_Brekke_1998SoPh..182...73K, Ugarte-Urra_etal_2009ApJ...695..642U, Hinode_10.1093/pasj/psz084}. The filamentary structure is linked with the loop sub-structure, with the concept of coronal strands introduced and usually assumed to be part of a larger entity (or coronal loop `bundle'). This multi-stranded vs. monolithic structure has been a long-standing debate in the solar community. Its importance stems from its strong link with the fundamental scales in which the heating operates in the solar atmosphere, either directly associated with the scales at granular level, where most of the energy originates \citep{Martinez-Sykora_2018ApJ...860..116M}, or indirectly, e.g. by modifying how MHD waves propagate and dissipate in inhomogeneous plasmas \citep{VanDoorsselaere_2014ApJ...795...18V, VanDoorsselaere_2020SSRv..216..140V}. 

Prior to the Solar Orbiter launch \citep{Muller2020}, Hi-C \citep[1st and 2nd flights][]{Kobayashi_2014SoPh..289.4393K,Rachmeler_2019SoPh..294..174R}, provided the highest spatial resolution observations in the EUV, namely in the \ion{Fe}{xii}~193~\AA\, line forming at $\approx1.5\times10^6$~K (1st flight) and in the \ion{Fe}{ix}~172~\AA\, line forming at $\approx10^{5.9}$~K (2nd flight). These observations indicate coronal strand widths on the order of a few hundred km  \citep{Peter_2013AA...556A.104P, Brooks_2013ApJ...772L..19B, Aschwanden_2017ApJ...840....4A, Williams_2020ApJ...892..134W}. On the other hand, such sub-structure does not appear to be present for all coronal loops and across different temperature regimes, as the above reports show. Sub-structure, such as coronal strands, may appear during the evolution of the loops, particularly their cooling, and thus may be linked to specific aspects of how the cooling happens. 

Coronal rain is the most dramatic display of cooling in the solar corona. It corresponds to cool ($10^3-10^5$~K) and dense ($10^{10}-10^{13}$~cm~$^{-3}$) plasma clumps appearing over a timescale of minutes in chromospheric and TR spectral lines in the solar corona, that preferentially fall towards the solar surface along coronal loops \citep{Kawaguchi_1970PASJ...22..405K,Leroy_1972SoPh...25..413L,Habbal_1985SoPh...98..323H,Foukal_1978ApJ...223.1046F,Wiik_1996SoPh..166...89W,Schrijver_2001SoPh..198..325S,DeGroof_2004AA...415.1141D,DeGroof05}. Coronal rain is closely related to prominences \citep{Vial_Engvold_2015ASSL..415.....V}, but high-resolution observations over the last decade with Hinode \citep{Kosugi2007,Antolin_2010ApJ...712..494A, Hinode_10.1093/pasj/psz084}, the Swedish 1-m Solar Telescope \citep[SST;][]{Scharmer_2003SPIE.4853..341S,Antolin_Rouppe_2012ApJ...745..152A}, the Goode Solar Telescope \citep[GST;][]{Goode_2003JKAS...36S.125G,Ahn_2014SoPh..289.4117A, Jing_2016NatSR...624319J}, the Solar Dynamics Observatory \citep[SDO;][]{Pesnell_2012SoPh..275....3P,Vashalomidze_2015AA...577A.136V} and the Interface Region Imaging Spectrograph  \citep[IRIS;][]{DePontieu2014,Antolin_2015ApJ...806...81A, Schad_2017SoPh..292..132S, DePontieu_2021SoPh..296...84D} have shown that coronal rain presents unique features in terms of its morphology and kinematics. At the smallest scales, coronal rain appears to be composed of clumps, which seem to also constitute the coolest and densest part. The widths (in the direction transverse to the flow) can be as low as 120~km \citep{Jing_2016NatSR...624319J} but generally are a few hundred km in H$\alpha$ \citep{Antolin_Rouppe_2012ApJ...745..152A} to $\approx500-600~$km in TR lines \citep{Antolin_2015ApJ...806...81A}, with little variation other than that expected by spatial resolution. On the other hand, they can extend greatly longitudinally (along the flow), with lengths about an order of magnitude or more. Recently, \citet{Sahin_2023} studied large-scale coronal rain over an active region (AR) with IRIS in chromospheric and TR lines, finding little variation in its morphological and dynamical properties over several hours and across the AR. The observed coronal rain strands appear to have very similar widths to the coronal strands observed by Hi-C, described above, which may either directly reflect a fundamental heating scale \citep{Jing_2016NatSR...624319J, Antolin_Froment_2022} or be associated with the cooling, as explained below. 

One of the most interesting aspects of coronal rain is that the clumps occur in tandem across a relatively large cross-section across the magnetic field (of a few Mm in width). This synchronicity and shared trajectory of clumps has led to the concept of rain shower, i.e. a larger-structure composed of coronal rain clumps \citep{Antolin_Rouppe_2012ApJ...745..152A}.  \citet{Sahin_2022ApJ...931L..27S} have shown that rain showers match well with cooling coronal loops observed in EUV, thereby helping to identify these in the large superposition (leading to line-of-sight confusion) of the optically thin solar corona \citep[what][refer to as `coronal veil']{Malanushenko_2022ApJ...927....1M}. 

There are currently 3 different kinds of coronal rain. The most commonly observed kind is known as `quiescent', and occurs preferentially in AR coronal loops. This kind does not require any specific magnetic topology (other than a loop-forming bi-polar structure). The second kind is linked to solar flares and is known as `flare-driven' coronal rain. It corresponds to the cool chromospheric loops (sometimes known as H$\alpha$ loops) appearing at the end of the gradual phase. The last kind is known as prominence-coronal rain hybrids, and involves a complex magnetic field with null-point topology at the top of loop arcades. A review of each can be found in \citet{Antolin_Froment_2022}. This work concerns the first kind, that is, the quiescent coronal rain of ARs. This kind is the most actively studied probably because of its strong link with coronal heating. 

Numerical work since the 90s have shown that complex magnetic topologies such as magnetic dips are not necessary for the generation of cool and dense, prominence-like structures in loops \citep{Antiochos_Klimchuk_1991ApJ...378..372A,Antiochos_1999ApJ...512..985A,Karpen_etal_2001ApJ...553L..85K}. Although we do not know what exactly the agents of coronal heating are (e.g. MHD waves or stress-induced magnetic reconnection), the generated spatial and temporal distribution of the magnetic energy along loops has unique consequences on the evolution of coronal loops, specifically on how they cool down. When the heating is strongly stratified (also known as `footpoint concentrated'), even if constant over time, the loop is often unable to reach thermal equilibrium and enters a state of thermal non-equilibrium (TNE). Its thermodynamic evolution undergoes cycles of heating and cooling, generally referred as TNE cycles, also known as evaporation-condensation cycles \citep{Kuin_1982AA...108L...1K,Mok_etal_1990ApJ...359..228M,Reale_1996AA...316..215R,Muller_2003AA...411..605M,Mendozabriceno_2005ApJ...624.1080M,Susino_2010ApJ...709..499S,Luna_etal_2012ApJ...746...30L}. This is true as long as the repetition frequency of the stratified heating events is faster than the radiative cooling time of the loop \citep{Johnston_2019AA...625A.149J}. \citet{Klimchuk_2019ApJ...884...68K} have quantified some of the requirements needed for TNE, and found that a volumetric heating ratio between apex and footpoint below 0.3 and a heating asymmetry between both footpoints under 3 ensures TNE. 

\begin{figure*}%
    \centering
    \subfloat{{\includegraphics[width=0.45\textwidth]{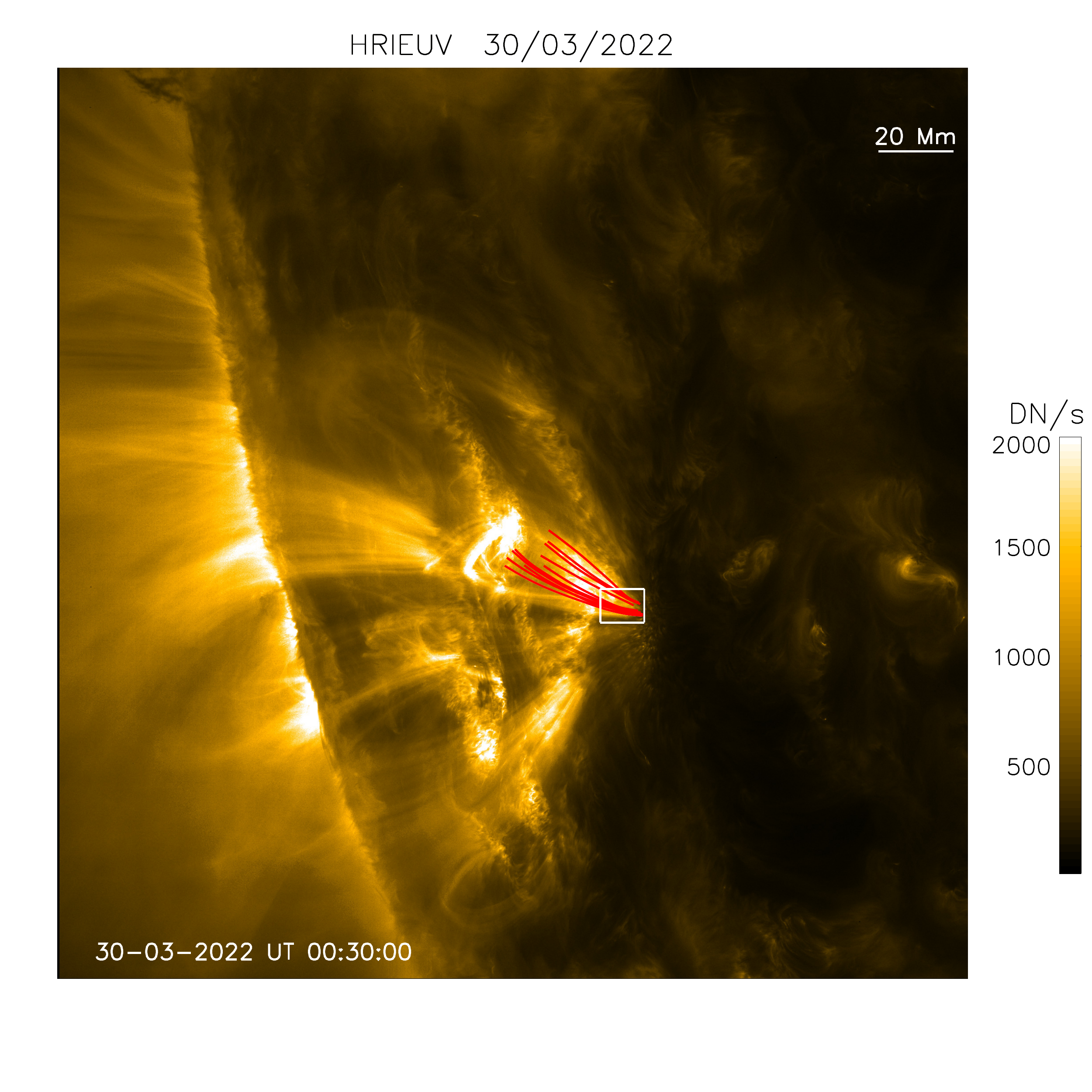} }}%
    \qquad
    \subfloat{{\includegraphics[width=0.45\textwidth]{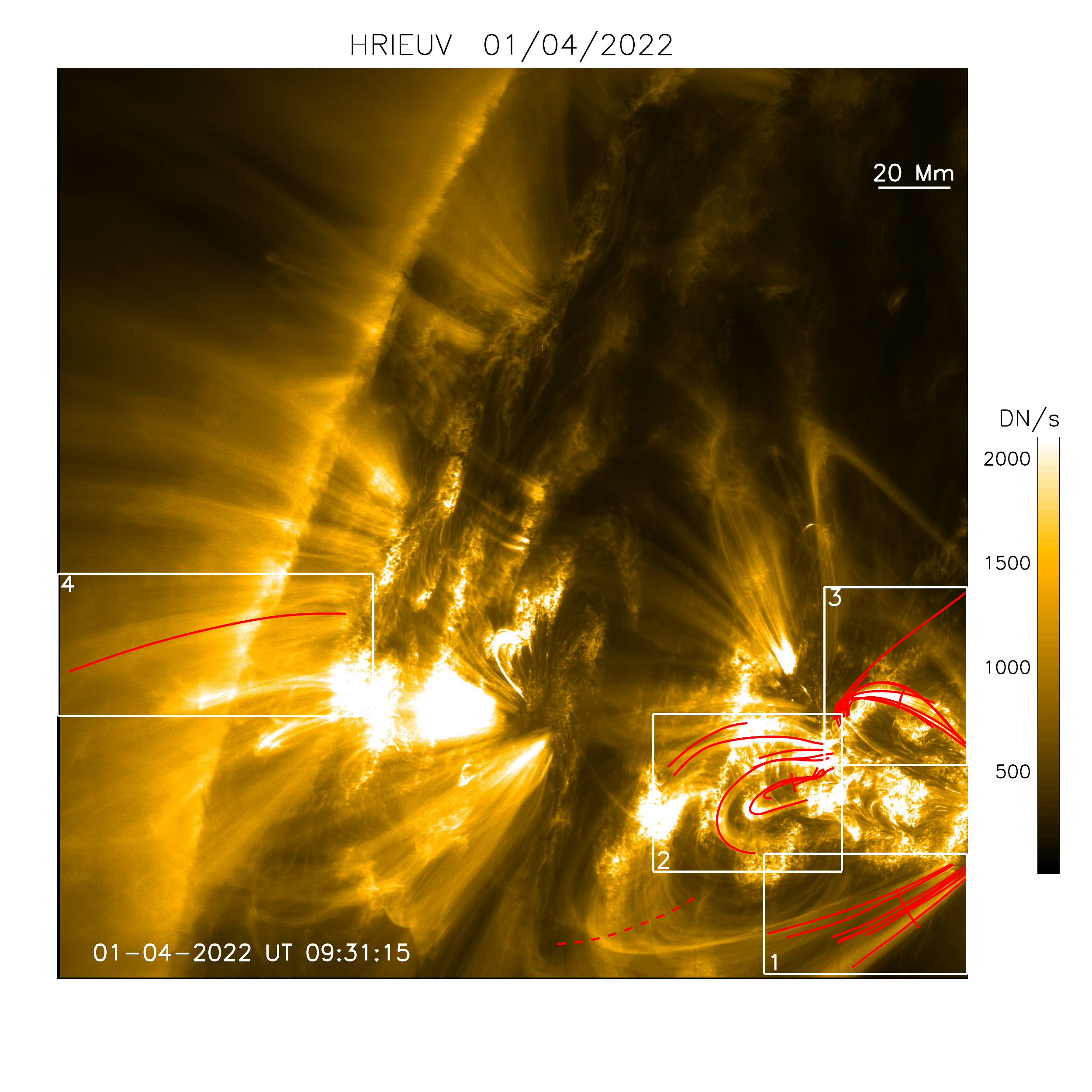} }}%
    \vspace{-0.5cm}
    \caption{Full FOV of \hrieuv for the 2022 March~30 (left) and April~1 (right) datasets analysed in this work. The inner white rectangles show sub-FOVs shown in later figures. The solid {red} curves denote the trajectories of some of the observed coronal rain clumps.}%
    \label{fig:fullFOV}%
\end{figure*}

In a nutshell, with a strongly stratified heating, the loop apex relies on thermal conduction for its heating. However, this spatial distribution leads to an essentially flat temperature profile along the loop length, or even a temperature dip at the apex, thus making conduction inefficient. Furthermore, the footpoint heating is very efficient at injecting material upwards (through chromospheric ablation / evaporation), making the loop overdense relative to the RTV scaling law \citep{Rosner_1978ApJ...220..643R}. The apex ends up having a thermal imbalance, which results in a runaway or catastrophic cooling due to the higher efficiency for plasmas at radiating their energy away at lower temperatures (in the TR - corona temperature range). While the heating can be very rapid (with the loop essentially empty) and therefore very hard to observe, the cooling progresses over a timescale of tens of minutes to hours, depending on the loop conditions. These long cooling times can lead to very long loop lifetimes \citep{Lionello_2016ApJ...818..129L}. The loop eventually evacuates and the cycle restarts if the heating conditions do not change, hence leading to cycles of heating and cooling. During the initial stages of the cooling, and when the cooling time is long enough relative to the cadence of the observations, the loop is expected to sequentially appear in cooler coronal passbands with specific non-zero timelags \citep{Kamio_etal_2011AA...532A..96K,Viall_Klimchuk_2012ApJ...753...35V, Viall_2021}. The cyclic pattern can lead to highly periodic EUV intensity pulsations on the order of hours, recently detected with EIT \citep{Auchere_2014AA...563A...8A} and the Atmospheric Imaging Assembly \citep[AIA;][]{Lemen_2012SoPh..275...17L,Froment_2015ApJ...807..158F}, indicating a heating function that is stable over days (and up to a week). At the end of the cooling part of the TNE cycle accelerated cooling is observed and coronal rain appears, which can therefore also occur periodically \citep{Auchere_2018ApJ...853..176A,Froment_2020AA...633A..11F}. This acceleration in the cooling rate and also the spatial and temporal character of coronal rain have been interpreted as a product of thermal instability (TI), but a debate exists in the community \citep{Klimchuk_2019SoPh..294..173K,Antolin_2020PPCF...62a4016A}. The essential role that TI may play in the observed coronal rain phenomena (and probably the long-period intensity pulsations as well), has led to the cycles being known as TNE-TI cycles \citep{Antolin_Froment_2022}. 

Thermal instability is a fundamental MHD process \citep{Parker_1953ApJ...117..431P,Field_1965ApJ...142..531F,Waters_2019,Claes_Keppens_AA624_2019}. Besides coronal rain, it has been invoked to explain phenomena from the short laboratory scales \citep{Lipschultz_1987JNuM..145...15L,Stacey_1996PhPl....3.2673S}, to very large intracluster medium scales \citep{White_Rees_1978MNRAS.183..341W,Cavagnolo_2008ApJ...683L.107C,Sharma_2013ASInC...9...27S}. In the context of the solar corona, thermal instability is less straightforward to apply {because the corona is very dynamic and is out of hydrostatic equilibrium \citep{Aschwanden_2001ApJ...550.1036A}}. Nonetheless, various analytic studies have argued that given the long timescales of TNE cycles, TI theory still holds, given the local and short timescale characteristics of its occurrence \citep{Claes:2021wg}. {\citet{Antolin_Rouppe_2012ApJ...745..152A} and \citet{Antolin_2015ApJ...806...81A} have} argued that TI may act as a synchronising mechanism for catastrophic cooling to occur simultaneously across a loop in TNE, thereby providing an explanation for rain showers {\citep[see also ][]{Froment_2020AA...633A..11F, Antolin_2020PPCF...62a4016A}}.  \citet{Sahin_2022ApJ...931L..27S} have used this link to unlock a way to estimate the TNE volume over an AR. By calculating the number of rain showers and their properties, they have estimated that at least 50\% of the AR is subject to TNE.

\begin{figure*}%
    \centering
    \includegraphics[width=1\textwidth]{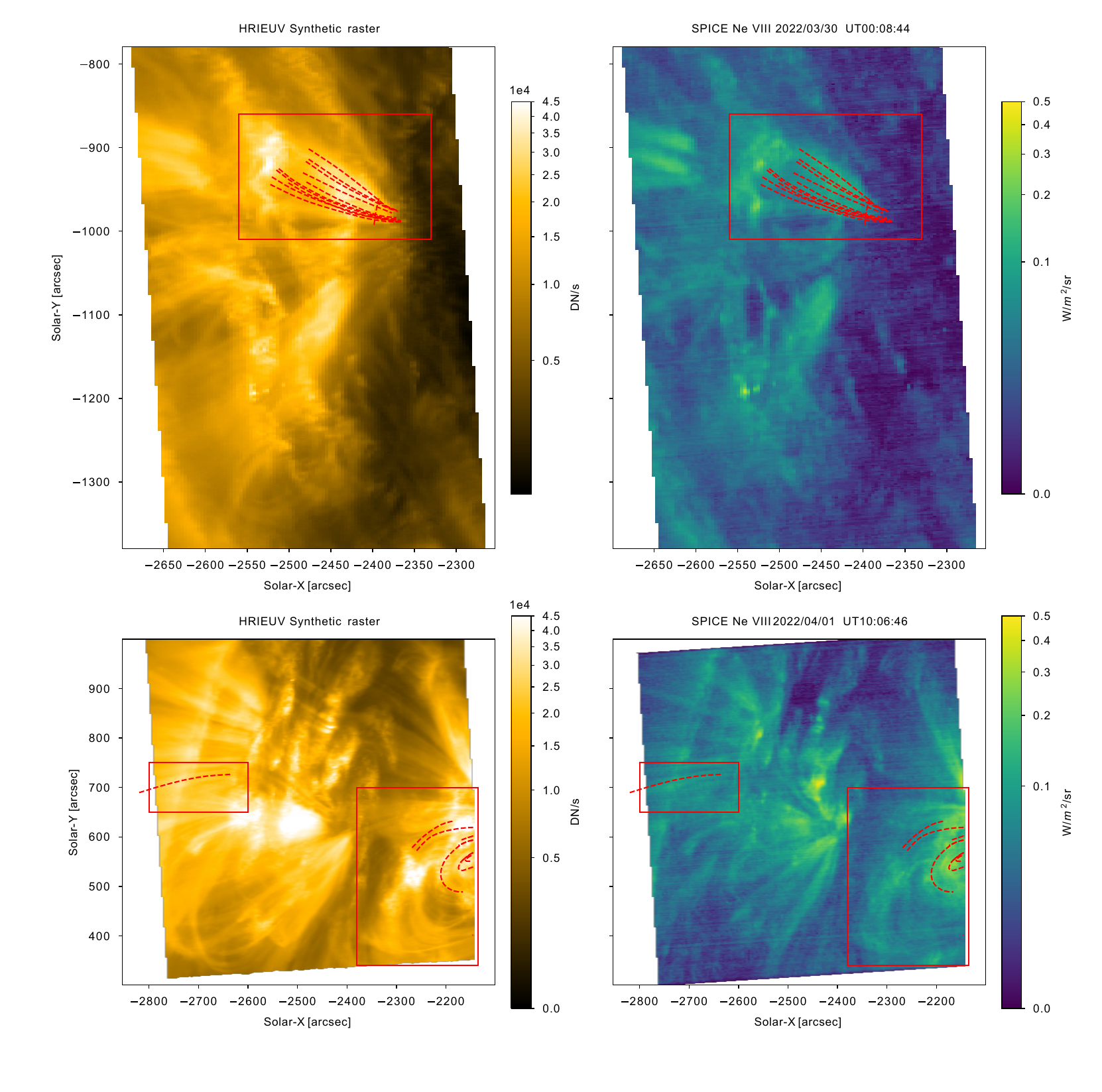}%
    \caption{{Co-aligned \hrieuv (left) and SPICE (right) full FOV for March~30 (top) and April~1 (bottom). The SPICE images correspond to a 96- and 160-step rasters for March~30 and April~1, respectively, and show the total intensity over the \ion{Ne}{VIII} line. The corresponding \hrieuv images are synthetic rasters that match the time and space distribution of data acquisition with SPICE. A spatial binning has been applied to \hrieuv to match the platescale of SPICE. The red rectangles correspond to regions of interest (ROI) in each FOV with overlaid rain paths in red identified with \hrieuv. The ROI on March~30 is shown in Figure~\ref{fig:reg1_mar30_spice}. The ROIs on April~1 to the West and East are shown in Figures~\ref{fig:spice_april1_regs1-2} and \ref{fig:spice_apr1_reg4}, respectively. See text for further details.} }%
    \label{fig:spicefullFOV}%
\end{figure*}

2.5-D radiative MHD simulations by \citet{Antolin_2022ApJ...926L..29A} have shown that the TI-driven catastrophic cooling leads to the formation of cool and dense cores at the head of the rain emitting in chromospheric lines, surrounded by thin but strongly emitting shells in TR lines that elongate in the wake of the rain, in what is known as the Condensation Corona Transition Region (CCTR). These structures are seen clearly in the \ion{Si}{IV}~1402 line observed by IRIS (forming at $\approx10^{4.8}~$K), and the \ion{Fe}{xii}~171 line observed by AIA~171 (forming at $\approx10^{5.8}$~K), and could therefore explain the filamentary/stranded morphology and dynamic nature of the corona seen in these lines \citep{Ugarte-Urra_etal_2009ApJ...695..642U}, as well as the common widths between coronal rain strands and coronal EUV sub-structure. High-resolution observations in the EUV are needed to confirm this hypothesis. Furthermore, \citet{Antolin_2022ApJ...926L..29A} have shown that TI produces a local enhancement of the magnetic field strength, due to the gas pressure loss during TI and frozen-in conditions. Due to the compression ahead of the rain (downstream) as it falls, a strong UV and EUV enhancement is obtained over the last few Mm prior to impact with the chromosphere. Furthermore, a strong rebound shock and upward flow is obtained from the impact, propagating at different speeds (corresponding to the tube speed and flow). These features have remained elusive in observations until now. 

Solar Orbiter was launched in early 2020 on a highly elliptical orbit around the Sun. By now, it has concluded two science close perihelia in its nominal mission phase, where the distance to the Sun was less than 0.32~au. In this work we analyse data from one of the two High Resolution Imagers (HRI) of the Extreme Ultraviolet Imager (EUI) telescopes on board Solar Orbiter \citep{Rochus2020}, corresponding to the first close perihelion, which happened between March and April 2022 \citep{Berghmans_etal_subm_2023}. The \hrieuv is an EUV imaging instrument in the 174~\AA~ passband, which is dominated by the \ion{Fe}{ix} (at 171.1\,\AA\,) and \ion{Fe}{x} (at 174.5\,\AA\, and 177.2\,\AA\,) emission forming at $\approx10^{6}~$K \citep{Chen_2021AA...656L...7C}. The importance of the high resolution achieved by this instrument was already evident in \hrieuv data from May 2020, when the spacecraft was at 0.556\,au, which revealed small EUV brightenings at spatial and temporal resolutions that could be barely detected by SDO/AIA observations \citep[known as `campfires';][]{Berghmans_2021AA...656L...4B}. 

{EUI also includes a high-resolution imager \hrilya, whose bandpass is centered at 121.6 nm and is dominated by the Lyman-$\alpha$ line of hydrogen. We do not use\hrilya in this work due to the degradation issues the telescope suffered during the perihelion approach \citep[see][ for details]{Berghmans_etal_subm_2023}.}

In this work we analyse \hrieuv data and identify several coronal rain events at unprecedented spatial resolution of $\approx240$~km in the EUV that allow {us} to clearly identify the EUV variability and morphology associated with coronal rain. The paper is organised as follows. The \hrieuv observations are presented in Section~\ref{sec:obs}. The methods are presented in Section~\ref{sec:methods}. Results can be found in Section~\ref{sec:results}, and discussion and conclusions in Section~\ref{sec:conclusions}.

\section{Observations}
\label{sec:obs}

The observations analysed in this work belong to the first perihelion passage of Solar Orbiter in March and April 2022. Specifically, we analyse two datasets of \hrieuv at 174~\AA, one of NOAA 12974 on 2022 March 30, and the other of ARs NOAA 12975 and 12796 on 2022 April 1, both on-disk \citep[][]{euidatarelease5}. At this time, Solar Orbiter was near quadrature with Earth (with separation angles between Solar Orbiter and the Sun-Earth line of $91.88^{\circ}$ and $102.02^{\circ}$ for March 30 and April 1, respectively). 

On March 30 and April 1, Solar Orbiter was at 0.33\,au and 0.34\,au, respectively. With an \hrieuv plate scale of 0.492\arcsec, the spatial resolution of these observations is estimated to be close to the Nyquist limit of $2\times0.492\arcsec$ \citep{Berghmans_etal_subm_2023}, corresponding to 237~km and 247~km, approximately. HRI images have $2048\times2048$~pixels, leading to a $17'\times17'$ field-of-view (FOV). The full FOV for each date can be seen in Figure~\ref{fig:fullFOV}. The observations on March~30 and April~1 are part of the R\_BOTH\_HRES\_HCAD\_Nanoflares and R\_SMALL\_MRES\_MCAD\_AR-Long-Term SOOPs \citep[][]{Zouganelis2020}, respectively, which operated the \hrieuv camera at a cadence of 3~s over a duration of 45~min (UT 00:03 -- 00:48) on March 30, and at a cadence of 10~s over a duration of $\approx75~$min (UT 09:19 -- 10:34) on April 1.

EUI is equipped with software controlled onboard calibration electronics to correct the images pixel-wise for offset and flat field before compression. The images are then prepped with the euiprep routine to level 2, which reduces the jitter and pointing error. However, significant jitter still remains that needs to be removed. To this end, we apply a cross-correlation technique to align the images as described in \citet{Chitta2022}.


For better visualisation of the fine structure in the \hrieuv images, we have applied the wavelets-optimised whitening (WOW) enhancement technique described in \citet{Auchere2023}. This method works by equalizing the variance at all scales and locations in the image, thereby reducing the large scale gradients and conversely enhancing the small scale structures.

{We have also checked \hrilya for the presence of rain in the Lyman-$\alpha$ line. However, none could be detected probably due to the problem affecting the resolution of the instrument, as discussed in \citet{Berghmans_etal_subm_2023}.}

 {Solar Orbiter also carries the Spectral Imaging of the Coronal Environement \citep[SPICE, ][]{Anderson2021}, as part of the remote sensing payload. For March 30, the SPICE data analyzed (data release 3.0\footnote{\url{https://doi.org/10.48326/idoc.medoc.spice.3.0}}),
 is the 96-step raster starting at 00:00:32 UTC with a field of view of $384^{\prime \prime}\times 914^{\prime \prime}$. It has a duration of 16~min 24~s and an exposure time of 10~s. The selected slit is 4\arcsec wide, while the data spatial pixel size is 1.098\arcsec along the slit. The temperature coverage of the spectral windows was from the chromosphere to the corona through the observation of the following spectral lines: \ion{H}{I} Ly$\beta$ 1025.72 \AA~(Log $T_e$ = 4.0 K); \ion{C}{III} 977.03 \AA~(Log $T_e$ = 4.8 K); \ion{S}{V} 786.47 \AA~(log $T_e$ = 5.2 K); \ion{O}{IV} 787.72 \AA~(log $T_e$ = 5.2 K); \ion{O}{VI} 1031.93 \AA~(log $T_e$ = 5.5 K); \ion{Ne}{VIII} 770.42 \AA~(log $T_e$ = 5.8 K ); \ion{Mg}{IX} 706.02 \AA~(log $T_e$ = 6.0 K).  
 For April 1st, we analysed five 160-step rasters which use the 4$^{\prime \prime}$ slit each producing a field of view of $640^{\prime \prime}\times 911^{\prime \prime}$. They run from 09:15:36 to 10:15:37 UTC. The rasters duration is 14~min with the exposure time of 5~s. The spectral windows of April 1$^{\mathrm{st}}$ study covered similar lines as the March 30 raster, with the exception of the \ion{S}{V} and \ion{O}{IV} lines. These two lines were replaced by \ion{N}{IV} 765.15 \AA~(Log $T_e$ = 5.2 K). For both datasets we used L2 data which are original data corrected for dark current, flat-field, geometrical distortion. An additional step in the data processing was the application of the radiometric calibration.}
 
{The pointing information in SPICE L2 headers is not accurate and the SPICE rasters need to be co-aligned with the \hrieuv sequence. We started by binning the  \hrieuv images to the same pixel size of SPICE. We then built a \hrieuv synthetic raster by selecting, from the \hrieuv time sequence, the image closest in time to each SPICE exposure making the raster. For each SPICE exposure, the SPICE pixel positions along the slit make an irregular grid in Helioprojective coordinates. The \hrieuv image closest in time to this exposure is reprojected into this grid. We then made SPICE images in \ion{Ne}{VIII} intensity by spectrally summing over the 32 pixels window. The \ion{Ne}{VIII} and \hrieuv 
images are, in fact, comparable in terms of plasma temperature coverage. Finally, the SPICE images are co-aligned with the \hrieuv synthetic raster using a cross-correlation technique. The SPICE FOV for March~30 and April~1 in the \ion{Ne}{VIII} line co-aligned with \hrieuv can be seen in Figure~\ref{fig:spicefullFOV}. Snapshots of each FOV in all the other spectral lines are shown in Figure~\ref{fig:spice_mar30_full} for March~30 and Figure~\ref{fig:spice_april1_full} for April~1.}

In addition to EUI images we also briefly analyse images from the Atmospheric Imaging Assembly \citep[AIA;][]{Lemen_2012SoPh..275...17L} on board the Solar Dynamics Observatory \citep[SDO;][]{Pesnell_2012SoPh..275....3P} to locate, if possible, the coronal rain events observed with \hrieuv. The AIA data correspond to level 2 data, processed through the standard SolarSoft packages. Since strict AIA-EUI co-alignment {at a pixel scale} is not needed for our purpose {(we do not need to identify specific rain trajectories across different viewpoints)}, we rely on co-alignment using header information through the JHelioviewer software \citep{Muller_etal_2009_jhelio}, which is sufficient to identify the large-scale structure common to both FOVs {(such as loops, rain showers, prominences etc.)}.

\section{Methodology}
\label{sec:methods}
\begin{figure}
    \centering
    \includegraphics[width=0.4\textwidth]{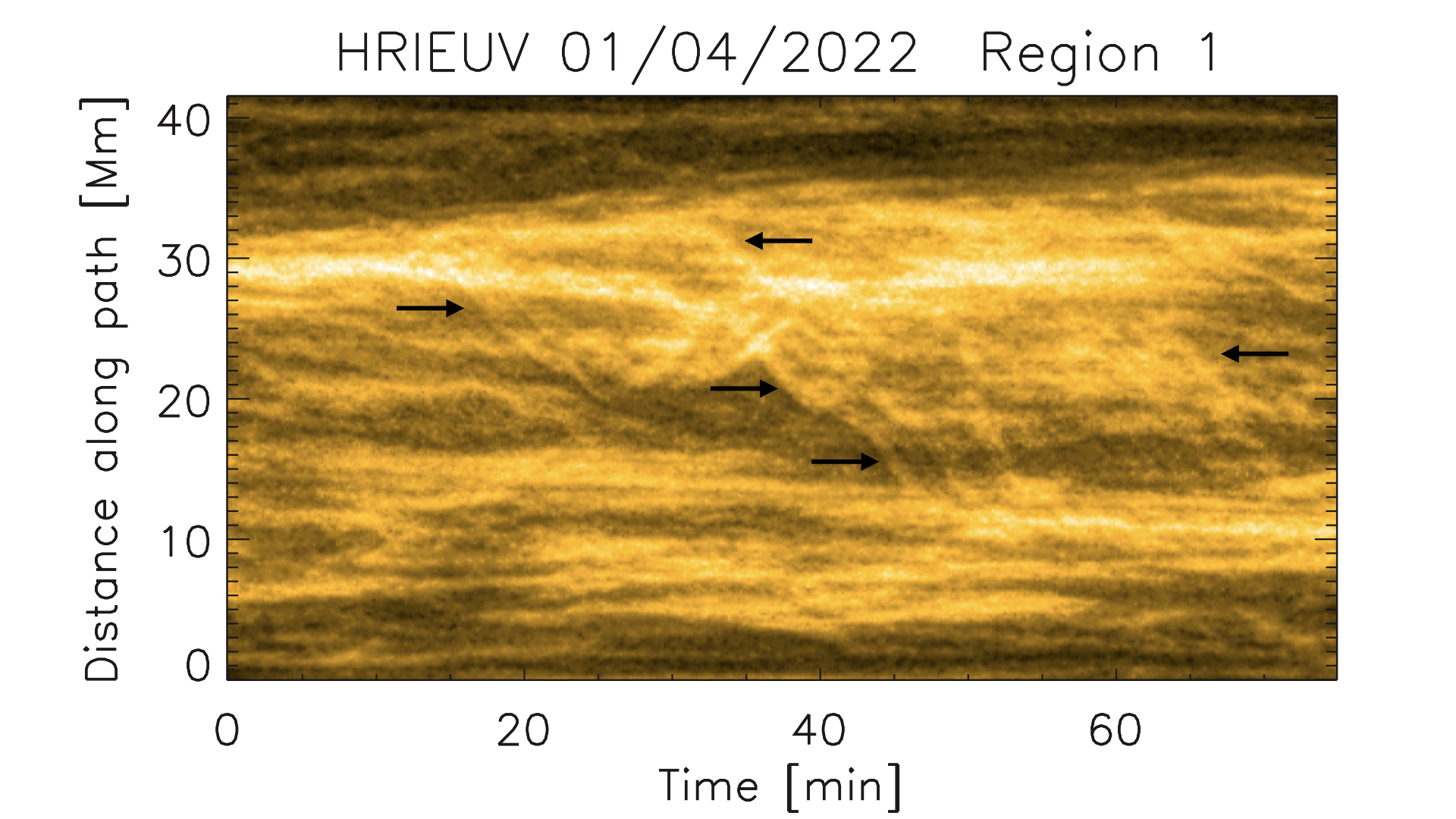}
    \caption{Time-distance diagram along the white dashed curve in Figure~\ref{fig:fullFOV} (right panel) for the April~1 \hrieuv observation, with distance `0' corresponding to the western end of the curve. The curve crosses a loop bundle at the apex, which is seen to undergo a large-scale reconfiguration (radially inward corresponding to shrinkage), as pointed by the arrows. {This time-distance diagram is made from images that have been processed with the wavelet-optimised whitening enhancement technique of \citet{Auchere2023}.}}
    \label{fig:reconf}
\end{figure}

\begin{figure}
    \centering
    \includegraphics[width=0.4\textwidth]{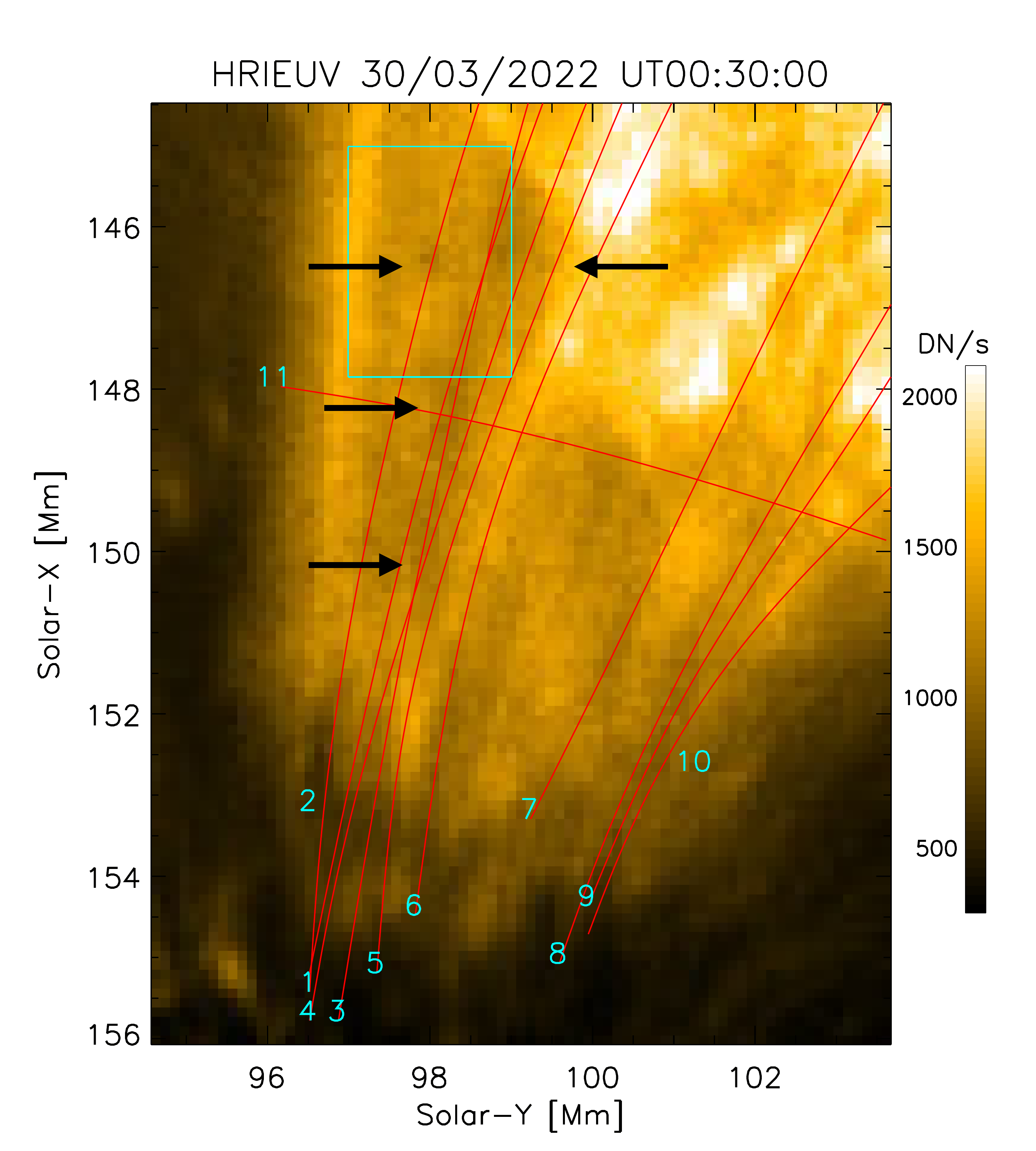}
    \caption{Close-up on the footpoint of the loop bundle where coronal rain is seen on March~30. The FOV corresponds to the white rectangle shown in Fig.~\ref{fig:fullFOV} (left), rotated by 90$^\circ$ so that the loop is orientated with the height of the page. The {red} and labelled curves vertically oriented denote several rain clump trajectories. The black arrows point to some of the clumps. The image corresponds to the average over 3 consecutive frames. The cyan rectangle in the figure corresponds to the FOV shown in Figure~\ref{fig:clump}. Path~11 is a transverse cut across the loop bundle. The accompanying animation runs from UT 00:23 to UT 00:37 and shows several coronal rain clumps in absorption falling towards the chromosphere (dark structure at bottom). Note the strong EUV variation associated with this event. The images composoing the movie are processed with the wavelet-optimized whitening enhancement technique of \citet{Auchere2023}. {The movie first runs without and then with the rain paths overlaid.} }
    \label{fig:March30all}
\end{figure}

\begin{figure}[h!]
    \centering
    \includegraphics[width=0.48\textwidth]{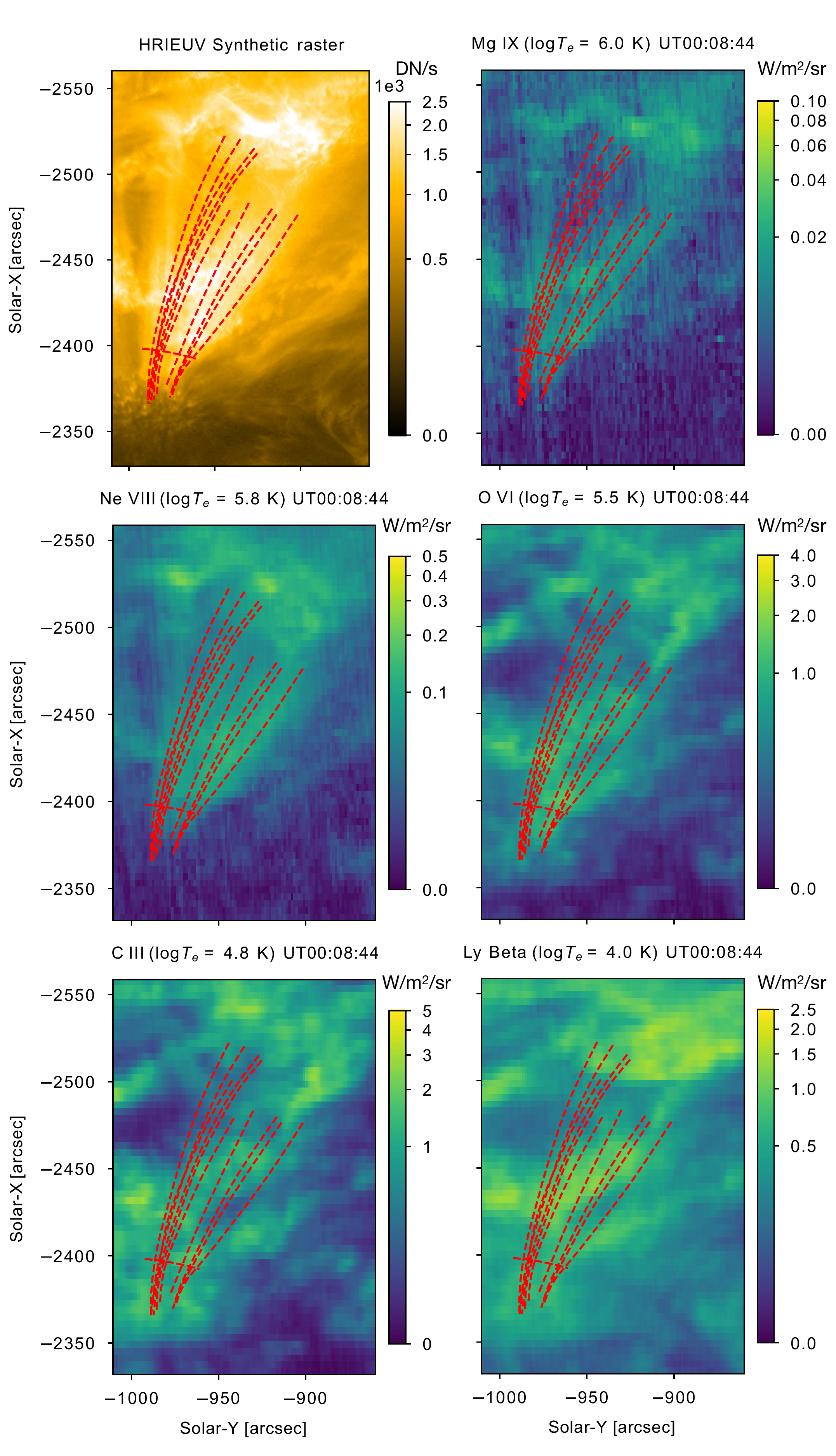}
    \caption{{SPICE multi-wavelength view on the loop bundle with rain seen on March~30. The FOV corresponds to the red rectangle shown in Figure~\ref{fig:spicefullFOV} (top) and the overlaid red curves mostly denote the rain paths seen with \hrieuv (see Figure~\ref{fig:March30all}). The \hrieuv panel corresponds to a synthetic raster matching the time of the SPICE raster (see text for details) but preserving the \hrieuv spatial resolution. Each SPICE panel shows the total intensity over a spectral line indicated in the subtitle, together with its temperature of maximum formation.}}
    \label{fig:reg1_mar30_spice}
\end{figure}

Coronal rain clumps and showers can be seen with a sharp eye without any image enhancement technique such as `WOW', but certainly become more discernible in the processed images. To analyse the morphology and dynamics of several of these events we start by tracking several rain clumps manually {in the image sequences}, with the help of CRISPEX (CRisp SPectral EXplorer), a widget-based tool programmed in the Interactive Data Language (IDL), which enables the easy browsing of the data, the determination of loop paths, extraction, and further analysis of space-time diagrams \citep{Vissers_Rouppe_2012ApJ...750...22V}.

\begin{figure*}
    \centering
    \includegraphics[width=0.92\textwidth]{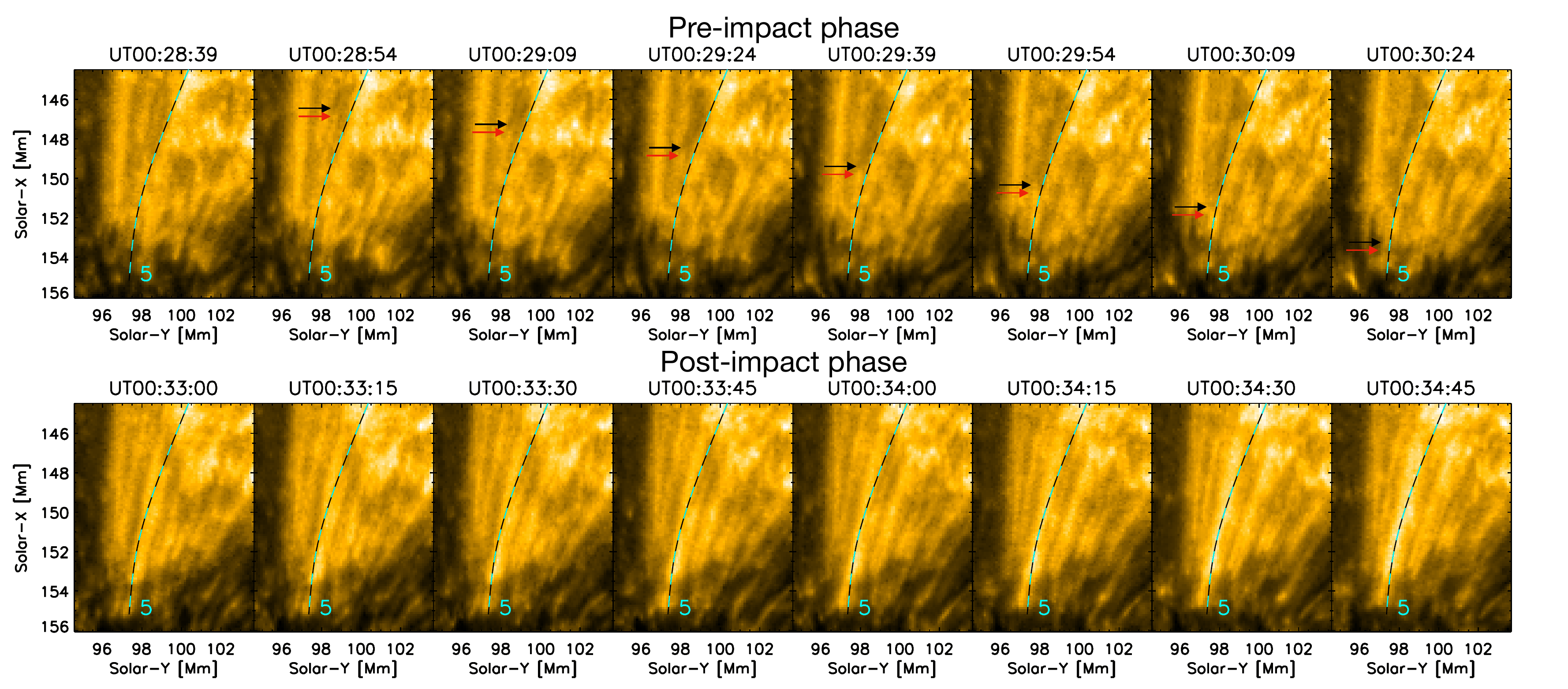}
    \caption{Sequence of 8 snapshots separated by 15~s showing the pre- (top) and post-impact (bottom) phases of a coronal rain shower from the event of March~30. The black arrows on the top panels show the head of a rain clump as it falls (note the bright tip indicated by red arrows, followed by a dark elongated structure). For reference, path~5 is shown in the Figure with a {cyan-black} dashed curve (same labelling as in Fig.~\ref{fig:March30all}). The lower panel shows a {bright upward propagating feature that we interpret as a combination of a rebound shock and flow produced by the impact of the rain shower. These images have been processed with the wavelet-optimized whitening enhancement technique of \citet{Auchere2023}}.}
    \label{fig:march30multi}
\end{figure*}

The determination of projected velocities in the plane-of-the-sky (POS) is done by calculating slopes in resulting rain clump tracks in the time-distance diagrams obtained from CRISPEX. These tracks define $(x,y,t)$ positions of the rain. The errors incurred in this process depend on the length of the tracks in the time-distance diagrams. By varying slightly the spatial and temporal end points of the tracks it is possible to estimate the errors in each calculation. {The availability of AIA in quadrature would allow to estimate the total speed of a rain event through 3D stereoscopy. However, this is beyond the scope of the present manuscript. }

{No rain is detected in the time-distance diagrams (through dark or bright propagating features) without prior check with the image sequences. This is necessary since several effects (such as lateral motions of loops) can produce apparent dark or bright propagating features in time-distance diagrams.}

Regarding the morphology of the rain, we are primarily interested in the observed widths. For this purpose, we apply the same semi-automatic routine as in previous work \citep[for more details see][]{Antolin_2015ApJ...806...81A}. For a given clump path and a given track in the time-distance diagram corresponding to this path, the routine takes a cut perpendicular to the trajectory at every $(x,y,t)$ location defined by the track and fits a single Gaussian over the interpolated profile. The Full-Width at Half-Maximum (FWHM) is then taken as the width of the clump at that location. Several checks are performed to reduce errors in the fitting. {We first calculate the average intensity profile of a feature for the time range in which it is seen (as defined by a slope in the time-distance diagram) and require a difference of at least 100~DN~s$^{-1}$ between its intensity (which can be in emission or absorption) and the background (which is also calculated over the same path but at times without rain, immediately before or after the rain episode). We then require the contrast between the feature's average intensity and the background to be at least 10~\%. Lastly, we also require that the Gaussian fit of the rain feature is good enough. The latter is ensured by requiring that the $\sigma-$error resulting from the fit is below 0.4 and that the total of the residuals fromt he fit is low, that is $\sum|I_{x,y}(x_{\perp},y_{\perp},t)-f(x_{\perp},y_{\perp},t)|<0.75$, where $(x_{\perp},y_{\perp})$ denotes the transverse cut to the path at point $(x,y)$, $I$ is the intensity along this cut, and $f$ denotes the Gaussian fit to $I$. Visual inspection of the fittings indicate that these values ensure an accurate calculation of the rain width while avoiding artefacts due to cosmic rays or other features.} The width of a clump corresponds to the average over all measurements performed for the track in the time-distance diagram corresponding to that clump. A measure of the error in the width calculation is given by the standard deviation over all the measurements for a given track. 

\section{Results}
\label{sec:results}
\subsection{Coronal loop bundles}

We analyse several coronal rain events observed with \hrieuv. On March~30, the event occurs in a coronal loop bundle belonging to AR NOAA 12974 in the southern hemisphere, which is on-disk but near the limb as seen in Figure~\ref{fig:fullFOV} (left panel). The rain is seen to fall onto the leading polarity, onto a region with abundance of dark jet-like structures fanning outwards. No sunspot is seen in the vicinity, suggesting that this corresponds to a decaying AR. Several rain clumps were followed, whose trajectories can be seen in the figure. 

On April~1, \hrieuv observed a wide region of activity composed of 2 ARs, NOAA 12975 and 12976, in the northern hemisphere, also not far from the limb as seen in Figure~\ref{fig:fullFOV} (right panel). Coronal rain is seen much more pervasively compared to the other \hrieuv observation. In particular, we analysed {four} different regions, labelled in the panel, where several coronal rain events can be seen, as indicated in the figure. 

Region 1 focuses on the West footpoint of a very large coronal loop bundle that is seen to undergo a wide-scale reconfiguration. Some of this reconfiguration is also associated with a bundle of loops that are rooted closer to the lower-right corner of box-2 in the right panel of Figure\ref{fig:fullFOV}. To see this more clearly we take a transverse cut at the apex of the loop bundle, as shown by the dashed curve in the right panel, and plot the time-distance diagram in Figure~\ref{fig:reconf} (distance `0' in the diagram denotes the western end of the dashed curve, as seen in Figure~\ref{fig:fullFOV}). In the diagram we indicate with arrows several instances of large-scale motions of individual coronal strands directed radially inwards, suggesting a shrinkage. This process is also accompanied by large-amplitude transverse oscillations that can also be identified in the figure. At the same time, large amounts of coronal rain are observed to fall along the leg captured by Region 1.

Region 2 on April~1 focuses on a region with different polarity compared to Region 1, where the other footpoint of the loop bundle appears to be rooted. Region 2 shows stronger activity (pores, moss, light-walls and jets) and a more complex magnetic topology as discussed in the accompanying paper \citep{Berghmans_etal_subm_2023}. Between regions 1 and 2, a highly twisted filament is seen, whose eruption was observed by EUI and SPICE on the following day, and is discussed in \citep{Berghmans_etal_subm_2023}. 

Region 3 on April~1 is located at the North-West of the AR. A different bundle of loops is seen, with footpoints close to those in Region 2, and therefore also in a high-activity region. 

{Region~4 on April~1 is located on the East limb and captures part of a long loop that is rooted in the trailing AR (NOAA~12976). }

\subsection{{March 30 loop bundle as seen with SPICE}}
\begin{figure}
    \centering
    \includegraphics[width=0.4\textwidth]{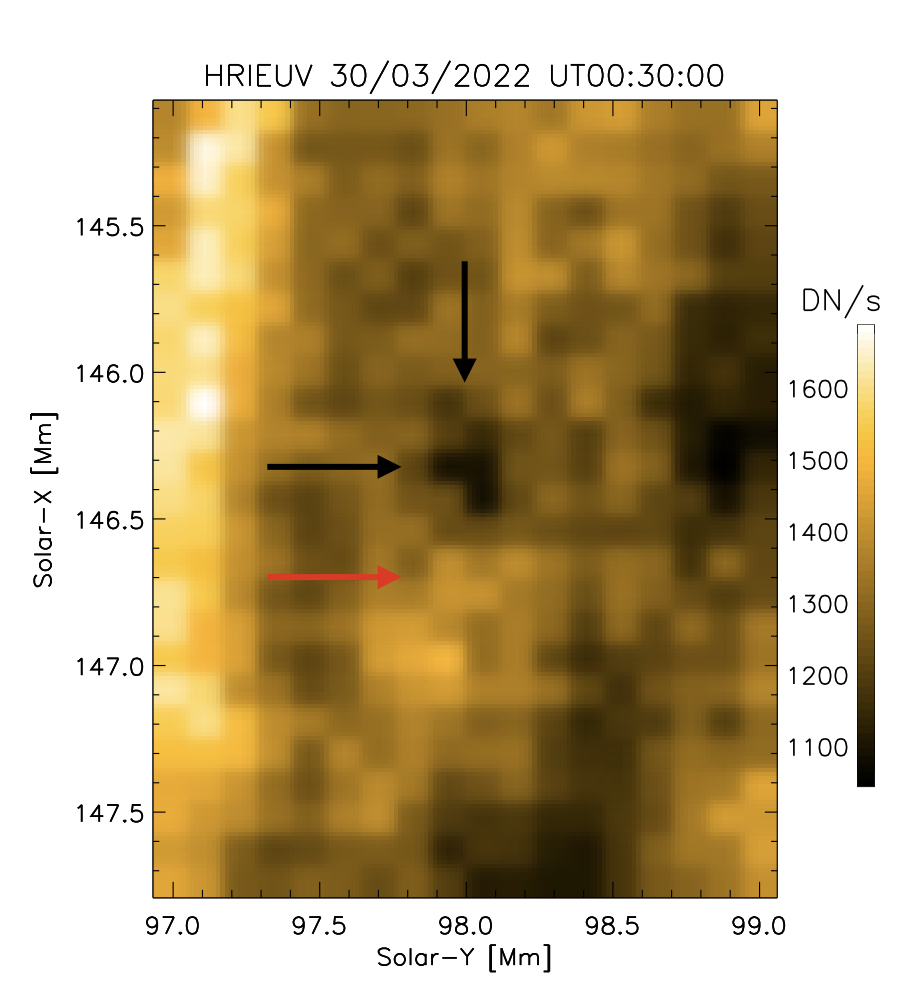}
    \caption{Close-up of the sub-FOV indicated by the white rectangle in Figure~\ref{fig:March30all}. The dark feature indicated by the arrows corresponds to the EUV absorption produced by a rain clump. {We interpret} the bright feature below (downstream of) the rain clump indicated by the {red} arrow as compression {and heating} produced by the rain clump as it falls. }
    \label{fig:clump}
\end{figure}

\begin{figure}
    \centering
    \includegraphics[width=0.48\textwidth]{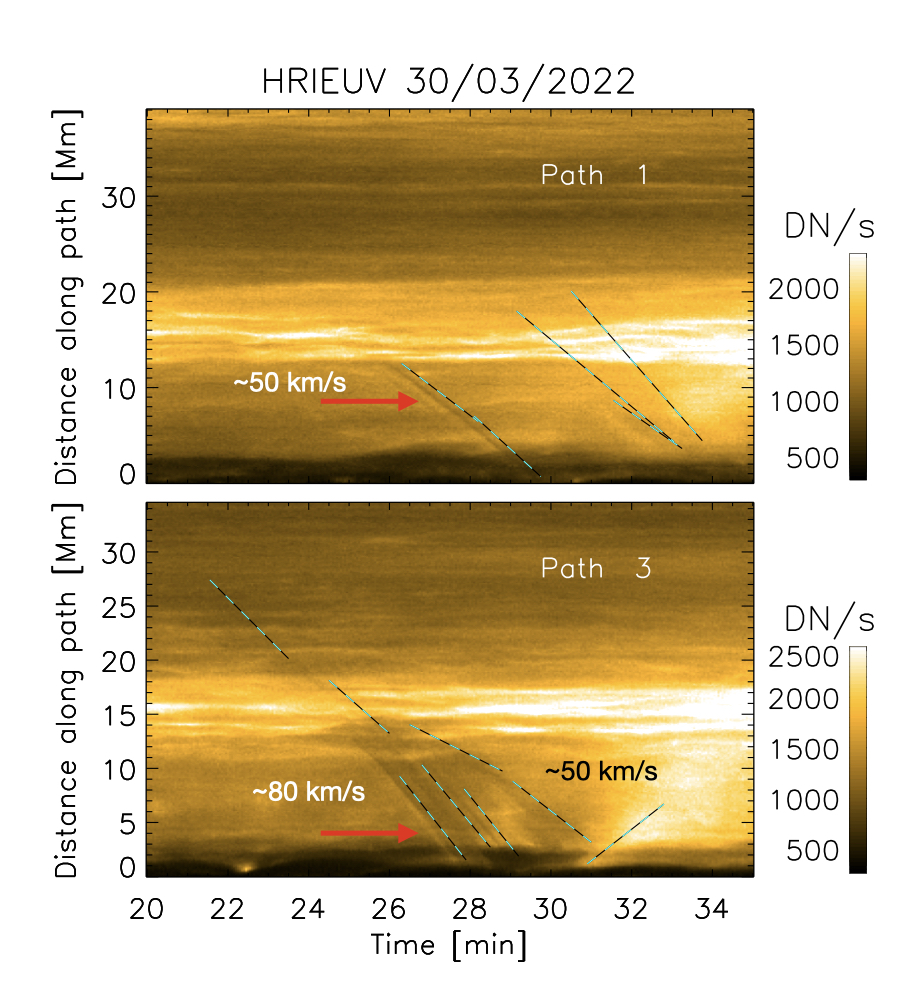}
    \caption{Time-distance diagrams along paths {1} and 3 indicated in Figure~\ref{fig:March30all}. Distance increases with height along the paths. The dark tracks traced by {cyan-black} dashed lines correspond to EUV absorption from falling rain clumps. {The dashed lines are offset by a fraction of a minute in order to better see the rain features}. An estimate of the projected velocity is given for the closest slope to each value. Note the bright tracks indicated by the {red} arrows immediately beneath the first dark tracks in each diagram. {We interpret} this feature as compression {and heating} from the clump. {We interpret} the upward bright and diffuse slope at the end of the time sequence ($t\approx30-32~$min) in Path~3 as a {combination of a} rebound shock and flow produced by the rain impact on the chromosphere. }
    \label{fig:td12march30}
\end{figure}
{Several of the loop bundles seen in \hrieuv can also be seen in SPICE in the \ion{Ne}{VIII} line, as expected from the similar formation temperature. In particular, the loop bundle where rain is observed with \hrieuv can be seen with SPICE. In Figure~\ref{fig:reg1_mar30_spice} we show a close-up on the loop bundle, which includes the FOV shown in Figure~\ref{fig:March30all}. The raster through this region occurred at UT00:08:44, which is roughly $20-25$~min prior to the main rain shower. The loop can be seen in the \ion{Mg}{IX} coronal line and in the upper transition region lines (\ion{Ne}{VIII} and \ion{O}{VI}), suggesting that it is in a state of cooling. However, emission in the lower temperature lines seems to be dominated by the background and we cannot detect any clear chromospheric emission from coronal rain. This could also be due to timing, since the main rain shower happens after the raster.}

{SPICE also executed high cadence 3-step rasters at various times during the \hrieuv observing window, including the rain shower time. However, the slit crosses higher up along the loop at $X\approx-2500\arcsec$ (barely crossing a few of the longer rain paths), coinciding with a strong background emission. We could not find any clear rain signatures in these fast rasters.}

\subsection{Two-pixel wide coronal rain clumps in absorption, and downstream compression and heating}
In Figure~\ref{fig:March30all} we show several coronal rain paths identified for the event of March~30. The coronal rain clumps can be seen in the figure and corresponding animation as dark features, produced by EUV absorption from neutral hydrogen, neutral and singly ionised helium \citep{Anzer_Heinzel_2005ApJ...622..714A}. In Figure~\ref{fig:march30multi} (top panel) we show several snapshots separated by 15~s each, where large and small absorption features can be seen falling. 

For better visualisation of the fine-scale structure we show in Figure~\ref{fig:clump} a sub-FOV of only $2\text{~Mm}\times3$~Mm centred on a dark absorption feature produced by a clump (white rectangle in Figure~\ref{fig:March30all}). Note that it is barely 2 pixels wide (i.e. $\approx240$~km), and is therefore the highest resolution of a rain clump in EUV absorption to date. Another interesting feature is the bright region downstream of the clump. The animation shows that this bright feature is always beneath the dark absorption feature from the clump. Similar features can be seen for other clumps, some appearing only moments prior to impact in the chromosphere. We interpret this feature as compression and heating produced by falling individual clumps.

To see the EUV variation produced by the rain more clearly, we plot in Figure~\ref{fig:td12march30} the time-distance diagrams corresponding to paths 2 and 3, shown in Figure~\ref{fig:March30all}. The dark tracks in this figure correspond to the EUV absorption produced by the rain as it falls. The observed slopes indicate average speeds of $70-80~$km~s$^{-1}$. Immediately below the first dark track, a parallel bright slanted track can be seen, corresponding to the compression and heating downstream of the rain clump. Note that several such bright tracks can be seen, but are all very small with lengths under 1~Mm \ (vertical axis in time-distance diagram).

Although we do not calculate accurately the lengths of the clumps in this work, a rough estimate is given by the size of the dark tracks (vertical distance) in the time-distance diagrams of Figure~\ref{fig:td12march30}, which can be seen to have $1-5~$Mm lengths.

\begin{figure}
    \centering
    \includegraphics[width=0.45\textwidth]{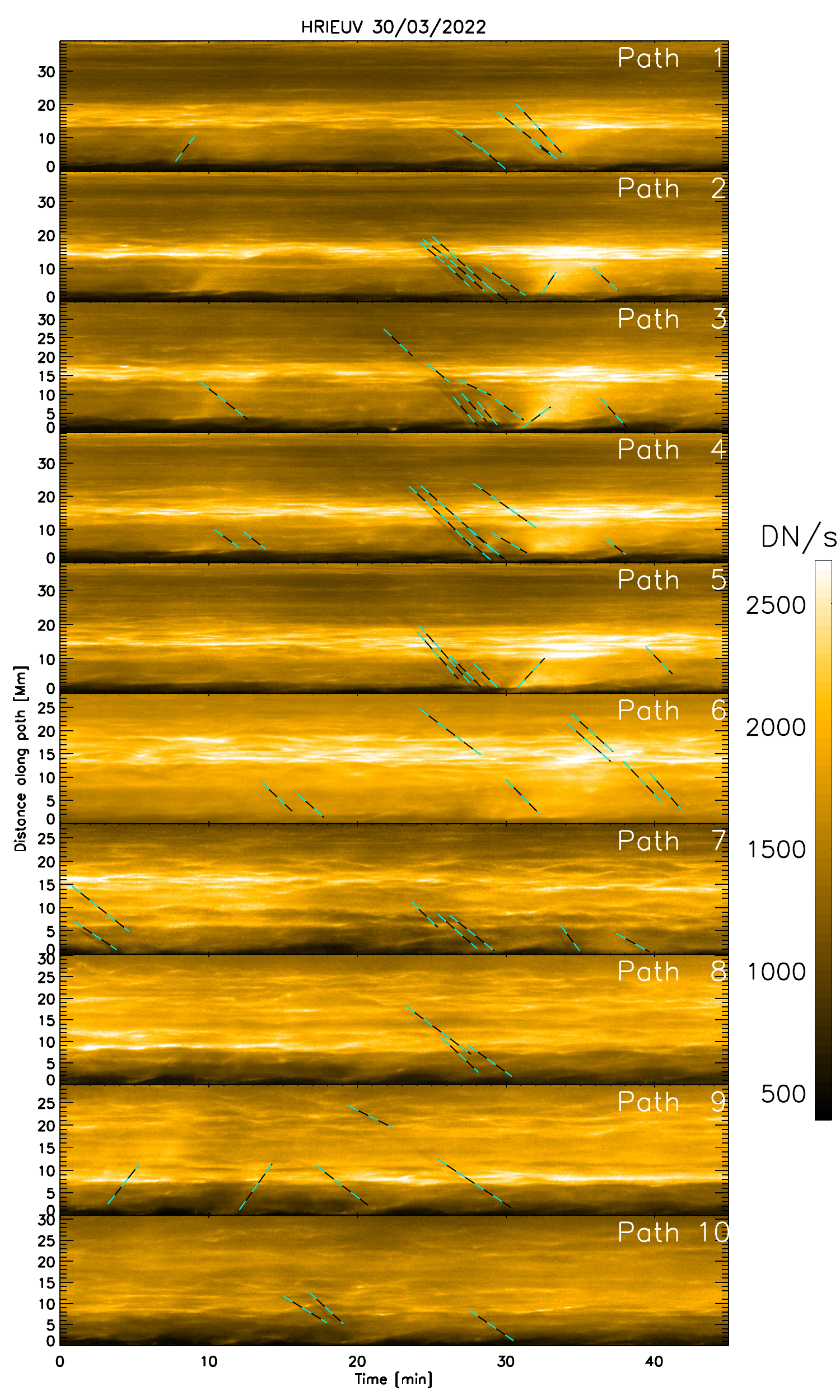}
    \caption{Time-distance diagrams for all coronal rain paths shown in the left panel of Figure~\ref{fig:fullFOV}. The {cyan-black} dashed lines with negative slopes show {some of} the tracks produced by the rain clumps as they fall. {The lines are offset in time by 1~minute to better see the rain features.} Note the extent of the rain shower across all paths. Upward propagating features (positive slopes) can also be seen, particularly at the end of the rain shower ($t\approx30~$min). Zero distance corresponds to the {footpoint of the loop} (seen in Figure~\ref{fig:March30all}).}
    \label{fig:tdmarch30}
\end{figure}

\subsection{Rebound shock and flow}

Figure~\ref{fig:march30multi} (lower panel) shows another interesting feature of the event from March~30, also seen in the animation of Figure~\ref{fig:March30all}. Following the impact on the chromosphere of the rain shower a bright upward propagating feature can be seen. We interpret this as a rebound shock and flow produced  by the rain impact. To the best of our knowledge, this is the first time that such feature is observed, despite being predicted by every numerical simulation of coronal rain \citep[e.g.][]{Muller_2003AA...411..605M, Fang_2015ApJ...807..142F}. This feature can also be seen in each time-distance diagram of Figure~\ref{fig:td12march30} as a bright upward slope just after the end of the rain shower. The slope indicates speeds between $50-130$~km~s$^{-1}$. The lower end of this velocity range matches the expected value for upward flows, while the upper end matches the tube speed for a plasma temperature of $10^{5.8}$~K, which is close to the temperature at the peak of the \ion{Fe}{X}~174~\AA~ formation. This is also supported by numerical simulations \citep{Antolin_2022ApJ...926L..29A}. 

\subsection{Extent of the rain shower}

To examine the extent of the rain shower we plot in Figure~\ref{fig:tdmarch30} the time-distance diagrams corresponding to all the paths shown in Figure~\ref{fig:March30all}. Several clump tracks are shown by dashed lines. We can see that the main rain shower occurs in the time range $t=20-30$~min. Although it can best be seen in paths 2 to 5 we can still traces of it across all the paths. This indicates that the extent of the rain shower across the loop bundle is at least 15~Mm in the POS, and possibly larger given the observed expansion of the loop bundle seen in Figure~\ref{fig:fullFOV} (left panel). {This is supported by the SPICE observations in Figure~\ref{fig:reg1_mar30_spice}, which show cool transition region emission over a width larger than that set by the rain clumps detected by \hrieuv.} Along the loop, the clumps can be tracked for up to 25~Mm. Note that many clumps are only clearly visible in the last 10 Mm, suggesting that the catastrophic cooling is non-uniform, with accelerated cooling rates down to chromospheric temperatures being more confined in the transverse direction. This effect may also be due to the line-of-sight, as shown in section~\ref{sec:aia}.

Figure~\ref{fig:tdmarch30} also shows that the rebound shock and flow occurs across a wide expanse, but appears more concentrated than the rain shower and can only be clearly seen in paths $1-6$.

\subsection{Region 1 of April~1 - a large coronal rain event}
\begin{figure}
    \centering
    \includegraphics[width=0.5\textwidth]{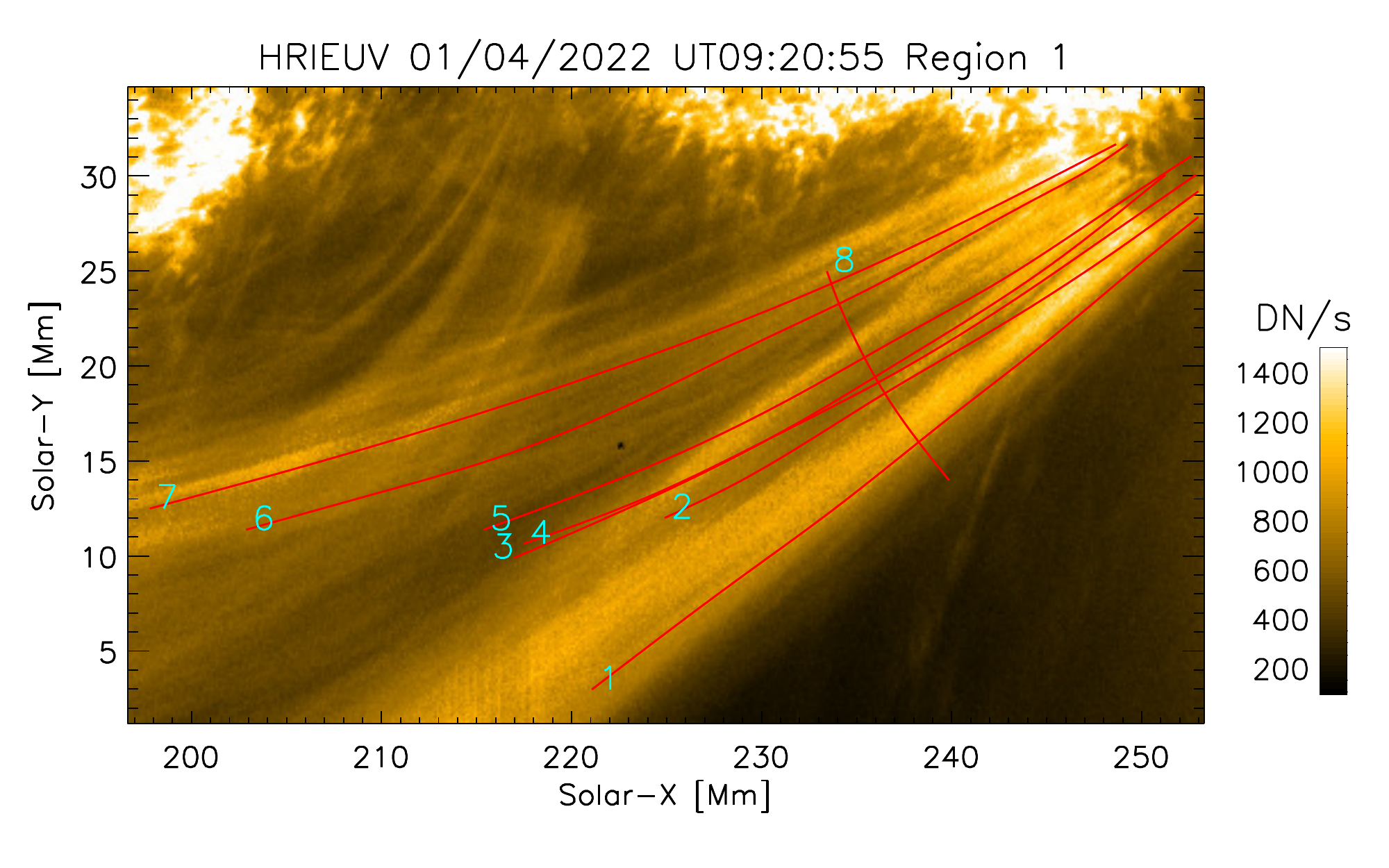}
    \caption{Close-up on the footpoint of a large coronal loop bundle observed on 2022 April~1 by \hrieuv. The FOV corresponds to that of Region~1 indicated by the white rectangle in Figure~\ref{fig:fullFOV}, {right} panel. Except for path 8 (which crosses the loop bundle), the red paths denote several rain paths. An animation of this figure is available, whose images have been processed with the wavelet-optimized whitening enhancement technique of \citet{Auchere2023}. It runs from UT09:19 to {UT10:34} and shows several rain clumps falling towards the chromosphere. {The movie first runs without and then with the rain paths overlaid.} }
    \label{fig:april1_reg1}
\end{figure}
We now turn our attention to some of the coronal rain events seen on the April~1 dataset (see Figure~\ref{fig:fullFOV}, right panel).

In Figure~\ref{fig:april1_reg1} we show the close-up on the footpoint of the large coronal loop bundle undergoing a global change (akin to shrinkage), denoted as Region~1  in Figure~\ref{fig:fullFOV} (right). We follow several rain clumps and plot in Figure~\ref{fig:april1st_reg1_td} the corresponding time-distance diagrams. A main shower event is seen in the time range $t=40-70$~min particularly along paths $2-4$, but signatures of another rain shower are also seen at the beginning ($t<20$~min), particularly along paths $6-7$. Also in this case, a combination of neighboring dark and bright paths can be seen. Although not as clear as for the March~30 event, some of the bright tracks may correspond to the downstream compression and heating, especially those that immediately precede the absorption feature. Note that some of these tracks only appear bright, for example the last track of Path~6. Furthermore, upward propagating features can also be observed, some of which do seem to appear just after rain impact. The observed morphology and speeds are similar to those seen for the March~30 event, all of which are calculated and presented in Section~\ref{sec:stats}.

The rain shower seen in Region~1 appears to be far wider in extent than that of March~30. As seen in Figure~\ref{fig:april1_reg1}, the transverse length across which the clumps are seen is at least 20~Mm, but certainly greater given the observed expansion. Furthermore, the clumps can be followed for longer lengths along the loop, with some being tracked for over 40~Mm. This suggests catastrophic cooling down to chromospheric temperatures over a larger coronal volume, {which is supported by SPICE observations}.

\begin{figure}
    \centering
    \includegraphics[width=0.5\textwidth]{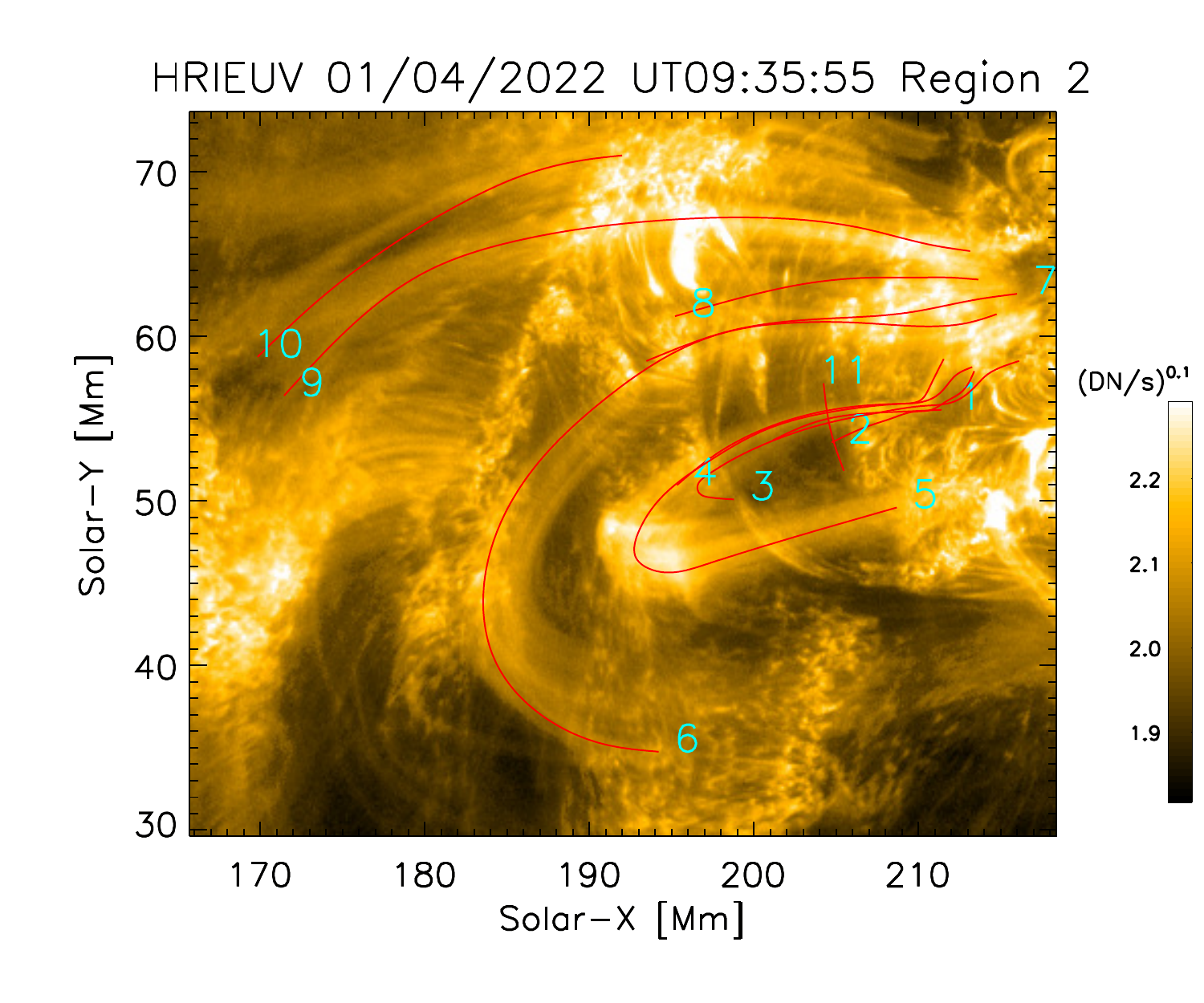}
    \caption{FOV corresponding to Region~2, shown in Figure~\ref{fig:fullFOV} (right panel). The {red} curves correspond to paths of some of the observed coronal rain clumps. {Note that intensities are scaled with a power of 0.1 to see a larger range of variations. An animation of this figure is available, whose images have been processed with the wavelet-optimized whitening enhancement technique of \citet{Auchere2023}. It runs from UT09:19 to UT10:34 and shows several rain clumps falling towards the chromosphere. The movie first runs without and then with the rain paths overlaid.}}
    \label{fig:april1_reg2}
\end{figure}

\begin{figure}
    \centering
    \includegraphics[width=0.5\textwidth]{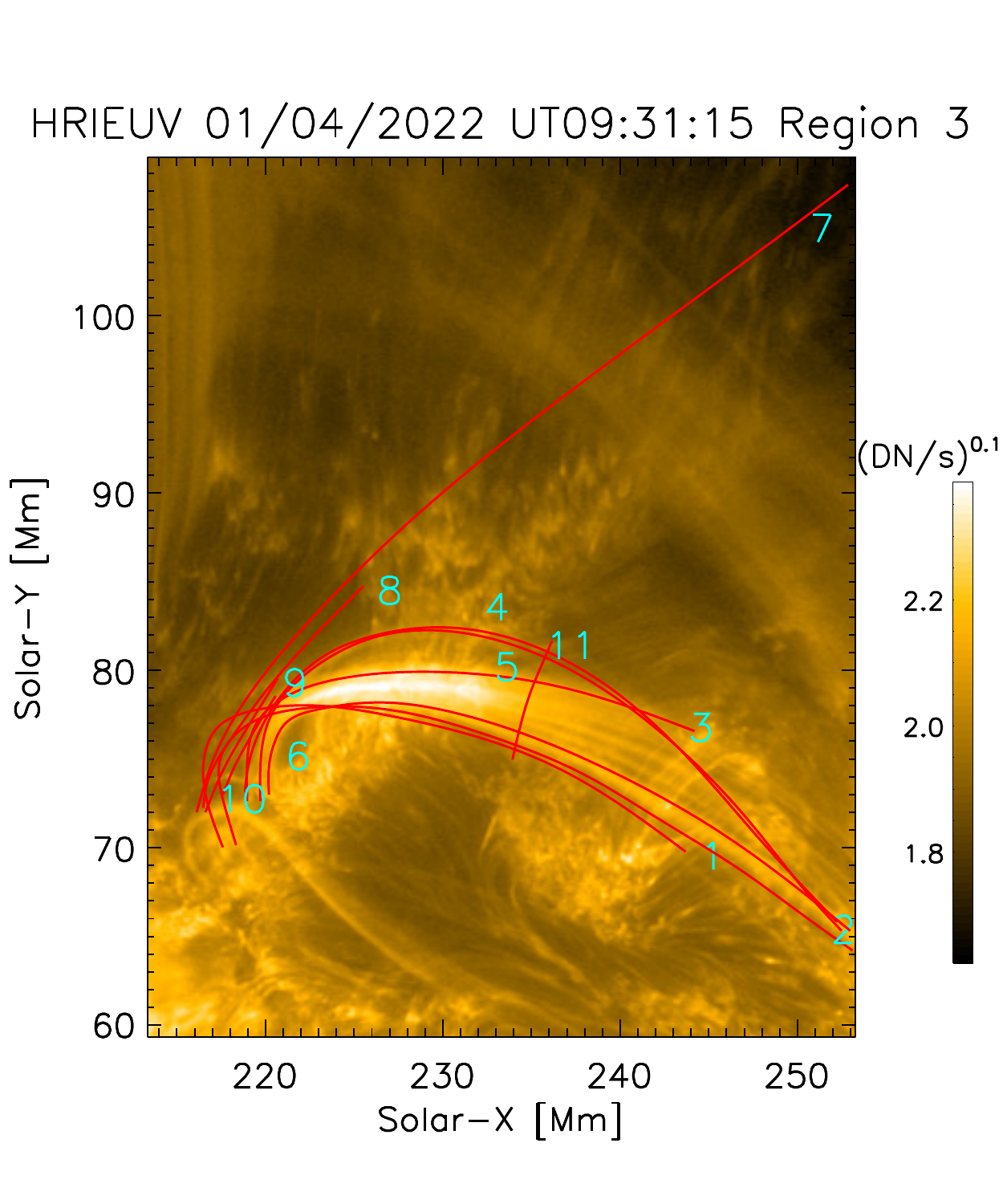}
    \caption{FOV of Region 3 shown in Figure~\ref{fig:fullFOV} (right panel). The {red} curves correspond to paths of some of the observed coronal rain clumps. {Paths 1 to 6 belong to the bright loop in the lower half of the image, while paths 7 to 10 belong to a very long loop going towards the top right corner of the image. Paths 9 and 10 are very short and parallel to each other. Note that intensities are scaled with a power of 0.1 to see a larger range of variations. An animation of this figure is available, whose images have been processed with the wavelet-optimized whitening enhancement technique of \citet{Auchere2023}. It runs from UT09:19 to UT10:34 and shows several brightenings followed by mostly bright rain clumps falling towards the chromosphere. The movie first runs without and then with the rain paths overlaid.}}
    \label{fig:april1_reg3}
\end{figure}
\subsection{Region 2 of April~1 - null-point topology at footpoint}\label{sec:nullpoint}

Region~2 on April~1 shows several coronal rain events belonging to different regions in the AR. In Figure~\ref{fig:april1_reg2} we show the FOV corresponding to Region~2, as shown in the full FOV of Figure~\ref{fig:fullFOV} (right panel), with several rain clump paths overlaid. The corresponding time-distance diagrams for these paths are shown in Figure~\ref{fig:april1_reg2_td}.

The most interesting rain shower is tracked by paths $1-5$ in these figures. First, a loop bundle appears, with the intensity all along the loop increasing in a uniform manner (but particularly at the apex) around UT 09:30 (about 10~min from the start of the observation). This brightening can be best seen along path~5 (white arrows in  Figure~\ref{fig:april1_reg2_td}). The intensity uniformly decreases along the loop over the next 10~min, after which, the first rain clumps appear, with the bulk of the rain seen after 20~min from the first intensity increase. Following the rain shower impact, the intensity increases strongly at the footpoint with some signatures of upward propagating features (red arrow in Figure~\ref{fig:april1_reg2_td})). As the rain falls, it is observed to strongly deviate from its path and spread in different directions, reminiscent of a null-point topology structure at the footpoint.

\begin{figure*}
    \centering
    \includegraphics[width=1\textwidth]{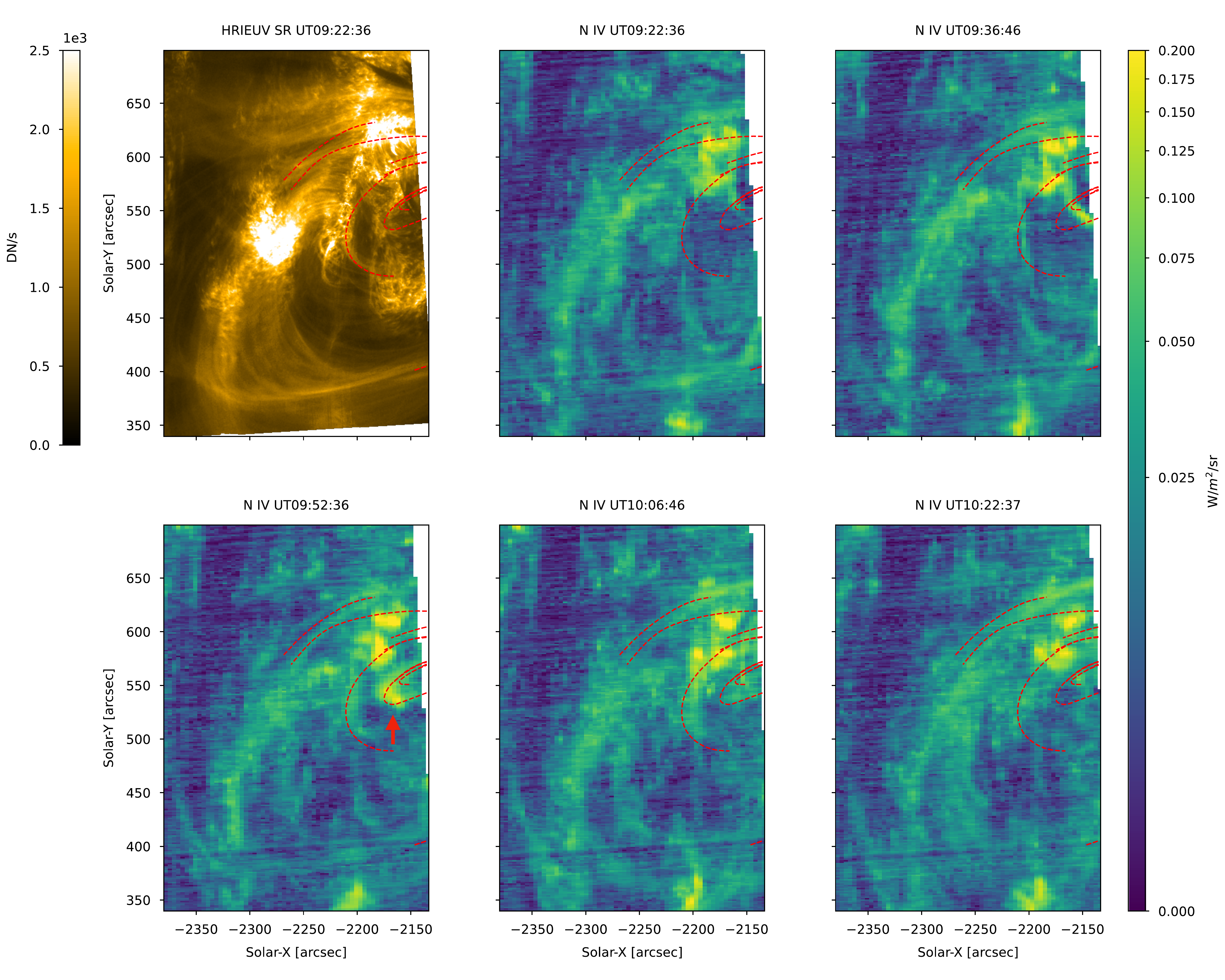}
    \caption{{SPICE rasters in the \ion{N}{IV} line ($\log T = 5.2$~K) over a region that overlaps with Regions 1 and 2 on April~1. The FOV corresponds to the red rectangle to the East shown in Figure~\ref{fig:spicefullFOV} (bottom) and the overlaid red curves  denote some of the rain paths seen with \hrieuv (see Figures~\ref{fig:april1_reg1} and \ref{fig:april1_reg2}). The \hrieuv panel corresponds to a synthetic raster matching the time of the first SPICE raster (see text for details) but preserving the \hrieuv spatial resolution. The SPICE panels show the total intensity integrated over the \ion{N}{IV} spectral line for each raster (time shown in subtitle). The brightening indicated by the red arrow in the UT09:52:36 panel corresponds to cooling through the passband.}}
    \label{fig:spice_april1_regs1-2}
\end{figure*}

The paths $6-7$ and possibly path~8 seem to correspond to another loop bundle that also experiences a similar uniform and global loop brightening as described above. In this case, the loop bundle brightens at the very start of the observation, and disappears after almost 1~hr. The rain is seen roughly 50~min after the start of the brightening, but is much fainter in terms of EUV absorption than for the previous case. This loop is rooted close to a pore, and periodic upward/downward ballistic motion is seen (particularly for paths~7 and 8), characteristic of Type~1 spicules \citep{Beckers_1964PhDT........83B} or the EUV counterparts of AR dynamic fibrils \citep[][]{Mandal2023}.  

Path $9$ (and possibly path~8 as well) may correspond to the other footpoint of the large loop bundle of Region~1. However, contrary to the other footpoint, the amount of rain that can be observed falling into this footpoint is minimal and is further very faint.

The last path~10 may also belong to the large loop bundle of Region~1, and the location of its footpoint is uncertain. Instead of the region where the previous paths are rooted, path~10 appears to be rooted in a moss. Minimal rain events are seen in this case. 

All rain tracks observed in Region~2 show mostly EUV absorption, with little EUV emission of the kind described earlier associated with compression. Also, most tracks are relatively short ($10-20~$Mm) when compared to Region~1, which may be due to the different inclination of the loops relative to the LOS. 

\begin{figure}
    \centering
    \includegraphics[width=0.5\textwidth]{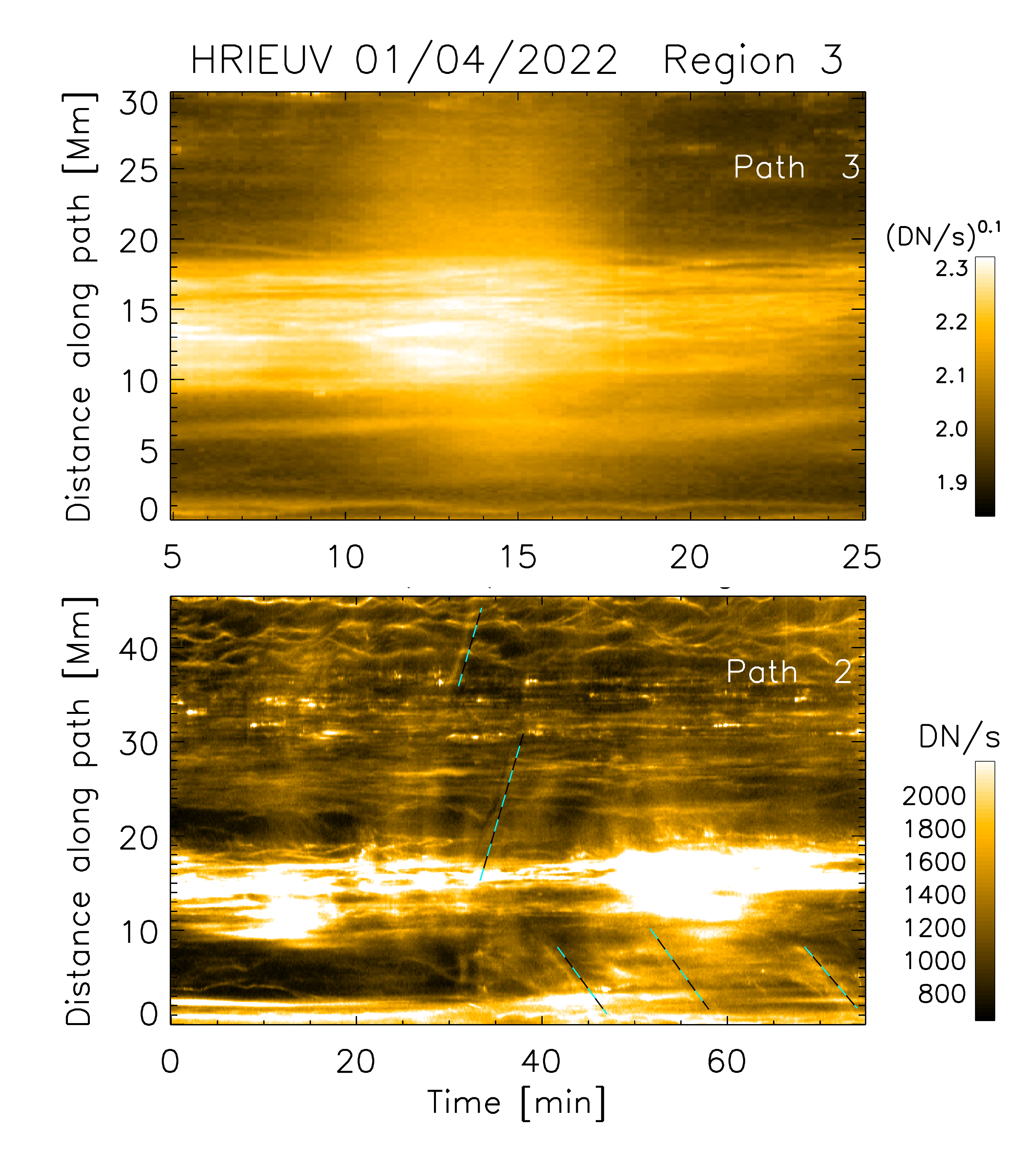}
    \caption{{Top: Loop brightening prior to a coronal rain event. This time-distance diagram (close up on the time range $t=[5,25]~$min) corresponds to path~3 in Region 3 on April~1 shown in Figure~\ref{fig:april1_reg3_td}. Note that the intensities have been scaled to the power of 0.1 to better see the large intensity variation. Bottom: Time-distance along path~2 in the same lop bundle. The white dashed lines correspond to dark and bright tracks from coronal rain.}}
    \label{fig:april1_reg3_sngl_3-2}
\end{figure}

\begin{figure}
    \centering
    \includegraphics[width=0.5\textwidth]{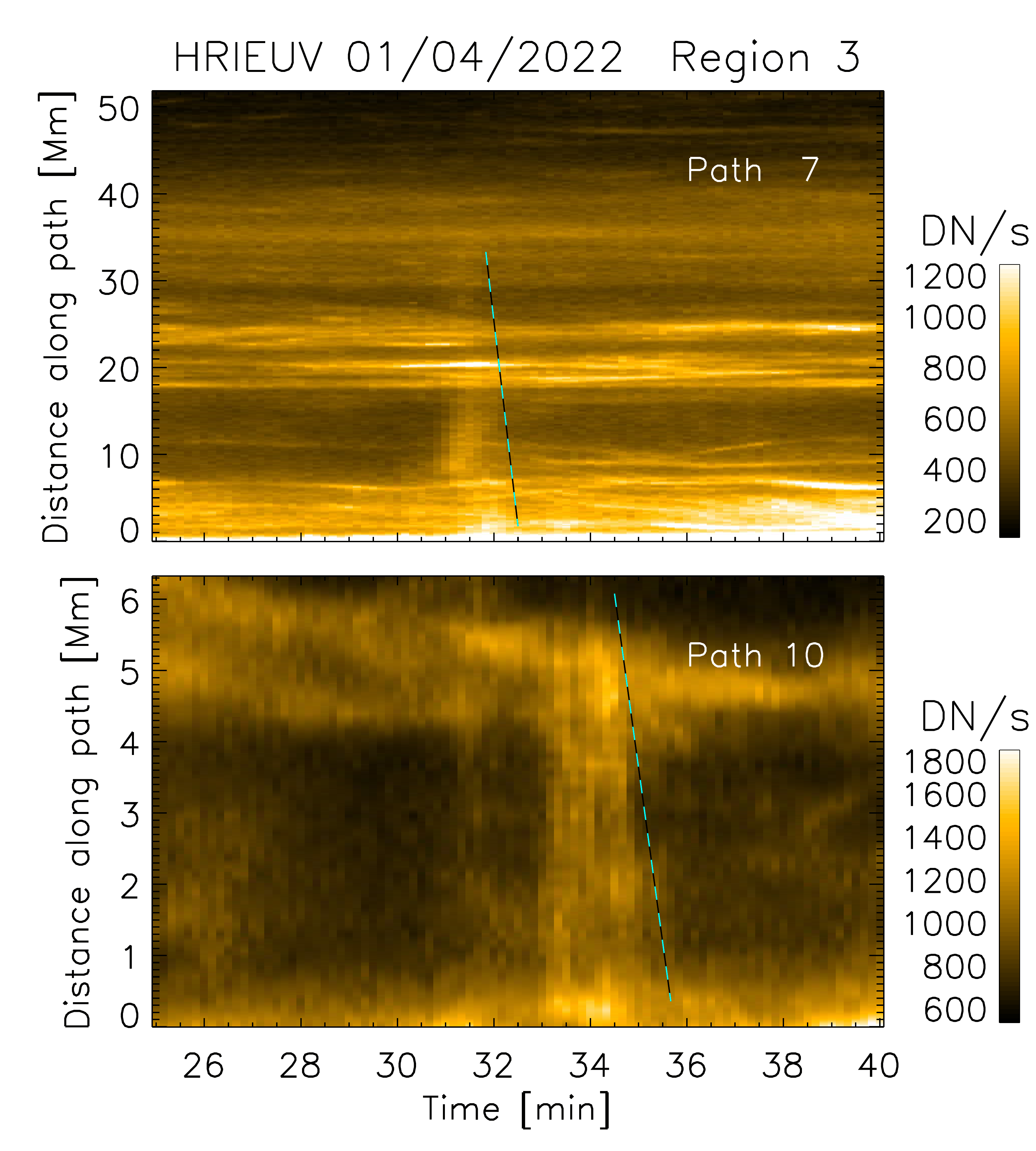}
    \caption{Close-up on the time range $t=[25,45]~$min in the time-distance diagram of paths 7 (top) and 10 (bottom) shown in Figure~\ref{fig:april1_reg3_td}. Note that the beginning of the brightening at $t\approx33$~min {starts} essentially {at the same time} along the path. The {cyan-black} dashed lines trace the outer envelope of the brightening indicating a speed of $\approx150~$km~s$^{-1}$.}
    \label{fig:april1_reg3_sngl10}
\end{figure}

\subsection{{SPICE view on Regions 1 and 2}}\label{sec:spice_april1_reg1_2}
{The SPICE rasters managed to capture part of the regions of interest observed with \hrieuv on April~1. This includes the top part of Region~1 and most of Region~2 (see Figure~\ref{fig:spicefullFOV}, bottom), which mostly correspond to the apexes of the loops with coronal rain. These loops are barely visible in the \ion{Mg}{IX} coronal line (Figure~\ref{fig:spice_april1_regs1-2_mg9}) but their filamentary structure are well seen in the upper transition region lines (\ion{Ne}{VIII}, \ion{O}{VI} in Figures~\ref{fig:spice_april1_regs1-2_ne8} and \ref{fig:spice_april1_regs1-2_o6}). Furthermore, emission similar in morphology and strong variability can also be seen in the lower transition region \ion{N}{IV} line (Figure~\ref{fig:spice_april1_regs1-2}). For instance, the large-scale brightening in the loop that corresponds to paths 1 to 5 in Region 2 (described in section~\ref{sec:nullpoint}) can be seen to peak in \ion{Mg}{IX} at UT09:36 (red arrow in the corresponding Figure), while in \ion{Ne}{VIII}, \ion{O}{VI} and \ion{N}{IV} the brightening peaks at UT09:52, and in \ion{C}{III} the peak is sseen at UT10:06 (with some hints of increased brightness in Lyman-$\beta$ as well but inconclusive due to the strong background). This strongly supports our interpretation of cooling through the passbands of \hrieuv and SPICE and subsequent appearance of rain in EUV absorption.}

\subsection{Region 3 of April~1 - {Localised and large-scale flash-like} EUV emission from rain}

Figure~\ref{fig:april1_reg3} shows Region 3, as denoted in the full FOV of Figure~\ref{fig:fullFOV} (right panel), with several rain paths overlaid. Most of the loops analysed in Region~3 show a very different inclination with respect to the LOS when compared to the previous cases. In this case, the LOS appears much less inclined with the loop plane, leading to a top view of the loop rather than a sideways view. Consequently, while the footpoint legs appear very short, the apexes appear long in the projected view.

Paths $1-6$ appear to correspond to the same loop bundle, although paths $4-5$ show slightly different trajectories relative to paths $1-3$, which may be indicative of braiding. This is further supported by the AIA~observations of the same event, as discussed in Section~\ref{sec:aia}. As for the loops in Region~2, a strong EUV enhancement uniformly along several coronal strands composing the loop bundle are seen roughly 10~min from the start of the observation. The coronal strands appear extremely thin, with sub-arcsecond widths {(see section \ref{sec:strands})}. This brightening can be clearly seen in the time-distance diagrams shown in Figure~\ref{fig:april1_reg3_td}. Most of the coronal strands disappear 20~min later. Both dark and bright tracks can be seen in most time-distance diagrams, indicative of flows in both directions (towards both footpoints). Some appear at the start of the global intensity enhancement and others appear $20-60$~min after. Note that despite the very close proximity of paths~1 and 2 and paths~4 and 5, they show different (dim) features in their evolution. AIA~304 confirms the presence of rain along this loop bundle (Section~\ref{sec:aia}).

Path~3 corresponds to one of the best visible coronal strands. As shown in the time-distance diagram, no clear bright or dark track can be seen. We select this path to more accurately investigate the uniform global intensity enhancement common to many of the strands in the loop bundle. In Figure~\ref{fig:april1_reg3_sngl_3-2} (top) we show the part of the time-distance diagram corresponding to the intensity enhancement for path~3. Compared to the background, the enhancement appears diffuse and seems to start close to the apex and propagates towards both footpoints in a couple of minutes. Overall, this global intensity enhancement over the strand lasts $\approx8~$min. In Figure~\ref{fig:april1_reg3_sngl_3-2} (bottom) we show a similar case for path~2. However, in this case about 4 intensity enhancements are observed and almost all are accompanied by  dark or bright propagating features.

Although many of the features in paths $1-6$ do not show the EUV absorption, but rather emission, we still associate them with coronal rain. Besides similar velocities (see Section~\ref{sec:stats}), the AIA observations of the same event provide conclusive proof (Section~\ref{sec:aia}). 

The last set of paths we analyse are paths $7-10$, which correspond to a different loop, apparently much larger in size. We were able to track a bright clumpy feature over 40~Mm, leading to path~7. The clump falls at speeds of $\approx$150~km~s$^{-1}$ in the POS, which, to our knowledge, is the fastest ever recorded speed for a falling clump with imaging devices (besides erupting fall-back). Just prior to impact, over the last $5-15$~Mm, several other clumps are seen along parallel paths (tracks in paths~8, 9 and 10), suggesting that the catastrophic cooling into chromospheric temperatures takes longer for these neighboring strands. 

In Figure~\ref{fig:april1_reg3_sngl10} we show a close-up of the time-distance diagram corresponding to paths 7 (top) and 10 (bottom) over the time where the falling clump is observed. The {bottom panel in the} figure shows a very interesting pattern. While the outer envelope (traced by the cyan-black dashed line in the figure) corresponds to the same speed as observed in the longer time-distance path of path~7 (top panel), there is an almost instantaneous intensity increase all along the path at time $t\approx33$~min. We believe that this feature is due to the compression of the rain downstream, that is, physically similar to the small brightening observed for the March~30 rain clumps. However, while the brightening for the latter is always immediately below the rain's head, for the present case it occupies a much larger longitudinal extent. This is probably due to a much stronger compression, {which is able to increase the temperature of the entire region below the rain to a temperature close to the emissivity peak of \hrieuv ($\log T=5.8-6$), thereby generating a flash-effect.}

\subsection{{Region 4 - Off-limb coronal rain}}
\begin{figure}
    \centering
    \includegraphics[width=0.5\textwidth]{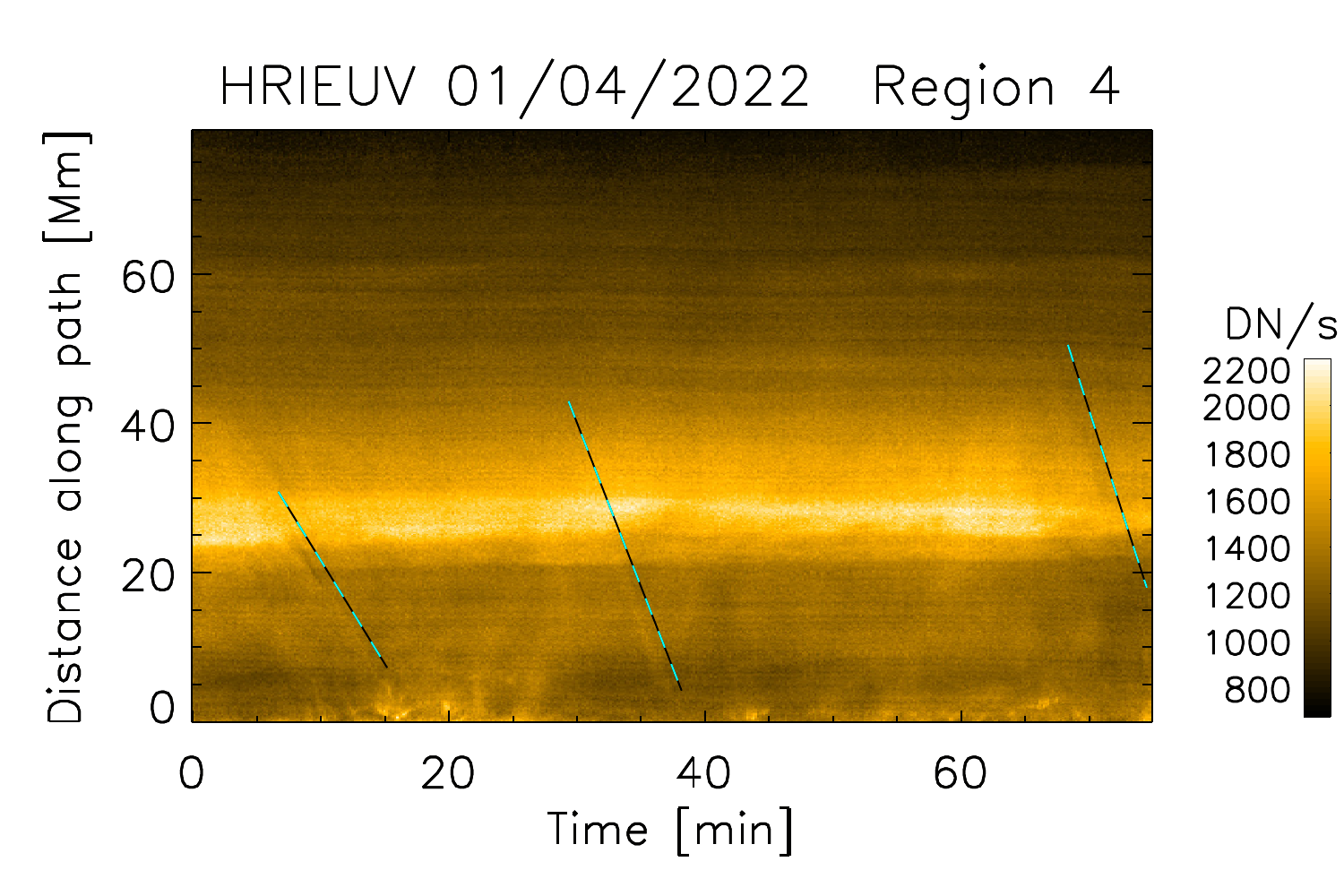}
    \caption{{Time-distance diagram along a loop observed partly off-limb by \hrieuv on April~1. The path corresponds to red curve in Region 4 shown in Figure~\ref{fig:fullFOV}. The cyan and black dashed lines correspond to dark and bright tracks from coronal rain (offset in time by 1~min in the figure for better visualisation).}}
    \label{fig:solo_apr1_reg4}
\end{figure}

\begin{figure}
    \centering
    \includegraphics[width=0.5\textwidth]{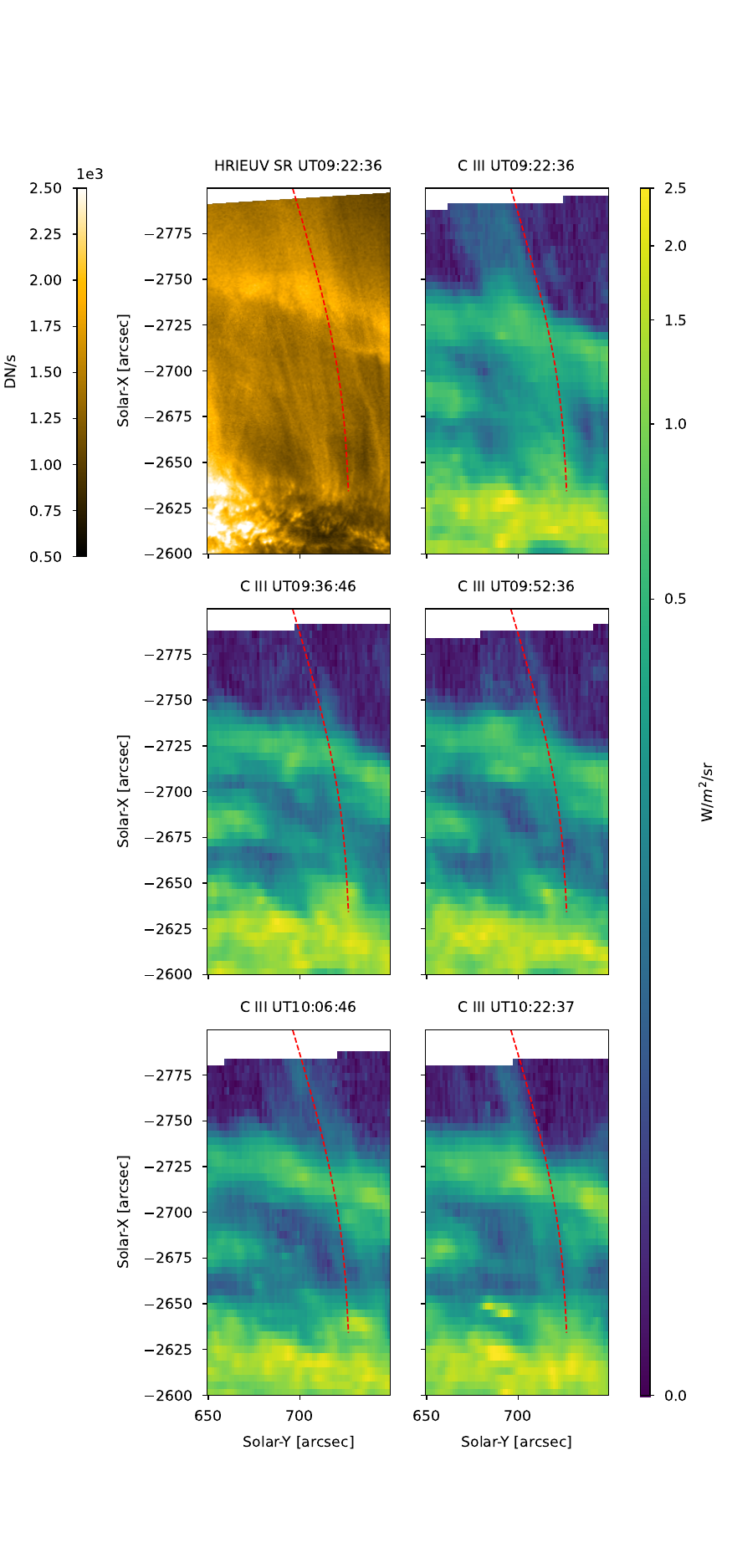}
    \caption{{SPICE rasters in the \ion{C}{III} line ($\log T=4.8$~K) over a region that overlaps with Region 4 on April~1. The FOV corresponds to the red rectangle to the East shown in Figure~\ref{fig:spicefullFOV} (bottom) and the overlaid red curves  denote some of the rain paths seen with \hrieuv (see Figures~\ref{fig:april1_reg1} and \ref{fig:april1_reg2}). The \hrieuv panel corresponds to a synthetic raster (SR) matching the time of the first SPICE raster (see text for details) but preserving the \hrieuv spatial resolution. The SPICE panels show the total intensity integrated over the \ion{C}{III} spectral line for each raster (time shown in subtitle).}}
    \label{fig:spice_apr1_reg4}
\end{figure}

{On April~1 \hrieuv captures various long loops rooted in the trailing AR closer to the limb. Small EUV absorption features falling towards the surface can be seen in one such loop, which we follow and show in red within Region~4 shown in Figure~\ref{fig:fullFOV}. The time-distance diagram along this path is shown in Figure~\ref{fig:solo_apr1_reg4}, where various characteristic dark and bright tracks of coronal rain can be seen falling at projected speeds of $50-90$~km~s$^{-1}$. This loop is also partly visible in the SPICE rasters of April~1. In Figures~\ref{fig:spice_apr1_reg4_mg9} to \ref{fig:spice_apr1_reg4_lyb} we show the evolution through the SPICE rasters of the emission in this region. In particular we notice strong variability in the upper and lower transition region (\ion{Ne}{VIII}, \ion{O}{VI}, \ion{N}{IV}) and chromospheric emission (\ion{C}{III} and Lyman-$\beta$) as shown in Figure~\ref{fig:spice_apr1_reg4}. Therefore, SPICE confirms the presence of coronal rain emitting at transition region and chromospheric temperatures in this loop.}

\subsection{Statistics}\label{sec:stats}

\begin{figure}
    \centering
    \includegraphics[width=0.5\textwidth]{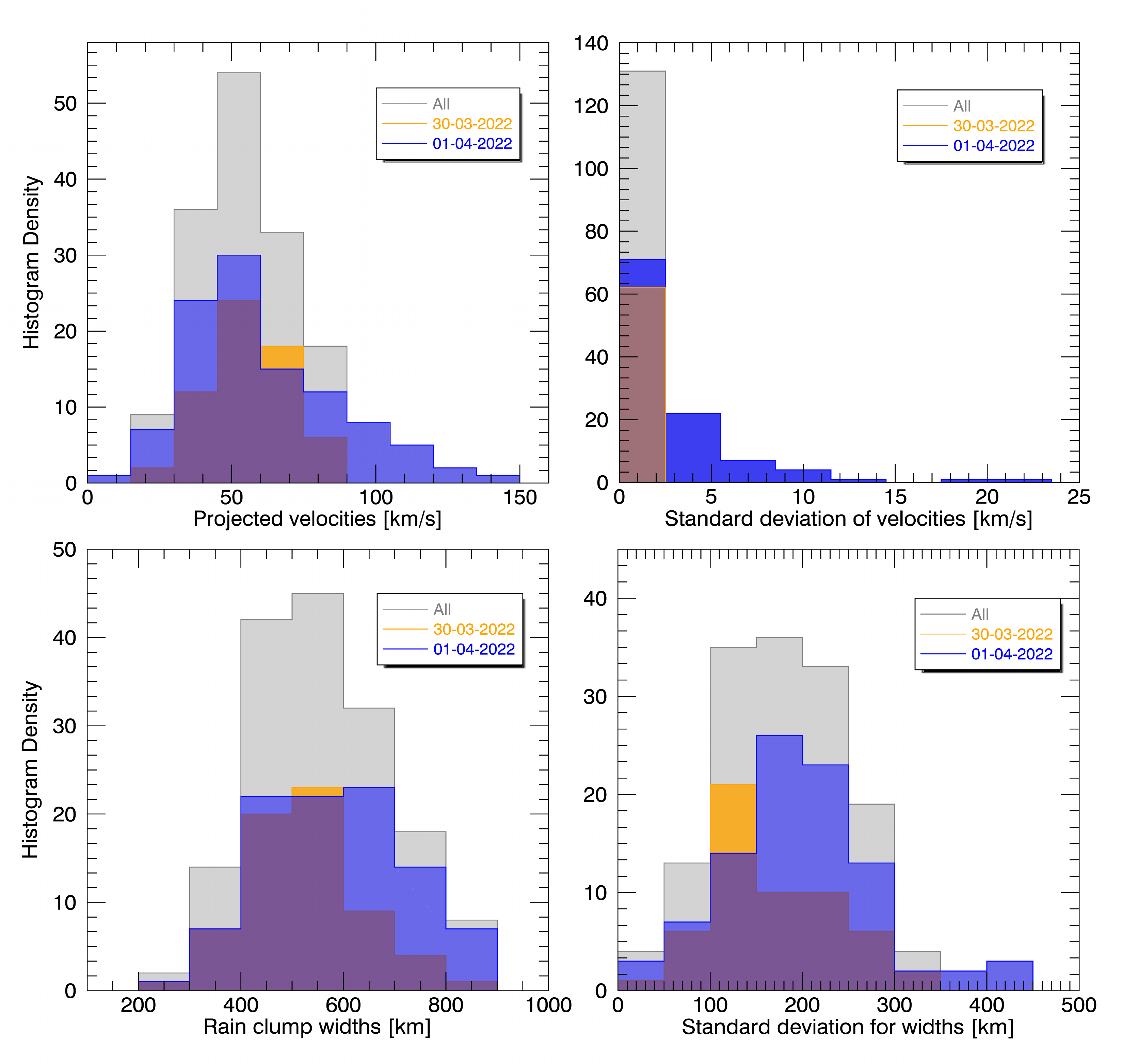}
    \caption{Projected velocities {(top) and widths (bottom)} and associated standard deviation {(corresponding right panels)} for all rain clumps tracked in this work. The colours {include transparency} and denote different dates (see legends).}
    \label{fig:vels_widths}
\end{figure}

In this section we provide statistics of all the velocity and width measurements on the coronal rain clumps (both in absorption and in emission). Please see Section~\ref{sec:methods} for the methods on how these quantities were calculated. 

In Figure~\ref{fig:vels_widths} we show a histogram of all the measured projected velocities for both datasets. We note that the peak of the distribution is between $40-60$~km~s$^{-1}$, with speeds as low as 10~km~s$^{-1}$ and as high as 150~km~s$^{-1}$. The latter high projected velocities are not common for coronal rain, and may well constitute the highest projected velocities to date (note that eruptive prominence fallback is not coronal rain). No major differences exist between both dates, except that April~1 presents a bigger spread, which is normal given the higher number of rain events analysed for that date. The bulk of the distribution matches well with previously reported speeds \citep{Antolin_Rouppe_2012ApJ...745..152A, Schad_2017SoPh..292..132S, Antolin_Froment_2022}, as well as those obtained in numerical simulations \citep{Fang_2015ApJ...807..142F, LiX_2022ApJ...926..216L}. As shown in the figure, the errors in the velocity measurements are generally small (on the order of 5~km~s$^{-1}$ for most).

Similarly, in Figure~\ref{fig:vels_widths} we show a histogram of all measured rain clump widths. The distribution peaks between $400-600$~km, but goes as low as 260~km and as high as 890~km. Both dates present {small} differences in terms of distribution shape. {While both datasets have relatively small pools, the dataset of Aptil~1 presents a broader distribution with values about 100~km larger.} The standard deviation figure indicates that there is a relatively large variation or error in the width measurement. This is not unexpected, since along a given track the rain clump's background varies significantly, leading to differences in the results of the Gaussian fits (despite efforts in reducing the background influence, see Section~\ref{sec:methods}). 

\begin{figure}
    \centering
    \includegraphics[width=0.5\textwidth]{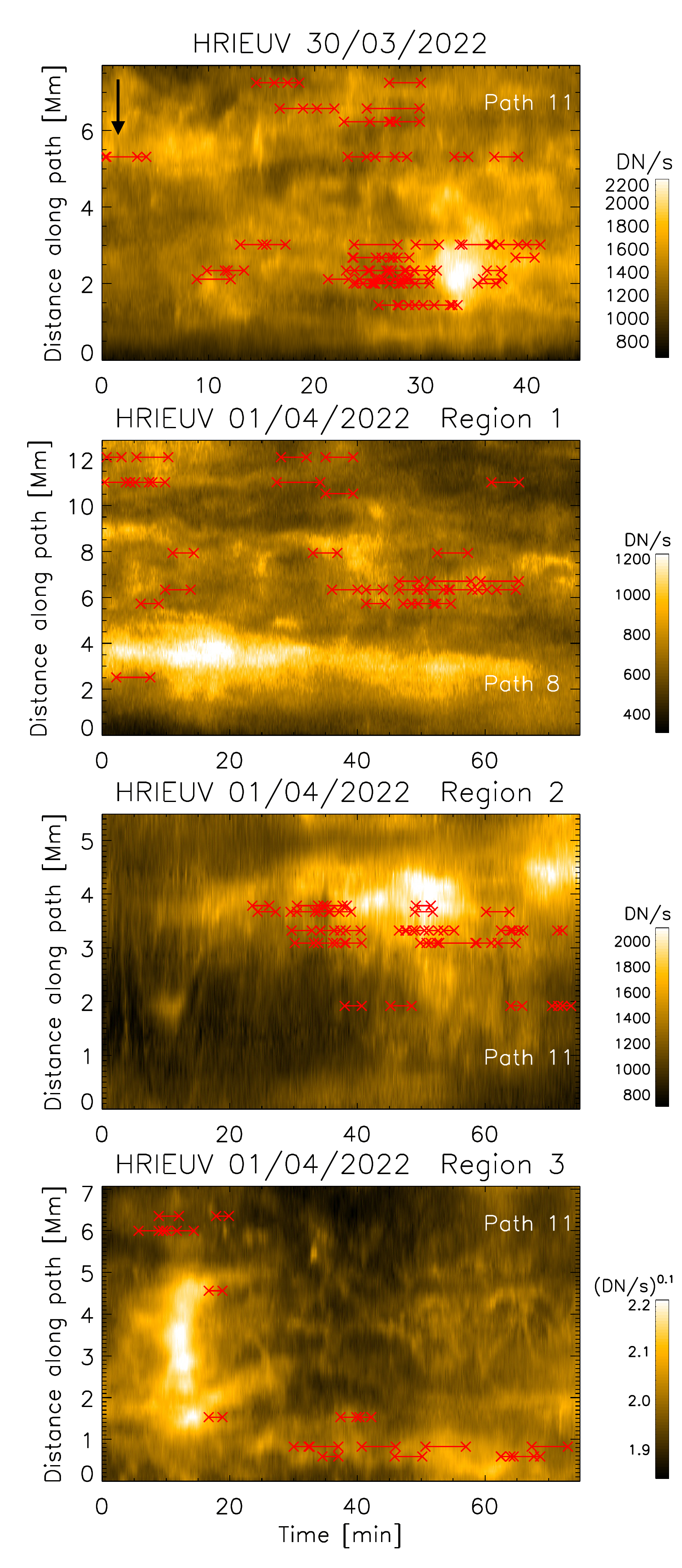}
    \caption{Time-distance diagram {along paths that cross several loop bundles: path~11 in Figure~\ref{fig:March30all} (top), path~8 in Figure~\ref{fig:april1_reg1} (second panel from top), path~11  in Figure~\ref{fig:april1_reg2} (third panel from top) and path~11 in Figure~\ref{fig:april1_reg3} (bottom)}. The red lines between crosses denote times and location at which rain clumps are observed. The vertical black arrow {in the top panel} indicates an example of an EUV absorption feature produced by a rain clump. {Note that the intensities in the bottom panel have been scaled to the power of 0.1 to better see the large intensity variation.}}
    \label{fig:across}
\end{figure}

\subsection{Coronal strands associated with coronal rain}\label{sec:strands}

To investigate more carefully a possible relation between the filamentary coronal structure (coronal strands) within loop bundles and the coronal rain they host, we take cuts across several of the analysed loop bundles, as shown in Figures~\ref{fig:March30all} (path~11), \ref{fig:april1_reg1} (path 8), \ref{fig:april1_reg2} (path 11) and \ref{fig:april1_reg3} (path 11). We show the resulting time-distance diagrams along these cross-cuts in Figure~\ref{fig:across}.

Several coronal strands can be seen in each Figure, some of which very likely belong to the loop bundle hosting the rain event. In many instances, as the rain crosses the transverse cuts it produces a small absorption feature (whose time duration depends on the clump's length as speed). One example of such a feature is shown by an arow in Figure~\ref{fig:across}. Note that it is preceded and followed by a bright EUV feature of roughly the same width as the rain clump. A large group of rain clumps is seen around $t=22-32$~min, followed by a bright feature at $t=32-40$~min. This bright feature corresponds to the rebound shock and upward flow produced by the rain shower impact.  

In Figure~\ref{fig:across} a group of strands can be seen appearing after $t=30~$min between distance 5 and 8 Mm across the transverse cut, which seems to coincide with the location in time and space of a group of rain clumps. Similarly, in the other panels in the Figure the rain clumps appear to be preferentially located in regions where coronal strands are observed. While sometimes a 1-to-1 association between a coronal strand and a rain clump can be made (particularly for the latter), it is not the general case. However, we can notice that the widths of coronal strands (around 500~km) are on average similar to rain clump widths.

\subsection{SDO/AIA observations at different line-of-sight and resolution}\label{sec:aia}

\begin{figure}
    \centering
    \includegraphics[width=0.5\textwidth]{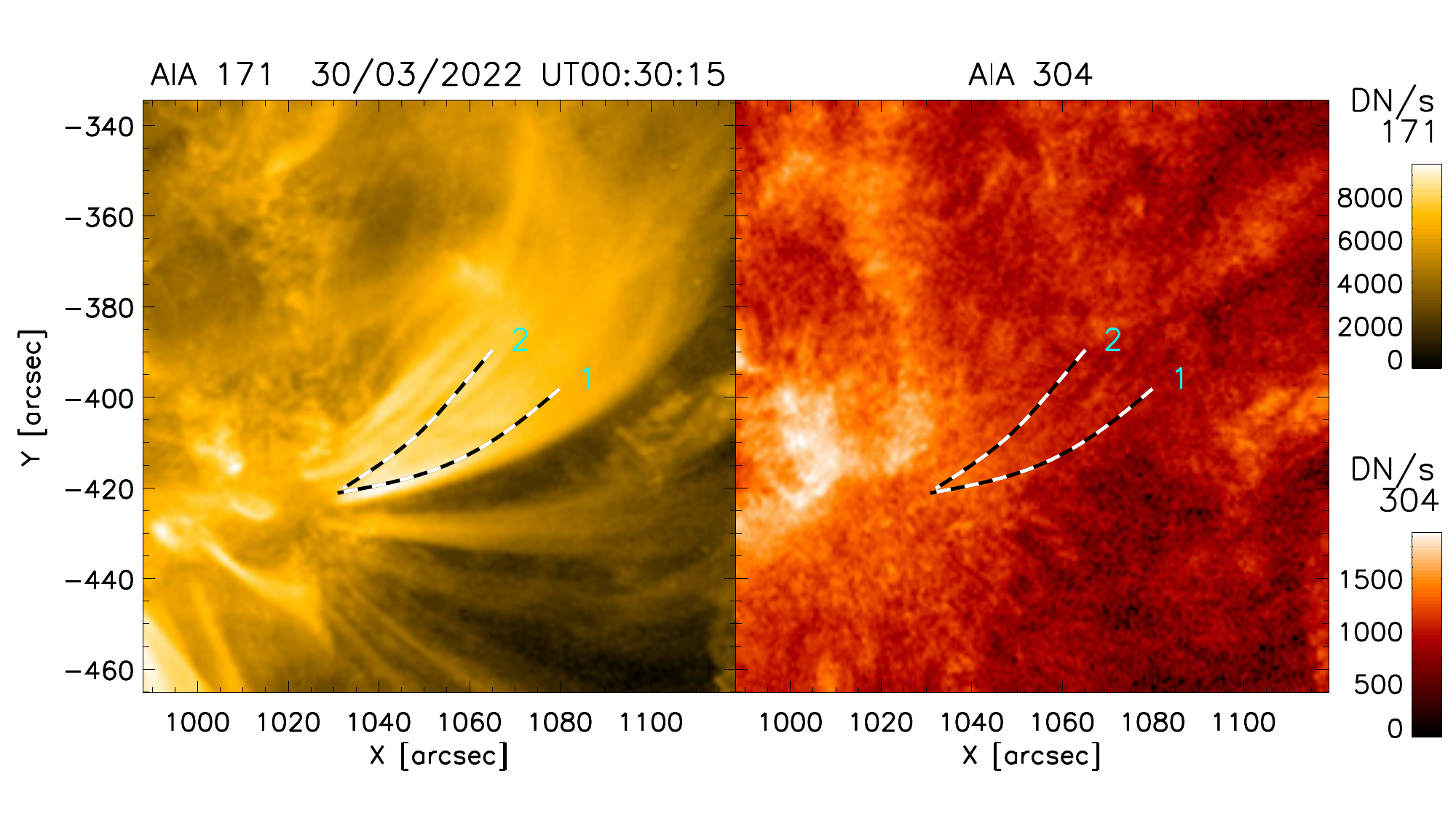}
    \caption{SDO/AIA observation in the 171 (left) and 304 (right) channels of the March~30 coronal rain event. AIA was roughly in quadrature with Solar Orbiter on this date. The large loop bundle observed in the AIA~171 map corresponds to the same loop bundle observed by \hrieuv in Figure~\ref{fig:March30all}. {The white-black dashed paths follow some rain clumps observed in AIA~304. An animation corresponding to this figure  is available. The animation shows several rain clumps in AIA 304 but no clear counterpart is seen in AIA 171, although strong intensity variations are observed at the same locations}. The animation {runs from UT00:02 to UT00:49} and comprises the time where coronal rain is observed with \hrieuv. It first runs without and then with the rain paths overlaid.}
    \label{fig:aia_march30}
\end{figure}

Our investigation on the coronal rain events with \hrieuv is complemented by co-temporal SDO/AIA observations. On these dates, Solar Orbiter was in quadrature with SDO, thereby offering a different view on the same ARs. In Figures~\ref{fig:aia_march30} and \ref{fig:aia_april1} we show snapshots for March~30 and April~1, respectively, with a FOV centred on the regions of interest. The same large-scale coronal structures can be identified across both instruments. 

\begin{figure}
    \centering
    \includegraphics[width=0.5\textwidth]{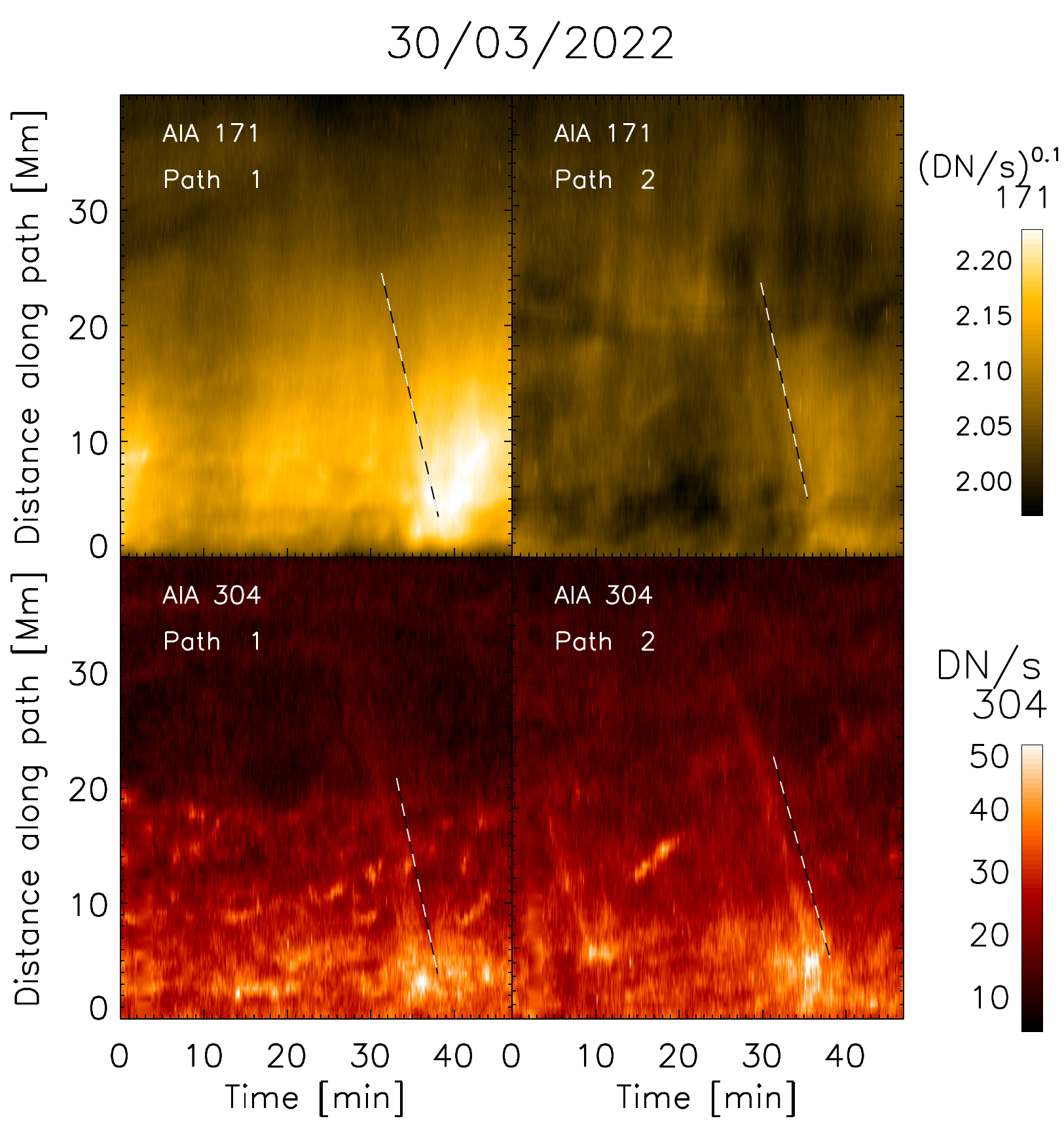}
    \caption{{Time distance diagrams along paths 1 and 2 shown in Figure~\ref{fig:aia_march30} on March~30, with AIA~171 (top) and AIA~304 (bottom). Dark and bright paths from coronal rain are highlighted with white-black dashed lines, which are offset by a fraction of a minute to better see the rain features. Zero distance corresponds to the footpoints of the loops. Note that the AIA~171 intensities have been scaled to the power of 0.1 to better see the large intensity variation.}}
    \label{fig:aia_tdmarch30}
\end{figure}

\begin{figure}
    \centering
    \includegraphics[width=0.5\textwidth]{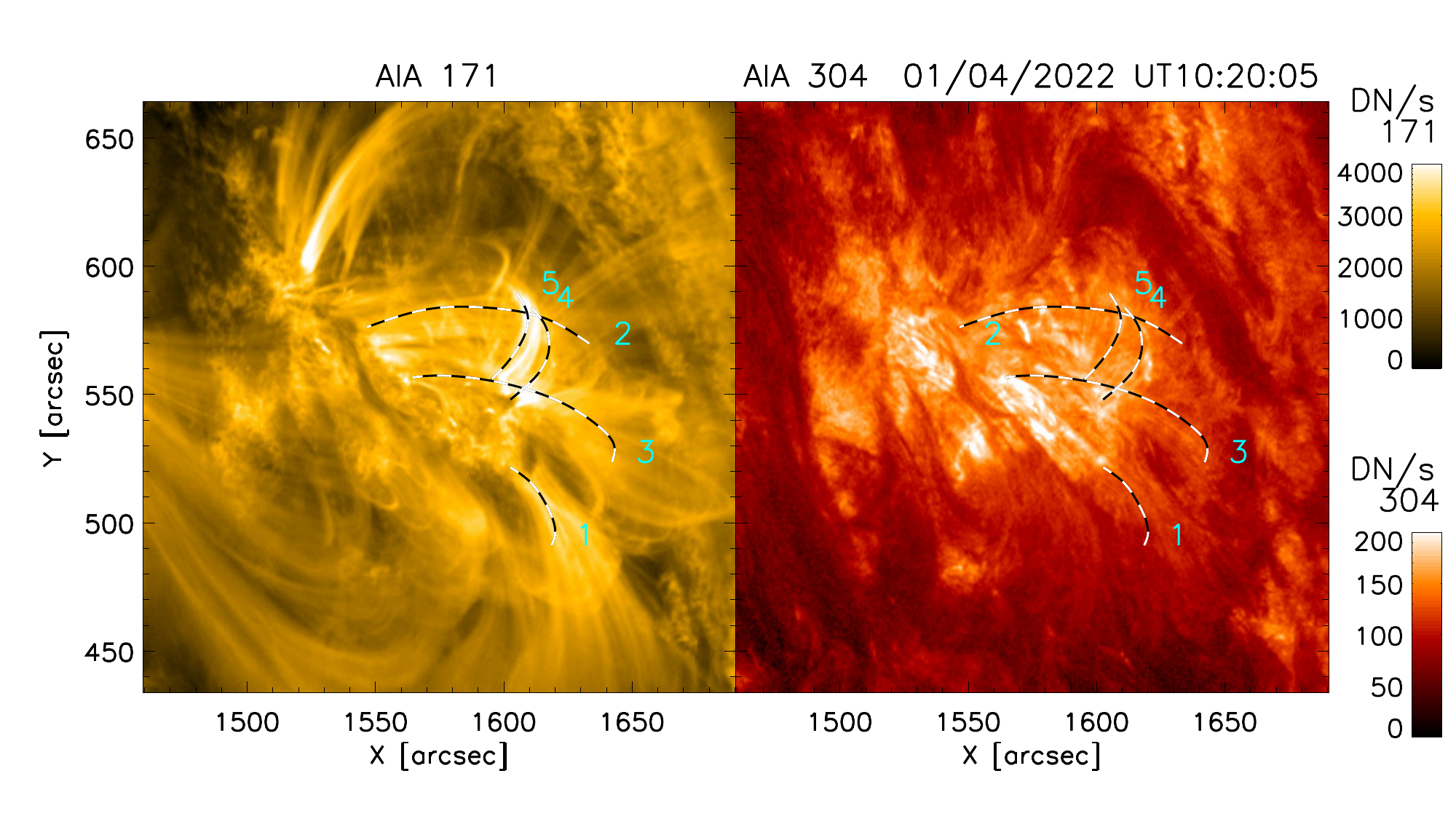}
    \caption{{SDO/AIA observation in the 171 (left) and 304 (right) channels of the April~1 coronal rain events. AIA was roughly in quadrature with Solar Orbiter on this date. Some of the large loop bundles observed by \hrieuv can be easily identified. The white-black dashed paths follow some rain clumps belonging to these loop bundles, observed in AIA~304. Path 1 follows the same loop bundle seen in Region~1 of Figure~\ref{fig:april1_reg1}. Paths 2 and 3 may correspond to the same rain events shown in paths 8 or 9 in Region~2 of Figure~\ref{fig:april1_reg2}. Paths 4 and 5 follow the same loop bundle shown in the lower part of Figure~\ref{fig:april1_reg3} (paths 1 to 6 in that Figure). An animation corresponding to this figure is available. The animation shows widespread rain in AIA 304 but the counterparts in AIA 171 can only clearly be seen for the loop around path 1, although strong intensity variations are observed at the same locations. The animation {runs from UT09:20 to UT10:32} and comprises the time where coronal rain is observed with \hrieuv. It first runs without and then with the rain paths overlaid.}}
    \label{fig:aia_april1}
\end{figure}

Despite the similarity between AIA~171 and \hrieuv in terms of the emission, and the large-scale nature of the event (occurring across at least 15~Mm in \hrieuv), no clear downflow can be observed in absorption or emission with AIA~171 on March~30 over the same time period. Sections of the loop bundle appear and disappear over the same duration, which likely correspond to the EUV variation associated with coronal rain observed with \hrieuv. Some upward flows, bright in EUV, are observed, which could correspond to the rebound shock and flow observed with \hrieuv. {To check this we have identified and tracked a few large rain clumps in AIA~304, whose paths are shown in Figure~\ref{fig:aia_march30}. The time-distance diagrams along these paths are shown in Figure~\ref{fig:aia_tdmarch30} and we can easily identify the characteristic rain slopes in AIA 304 (tracks in the time-distance diagrams). Note the brightenings at the times the rain impacts the chromosphere. With the help of~AIA 304 we are able to recognise the rain features in AIA~171, such as EUV absorption and in particular a rebound shock and flow feature following the rain impact. The strong EUV variation that is seen in the image sequence is therefore attributable to the rain episode. }

\begin{figure*}
    \centering
    \includegraphics[width=1\textwidth]{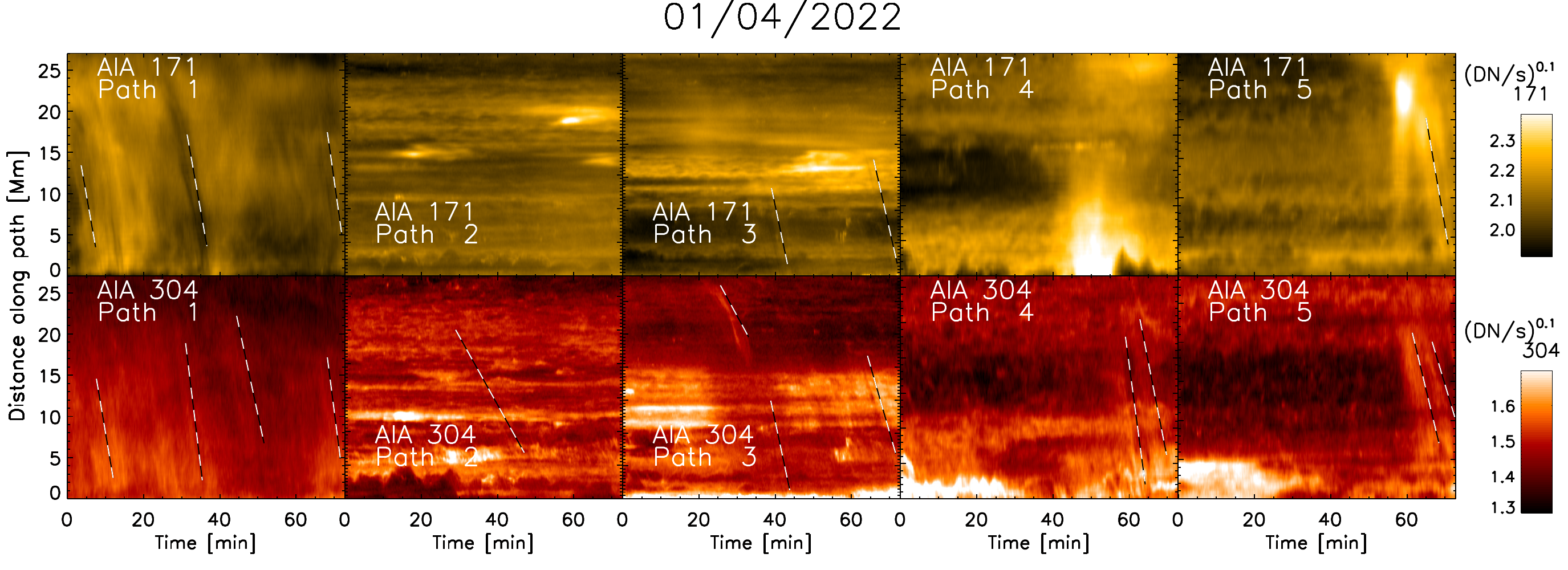}
    \caption{{Time distance diagrams along paths 1 to 5 shown in Figure~\ref{fig:aia_april1} on April~1, with AIA~171 (top) and AIA~304 (bottom). Dark and bright paths from coronal rain are highlighted with white-black dashed lines, which are offset in time by 1~min to better see the rain features. Zero distance corresponds to the footpoints of the loops. Note that the intensities have been scaled to the power of 0.1 to better see the large intensity variation.}}
    \label{fig:aia_tdapril01}
\end{figure*}

On April~1, only region~1 shows clear coronal rain in {the image sequence of} AIA~171, with the characteristic EUV absorption features downflowing along the loop. Regions~2 and 3 show very similar EUV variation as observed with \hrieuv, but as for March~30, we were unable to {directly} identify coronal rain downflows only based on EUV absorption features {in the image sequences}. However, the expected coronal rain is revealed in the 304 channel and we were able to roughly identify the large-scale coronal rain events for each of the loops. {As for March~30, we track several large rain clumps in each region (shown in Figure~\ref{fig:aia_april1}) and plot the time-distance diagrams in Figure~\ref{fig:aia_tdapril01}. Path~1 tracks a rain clump belonging to Region~1's loop, paths 2 and 3 belong to Region~2 and may correspond to the loops outlined by paths 8 and 9 in Figure~\ref{fig:april1_reg2}, and paths~4 and 5 follow clumps in the loop bundle outlined by paths 1 to 6 in Figure~\ref{fig:april1_reg3}. We are able to detect several rain tracks in the time-distance diagrams of AIA~304 and a very clear EUV absorption profile in AIA~171 for path~1. However, the signatures in AIA~171 are much harder to detect. In paths~2 and 4 we do not see any features of the rain falling in AIA~171, but a bright feature can be seen near the footpoint of path~4 that may correspond to the impact and rebound shock and flow from a rain shower. In paths~3 and 5 we are able to recognise a few EUV absorption and emission tracks co-temporal and co-spatial to the rain tracks in AIA~304. The bright EUV emission in 171 is similar to that seen in \hrieuv in paths~1 or 6 of Figure~\ref{fig:april1_reg3_td}, in that a large-scale emission is seen simultaneously all along the path that precedes the bright downflowing feature from the rain.} 

It is interesting to note that, as shown by the AIA~304 animation, there is widespread coronal rain over the AR {over the same FOV as \hrieuv and} including loops for which it is not observed with \hrieuv. {A large amount of coronal emission ahead of the rain along the LOS can easily decrease the contrast produced by EUV absorption or emission from the rain in the 171 or 174 channels, thereby making it undetectable. As we show here, high spatial resolution can help retaining this contrast, which is why we observe far more in \hrieuv than AIA~171. On the other hand, AIA~304 shows the emission/absorption from the rain more directly since no further emission is present ahead of the rain along the LOS. Nevertheless the rain can be hard to discern due to the very bright TR background in AIA~304.}

{The larger FOV provided by SDO/AIA sheds more light into the large-scale magnetic structure of Region~1 on April~1. AIA~171 reveals a topology akin to null-point topology, with open field lines (or much larger loops) above, with the loop arcade below the null-point. This provides an explanation to both the large-scale reconfiguration and the observed large-scale coronal rain event in the loop, with continuous coronal rain over a very wide loop cross-section observed for the entirety of the \hrieuv observation. Indeed, as discussed in \citet{LiL_2018ApJ...864L...4L,Mason_2019ApJ...874L..33M}, magnetic dips are often observed above null-point topologies, which act as mass reservoirs where large amounts of material can catastrophically cool down. The cool and dense material can then either spill sideways from the magnetic dip or flow through the null-point, facilitated by the reconnection process, downward into the newly formed reconnected loops. This process can be very long-lived \citep{Chitta_2016_aa587,ChenH_2022AA...659A.107C} and can be accompanied by a reconfiguration of the loop, similar to shrinkage (as in the standard flare model).  }

\subsection{{Widespread coronal rain as suggested by SPICE}}
\begin{figure*}
    \includegraphics[width=1\textwidth]{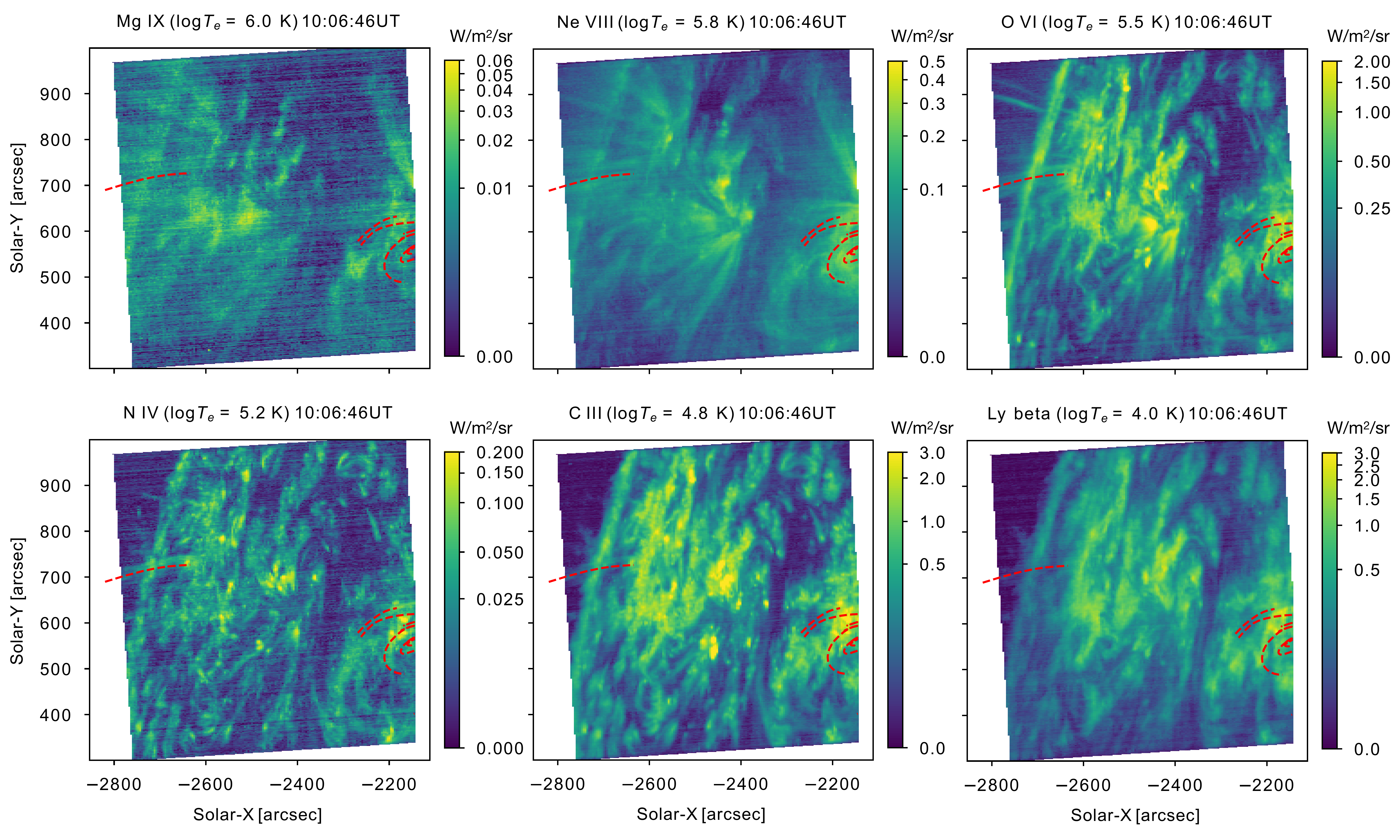}
    \caption{{SPICE multi-wavelength full FOV on April~1. The SPICE FOV is the same as that shown in Figure~\ref{fig:spicefullFOV} (bottom), co-aligned with \hrieuv. The overlaid red curves denote some of the rain paths seen with \hrieuv (see Figures~\ref{fig:april1_reg1} and \ref{fig:april1_reg2}). Each SPICE panel shows the total intensity over a spectral line indicated in the subtitle, together with its temperature of maximum formation.}}
    \label{fig:spice_april1_full}
\end{figure*}

{All the loops with coronal rain captured with \hrieuv show clear counterparts in SPICE in the upper and lower transition region lines. For the off-limb loop of Region 4 we were also able to capture clear emission in the chromospheric lines of SPICE, thanks to the lower background emission. In Figure~\ref{fig:spice_april1_full} we show a multi-wavelength view of the full FOV of SPICE for 1 raster. What is striking of this figure is that cool loops emitting in the upper and lower transition region lines, with similar features (morphology, variability) as those where we have detected coronal rain, are widespread in the FOV. Indeed, all the 5 rasters of this region show strong variability in these loops. This strongly suggests that at least for this AR coronal rain is widespread and that only a fraction of it is observed in EUV absorption with \hrieuv. This matches also the picture obtained with AIA~304. }

\section{Discussion and Conclusions}
\label{sec:conclusions}

In this work we analysed coronal rain at unprecedented spatial resolution in the EUV, provided by Solar Orbiter observations with \hrieuv during the 2022 spring perihelion. {Observations were complemented with SPICE and AIA, providing a multi-wavelength picture of the phenomenon in various active regions.} The resolution achieved is $\approx240~$km, which is only about two times lower than previous H$\alpha$ observations of the phenomenon with the SST \citep{Antolin_Rouppe_2012ApJ...745..152A}. The strong correlation between EUV absorption and H$\alpha$ emission, expected from radiative transfer theory \citep{Anzer_Heinzel_2005ApJ...622..714A}, is here confirmed down to the smallest resolved scales ever achieved, with clump FWHM widths as small as 260~km. This fine-scale structure is also reflected in the fact that many clumps only produce very faint EUV absorption features, thereby requiring high-resolution and sensitivity to detect them.

We selected 2 dates on which Solar Orbiter observed several ARs on-disk {and partly off-limb}. On March~30 we focused our attention on one coronal loop bundle and discovered new features of coronal rain dynamics. As the rain falls, the region immediately beneath the clump (downstream) is observed to brighten, leading to very fine light streaks in time-distance diagrams. {We interpret this phenomenon as being the result of compression and heating due to the relatively large momentum of the condensation.  \citet{Fang_2015ApJ...807..142F} report the formation of rebound shocks when coronal rain forms, produced by the localised loss of pressure driving strong flows that lead to the condensation. These rebound shocks may be thought as the source of the brightening that we see. However, the rebound shocks are then seen to propagate away at the sound speed from both sides of a condensation, much faster than the condensation’s falling speeds. There is no reason why these rebound shocks should propagate at the same speed. Therefore these rebound shocks cannot explain our observations. On the other hand, our interpretation as commpression and heating is supported by 2.5-D MHD simulations of coronal rain by \citet[][see their Figure 4]{LiX_2022ApJ...926..216L} and \citet{Antolin_2022ApJ...926L..29A}. This phenomenon is therefore similar to the fireball  phenomenon on Earth {linked to meteoric ablation}, with the region below {the clump being compressed and heated} as it falls.}  We do not observe this phenomenon for all coronal rain events, which suggests that not only a high spatial resolution is needed but also favorable line-of-sight relative to the rain trajectory. It is also possible that the compressed material exists in all cases but emits at different temperatures not sampled by \hrieuv. 

EUV enhancement associated with coronal rain is also observed during the fall ahead and at the wake of rain clumps. The latter can be seen in the time-distance diagrams along or across rain clump trajectories as an increase of the EUV intensity that follows the EUV absorption feature. This may correspond to a coronal strand and is likely due to the CCTR, which extends from the clump far into the wake, as shown by numerical simulations \citep{Antolin_2022ApJ...926L..29A, LiX_2022ApJ...926..216L}. {Observations with SPICE show the filamentary structure in the upper and lower transition region lines, thereby supporting this interpretation.} Ahead of the clump, besides the fireball feature, a second, more elusive kind of brightening is observed just prior to the rain impact (with the brightening spreading over $\approx6$~Mm), and manifests as a flash-like simultaneous intensity enhancement of the entire space between the clump and the chromosphere. We suspect that this is also due to the compression of the plasma ahead of the clump, as {suggested by} the same numerical simulations cited previously, {with a compression strong enough as to increase the temperature of the entire region below the rain close to the emissivity peak of \hrieuv ($\log T=5.8-6$), thereby generating a flash-effect.} However, we only found one clear occurrence of this larger-scale compression, for which the rain is extremely fast (with projected speeds of 150~km~s$^{-1}$). Such speeds may be supersonic, in which {case} the compression argument may not fully explain it. Indeed, if produced by compression, it means that sound waves are produced ahead of the clump and travel fast enough to compress the entire region donwstream (leading to the `flash-like' character of the brightening). However, this is not possible if the clumps are supersonic. 

Another new phenomenon is the rebound shock and upflow that follows the rain impact onto the chromosphere. The feature is detected best after the end of rain showers rather than individual rain clumps, and is characterised by a bright and diffuse  propagating EUV feature roughly along the same paths over which the cool downflows occur. The propagating speeds are between $50-130$~km~s$^{-1}$, depending on the selected path for the time-distance diagram. Based on multi-dimensional simulations of this phenomenon \citep{Antolin_2022ApJ...926L..29A}, we suspect that this difference is due to the combination of the rebound shock, which propagates at the tube speed of $\approx130~$km~s$^{-1}$ (for temperatures corresponding to the \ion{Fe}{X}~174~\AA~ formation temperature of $10^{5.98}$~K), and an upward flow produced by the impact (which are bound by gravity and thus slower). To the best of our knowledge, this is the first report of this effect, despite being predicted by numerical simulations for decades \citep{Muller_2003AA...411..605M, Mendozabriceno_2002ApJ...579L..49M, Antolin_2010ApJ...716..154A, Fang_2015ApJ...807..142F, LiX_2022ApJ...926..216L}. Indeed, we expect a response from the lower atmosphere in the form of an upflow or upward propagating wave along every magnetic field line and probably even over a wider region compared to that affected by the rain impact, due to the high-$\beta$ conditions and strong gas pressure increase at the chromospheric heights of rain impact \citep{Antolin_2022ApJ...926L..29A}. {For the March~30 event,} this upward propagating feature is seen to reheat the loop bundle and may correspond to the start of a new TNE cycle. {Simulations indicate that efficient refilling and reheating of the loop is only obtained in the case that the stratified heating is still on-going. In other words, if there is no continuous heating at the loop footpoint, the simple rebound shock and flow obtained from a rain shower is insufficient to bring the density and temperature back to usual coronal values.} The rebound shock and upflow can be seen mainly after rain showers rather than individual rain clumps, {suggesting an additional feedback effect from the large rain shower momentum}. Only a subset of rain showers show these features, despite their relatively large-scale nature, indicating that the conditions to observe this atmospheric response to the rain impact are stringent. This is further supported by the AIA~171 co-temporal observations on March~30 in quadrature with \hrieuv, where some hints of an upward propagating disturbance are observed but remain much dimmer and elusive.

As shown by \citet{Sahin_2022ApJ...931L..27S}, rain showers can help identify coronal loop entities (defined as a coronal volume evolving in similar way and therefore subject to similar heating conditions). This is particularly important given the optically thin nature of the corona, leading to great LOS superposition \citep[also referred to as `the coronal veil';][]{Malanushenko_2022ApJ...927....1M}. Furthermore, the occurrence of coronal rain points to a state of TNE, for which stringent coronal heating conditions are needed. {This is particularly the case of quiescent coronal rain, which occurs in usual coronal loops and is accompanied by long-period intensity pulsations for long-duration heating \citep{Froment_2020AA...633A..11F}. It is still unclear whether the TNE-TI scenario still applies to prominence-coronal rain hybrid structures \citep[although initial results indicate quasi-periodic occurrence that may correspond to long-period intensity pulsations][]{LiL_2019ApJ...884...34L} or even to flare-driven rain.} 

\citet{Klimchuk_2019ApJ...884...68K} have quantified the conditions for TNE, which include a strong heating stratification (ratio of apex-to-footpoint volumetric heating rate below 0.1) and not too high heating asymmetry between footpoints (below 3, to prevent siphon flows that effectively reduce the lifetime of the condensations in the corona relative to their falling time). \citet{Johnston_2019AA...625A.149J} further show that a high frequency rate for the heating events (with repeating time less than the radiative cooling time of the loop) is needed. Several of the observed rain showers in this work occur following a global brightening of the loop. This is expected from the TNE-TI cycle, in which the loops start at a hot, {indeterminate} temperature and radiatively cool down. What is observed then corresponds to the loop's temperature going through the temperature response range of the \hrieuv channel, which peaks at $\approx10^{6}$~K \citep{Rochus2020}. {Usual draining and cooling of loops would explain this global brightening feature, but would fail to explain the appearance of the clumpy downflows in EUV absorption or emission that follow the global brightening events \citep{Peter_etal_2012AA...537A.152P}. This is also the case in post-flare cooling, where simple draining and global cooling of loops is unable to explain flare-driven rain \citep{Reep_2020ApJ...890..100R}. Our cooling interpretation is further supported by the SPICE observations, which show that the brightening is delayed for cooler spectral lines at transition region and chromospheric temperatures.} 

It is interesting to note that the {observed global} brightening occurs fairly uniformly {and symmetrically} along the loop, with the apex brightening a few minutes prior to the rest of the loop. {To our knowledge, very few works investigate in detail (including the necessary forward modelling into e.g. AIA~171) the spatial distribution of the cooling during TNE cycles. The published literature indicate that we should expect symmetric global brightening in a channel such as AIA~171 during the cooling stage of TNE cycles when the heating is symmetric across both footpoints \citep{Muller_2003AA...411..605M, Peter_etal_2012AA...537A.152P, Winebarger_2018ApJ...865..111W, Johnston_2019AA...625A.149J}, while asymmetric heating or loop geometry may tend to produce asymmetric global brightening, with significant brightening only along one loop leg \citep{Mikic_2013ApJ...773...94M, Froment_2018ApJ...855...52F, Pelouze_2022AA...658A..71P}. This would suggest that both the heating along the observed loops with rain and the geometry of such loops are symmetric rather than asymmetric.} 

In our observations, the rain is observed to occur in some cases right after the loop {global} brightening (within 10~min or so), while other cases show longer waiting time (over 40~min), which indicates different average loop densities. While some clumps are seen to fall over large distances over 40~Mm, most only appear in the last $10-20$~Mm. {This spatial occurrence of coronal rain is often the case when the heating or loop geometry is asymmetric since such configurations generate siphon flows that reduce the lifetime of the cooling plasma in the loop, also leading to what is known as incomplete condensations that do not appear in chromospheric lines \citep{Mikic_2013ApJ...773...94M, Froment_2018ApJ...855...52F, Pelouze_2022AA...658A..71P}. Hence, contrary to the argument in the previous paragraph, this would suggest asymmetric heating or loop geometries. This contradiction  suggests another possibility. For instance, it is also possible that better visibility of the condensations with decreasing height is due to a simple LOS effect, since we expect a larger column mass of cool material along the LOS (and therefore stronger EUV absorption) closer to the footpoints. This is supported by the SPICE observations, which show cool emission down to low transition region temperatures or less along the loops with rain.}

{Another interesting point is that the height at which the condensations occur} seems to decrease the {farther} away we are from the largest/clearest clumps (those producing the clearest EUV absorption features). This behaviour
suggests progressive and non-uniform cooling across the loop, which may reflect slightly different conditions, such as field line length and differences in the heating that ultimately affect the character of the thermal instability \citep[as seen in numerical simulations, e.g.][]{Froment_2018ApJ...855...52F,Pelouze_2022AA...658A..71P}. 

We observe some showers that occur over a very wide volume, over $15-20$~Mm in the POS. This is a lower estimate, given that the loops are observed to expand and most rain clumps appear closer to the footpoint, over the last $10-20$~Mm lengths prior to impact. {The SPICE observations confirm this, since the loops appear wider in the cool transition region lines. }The footpoint of these loop bundles is relatively wide ($4-10~$Mm, as seen in Figures~\ref{fig:March30all}, \ref{fig:april1_reg1}, \ref{fig:april1_reg2}, and \ref{fig:april1_reg3}). This suggests similar heating conditions over a relatively wide region (several times the granular scale) and probably a locking / synchronising mechanism that can act over wide distances across the magnetic field. Thermal instability has been suggested for this synchronising role \citep{Antolin_Froment_2022}.

We observe a width distribution for rain clumps peaking at $\approx500\pm200$~km. At the smaller range of this distribution we have the widths observed in H$\alpha$ with the SST or GST  \citep{Froment_2020AA...633A..11F,Jing_2016NatSR...624319J}, while the larger range is common for the widths observed in chromospheric and TR lines with IRIS \citep{Antolin_2015ApJ...806...81A, Sahin_2023}. Not much variation is observed across different regions. This little variation has been reported in \citet{Sahin_2023} and suggests a more fundamental nature of plasma morphology in MHD. These widths could be governed by the underlying magnetic topology and/or by the length scales of the heating \citep{Antolin_2022ApJ...926L..29A}. However, they can also be associated with thermal instabilities \citep{Antolin_2015ApJ...806...81A,Claes_Keppens_AA624_2019, Claes_AA_2020}. We have shown that very sharp bright coronal strands appear co-located with the rain clumps within the loop bundle and exhibit very similar widths of $\approx500$~km, consistent with the widths of coronal sub-structure found with Hi-C \citep{Brooks_2013ApJ...772L..19B, Peter_2013AA...556A.104P, Aschwanden_2017ApJ...840....4A,Williams_2020ApJ...892..134W}. This similarity suggests that, (a) the sub-structure is similar for TNE and non-TNE loops (assuming that at least part of the investigated loops by Hi-C are not under TNE) and, (b) such morphology does not directly depend on the spatial scales of the heating in the lower atmosphere. For instance, we know that the latter determines the spatial distribution of the rain showers \citep{Sahin_2022ApJ...931L..27S}, but clearly the rain showers do not have the same widths as rain clumps (differing by an order of magnitude). Hence, a different mechanism may be responsible for the fine-scale structure for both the rain and the coronal strands. This mechanism may be the same in the case of TNE loops and may correspond to the CCTR produced by thermal instability, as shown in \citet{Antolin_2022ApJ...926L..29A}. 

The observed speeds for the coronal rain clumps exhibit a wide projected velocity distribution, as reported in the past \citep[e.g.][]{Antolin_Rouppe_2012ApJ...745..152A, Kohutova_2016ApJ...827...39K, Verwichte_2017AA...598A..57V,Schad_2017SoPh..292..132S}. The peak of the observed distribution is below 50~km~s$^{-1}$, with minima and maxima of $10-150$~km~s$^{-1}$. This peak contrasts with previously observed peaks of $80-100$~km~s$^{-1}$ \citep{Antolin_2020PPCF...62a4016A}. This may be explained by the fact that previous reports focus on off-limb coronal rain for which the FOV is small and only captures a small region around the footpoint (a constraint of current ground-based instrumentation that depends on AO locking). Here, we are able to detect the rain {closer to its formation time}  higher up along the loop, where the speeds are naturally lower. This is in agreement with a recent AR-scale study of coronal rain with IRIS by \citet{Sahin_2023} and with 2.5-D MHD numerical simulations \citep{LiX_2022ApJ...926..216L}. 

Most of the rain events we observe are rooted in moss, with strong jet activity at the footpoints. Previous studies have indicated that strong heating may occur in such regions \citep{Testa_2014Sci...346B.315T,Chitta_2018,Tiwari_2019ApJ...887...56T,Nived_10.1093/mnras/stab3277}, favorable for the onset of TNE-TI. In addition, we also observed a structure undergoing a large-scale reconfiguration on April 1 (Region 1). This topological change may play a major role in triggering TI due to the expected long-wavelength perturbations (through magnetic pressure for example). Through coronal rain tracking we were able to detect a null-point topology at the footpoint of one of the loops. Such structures are preferential locations for magnetic reconnection and therefore heating \citep{Chitta_2017,Priest_2018ApJ...862L..24P,Syntelis_2019ApJ...872...32S}. 

The availability of two co-temporal observations in similar TR lines but with very different LOS, provided by SDO and Solar Orbiter in quadrature, allows to disentangle to some extent the effect of LOS superposition and spatial resolution. Large-scale events such as those of March~30 and most of April~1 were not detected in AIA~171, suggesting a major role of LOS superposition. On the other hand, the various events observed on April~1 with \hrieuv pale in comparison to the spatial extent seen in AIA~304 for that day, {also indicated by the SPICE observations.} This suggests that although \hrieuv is a game changer for observing coronal rain on-disk (in terms of its fine-structure and the associated EUV changes) it is not the ideal channel for detecting how pervasive the phenomenon is in the solar atmosphere. Still, the \hrieuv observations show for the first time the extent of the EUV variation associated with coronal rain events. {We see EUV variation} from the small scales of rain clumps and fireballs to the large loop scales of CCTR-induced coronal strands and rebound shocks and flows {that partly} reheat the loop bundles. This supports previous suggestions based on numerical simulations \citep{Antolin_2022ApJ...926L..29A} that the TNE-TI scenario plays a major role in the observed filamentary morphology and high variability of the corona in TR {and low coronal} spectral lines \citep{Kjeldseth_Brekke_1998SoPh..182...73K,Ugarte-Urra_etal_2009ApJ...695..642U, Hinode_10.1093/pasj/psz084}. 

\begin{acknowledgements}
Solar Orbiter is a space mission of international collaboration between ESA and NASA, operated by ESA. The EUI instrument was built by CSL, IAS, MPS, MSSL/UCL, PMOD/WRC, ROB, LCF/IO with funding from the Belgian Federal Science Policy Office (BELPSO); Centre National d'Études Spatiales (CNES); the UK Space Agency (UKSA); the Deutsche Zentrum f\"ur Luft- und Raumfahrt e.V. (DLR); and  the Swiss Space Office (SSO). The building of EUI was the work of more than 150 individuals during more than 10 years. We gratefully acknowledge all the efforts that have led to a successfully operating instrument. {The development of SPICE has been funded by ESA member states and ESA. It was built and is operated by a multi-national consortium of research institutes supported by their respective funding agencies: STFC RAL (UKSA, hardware lead), IAS (CNES, operations lead), GSFC (NASA), MPS (DLR), PMOD/WRC (Swiss Space Office), SwRI (NASA), UiO (Norwegian Space Agency). SDO is a mission for NASA’s Living With a Star (LWS) program.}
The ROB co-authors thank the Belgian Federal Science Policy Office (BELSPO) for the provision of financial support in the framework of the PRODEX Programme of the European Space Agency (ESA) under contract numbers 4000112292, 4000134088, 4000134474, and 4000136424. P.A. and D.M.L. acknowledge funding from STFC Ernest Rutherford Fellowships No. ST/R004285/2 and ST/R003246/1, respectively. S.P. acknowledges the funding by CNES through the MEDOC data and operations center. L.P.C. gratefully acknowledges funding by the European Union (ERC, ORIGIN, 101039844). Views and opinions expressed are however those of the author(s) only and do not necessarily reflect those of the European Union or the European Research Council. Neither the European Union nor the granting authority can be held responsible for them. {This research was supported by the International Space Science Institute (ISSI) in Bern, through ISSI International Team project \#545 (`Observe Local Think Global: What Solar Observations can Teach us about Multiphase Plasmas across Physical Scales')}.

\end{acknowledgements}

\bibliographystyle{aa}
\bibliography{bibliography,patbib}

\appendix
\section{Multi-wavelength views with SPICE on March~30 and April~1.}

\begin{figure*}
    \centering
    \includegraphics[width=1\textwidth]{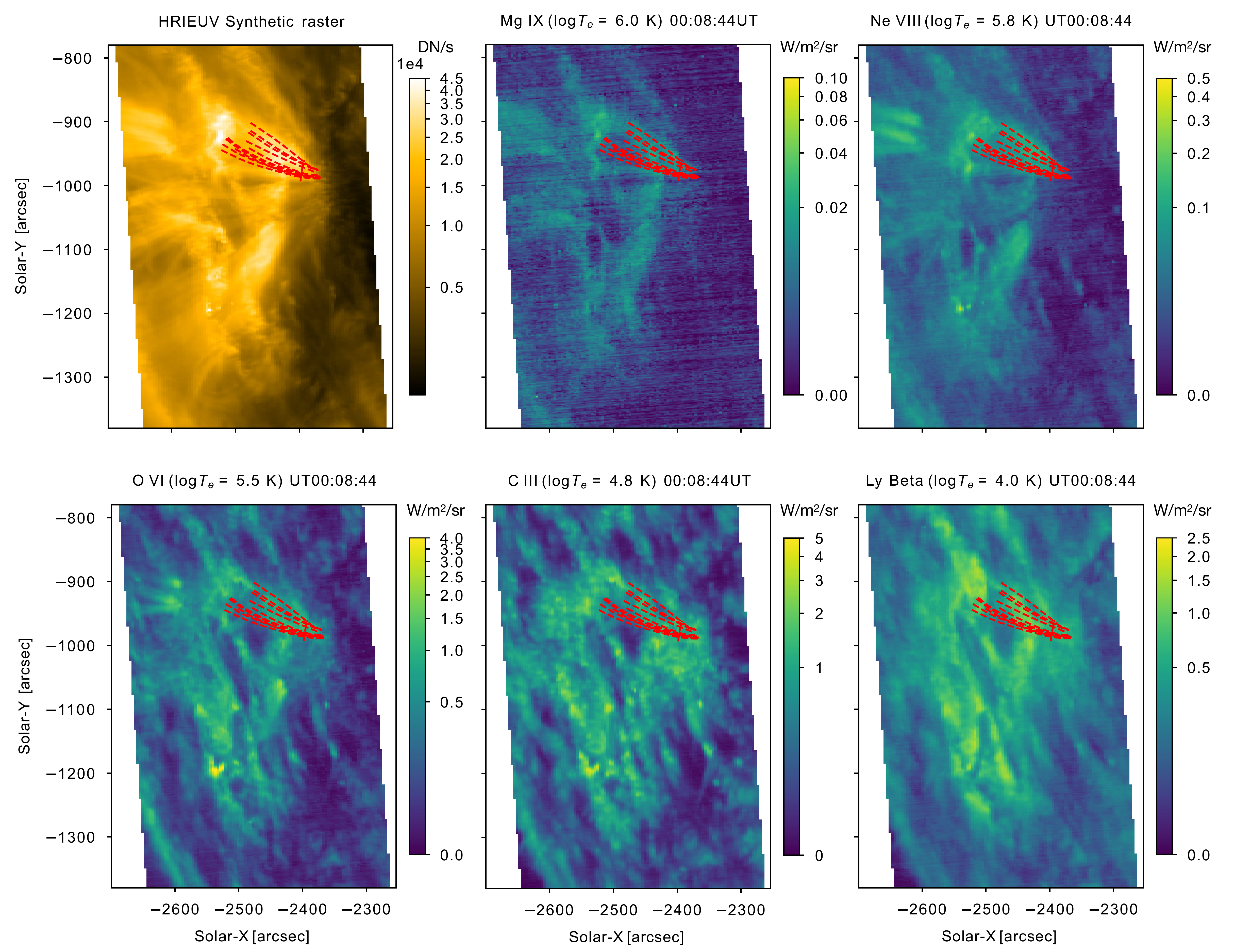}
    \caption{{SPICE multi-wavelength full FOV on March~30. The SPICE FOV is the same as that shown in Figure~\ref{fig:spicefullFOV} (top). The \hrieuv panel corresponds to a synthetic raster matching the time of the SPICE raster (see text for details) including spatial binning to match the SPICE platescale. Each SPICE panel shows the total intensity over the indicated spectral line, together with its temperature of maximum formation. The overlaid red curves denote the rain paths seen with \hrieuv (see Figure~\ref{fig:March30all}).}}
    \label{fig:spice_mar30_full}
\end{figure*}

\begin{figure*}
    \centering
    \includegraphics[width=1\textwidth]{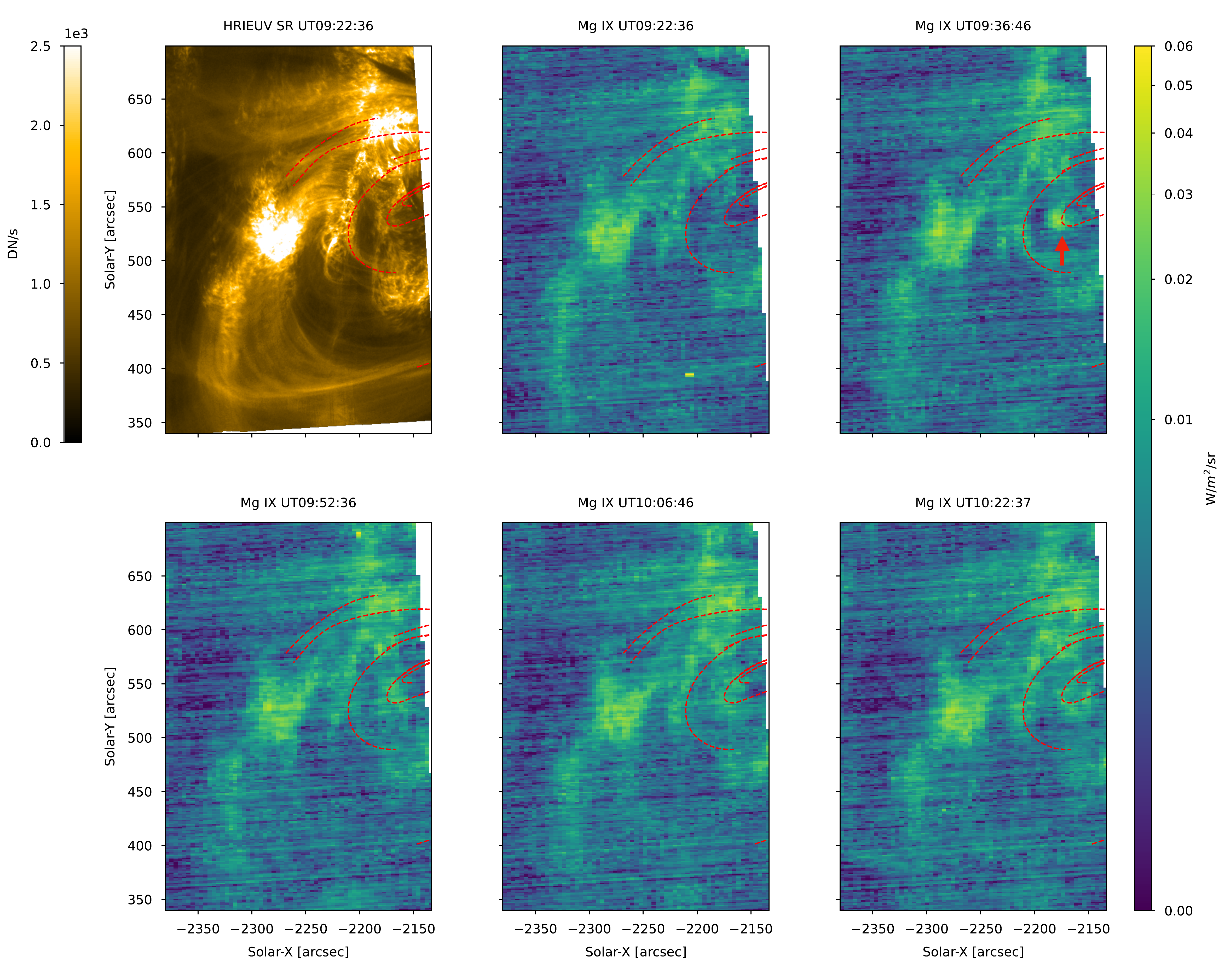}
    \caption{{Same as in Figure~\ref{fig:spice_april1_regs1-2} but for the \ion{Mg}{IX} line.}}
    \label{fig:spice_april1_regs1-2_mg9}
\end{figure*}

\begin{figure*}
    \centering
    \includegraphics[width=1\textwidth]{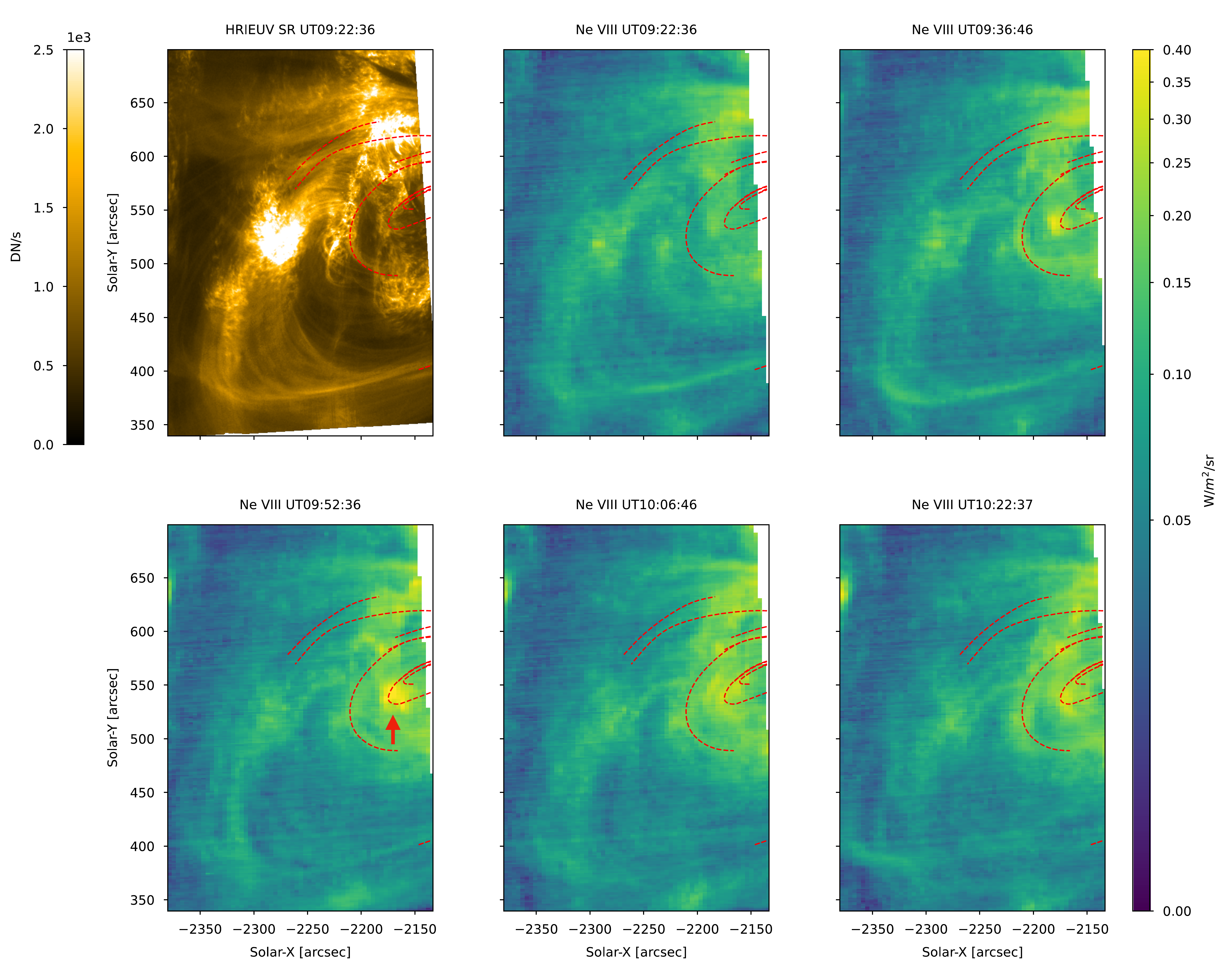}
    \caption{{Same as in Figure~\ref{fig:spice_april1_regs1-2} but for the \ion{Ne}{VIII} line.}}
    \label{fig:spice_april1_regs1-2_ne8}
\end{figure*}

\begin{figure*}
    \centering
    \includegraphics[width=1\textwidth]{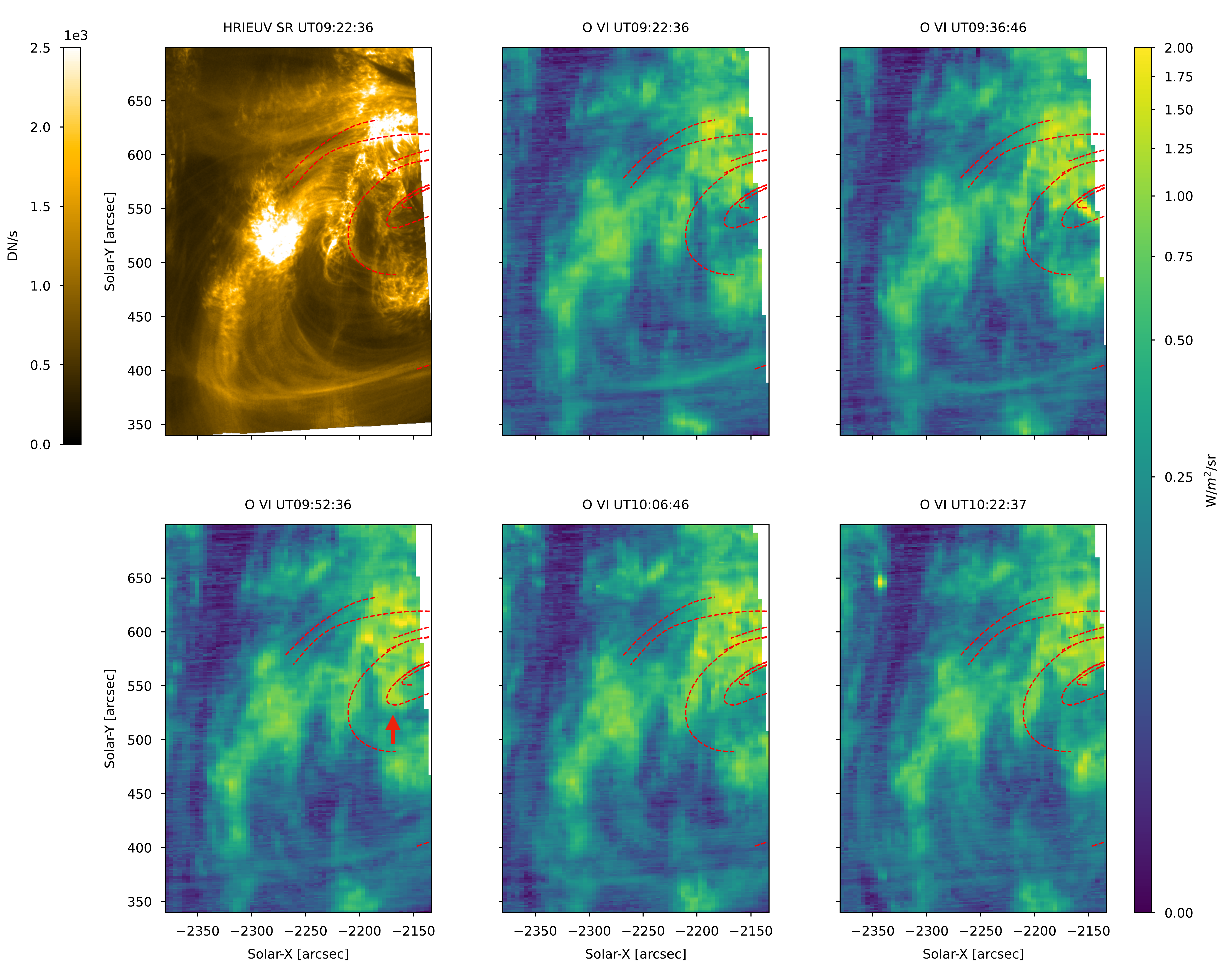}
    \caption{{Same as in Figure~\ref{fig:spice_april1_regs1-2} but for the \ion{O}{VI} line.}}
    \label{fig:spice_april1_regs1-2_o6}
\end{figure*}

\begin{figure*}
    \centering
    \includegraphics[width=1\textwidth]{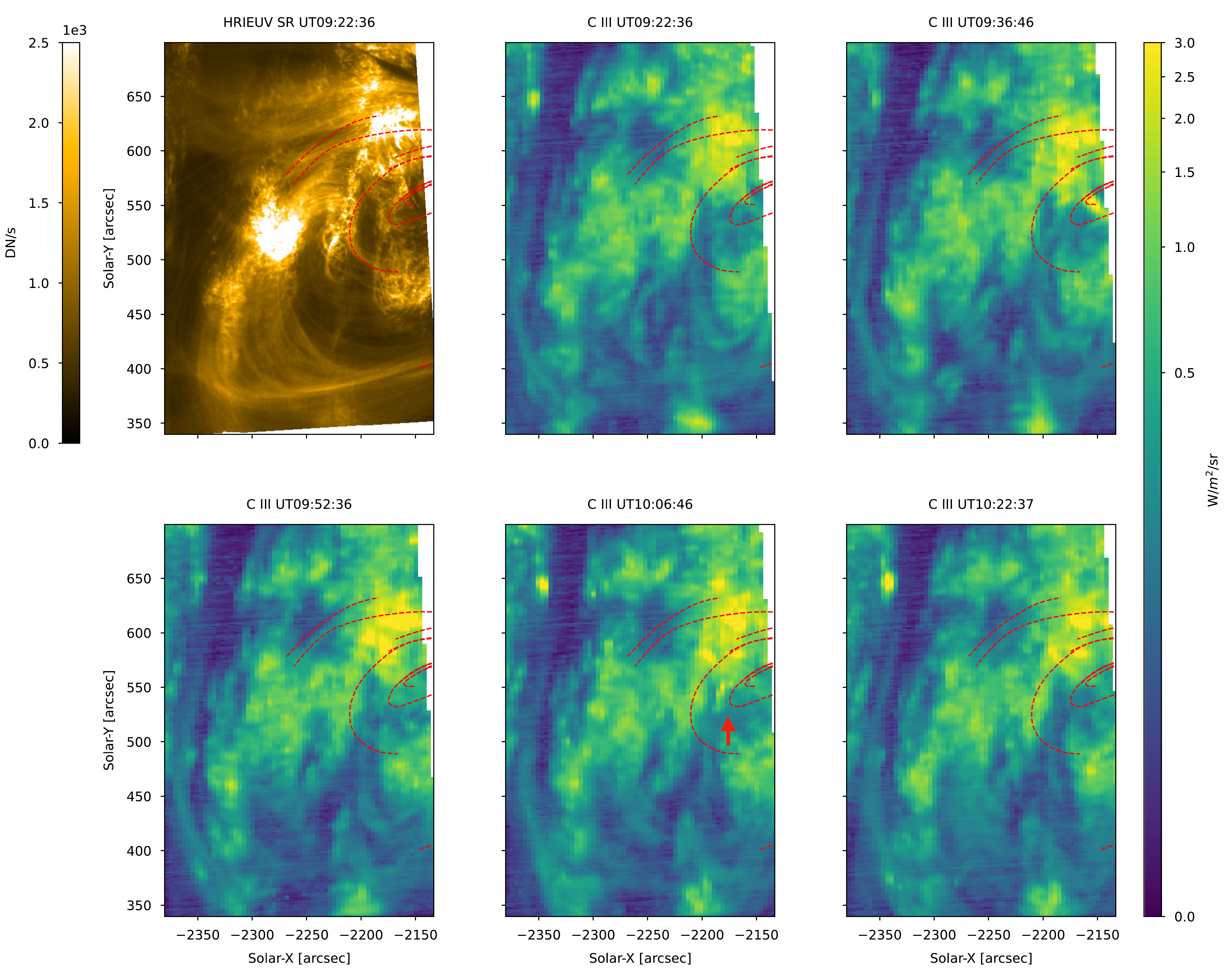}
    \caption{{Same as in Figure~\ref{fig:spice_april1_regs1-2} but for the \ion{C}{III} line.}}
    \label{fig:spice_april1_regs1-2_c3}
\end{figure*}

\begin{figure*}
    \centering
    \includegraphics[width=1\textwidth]{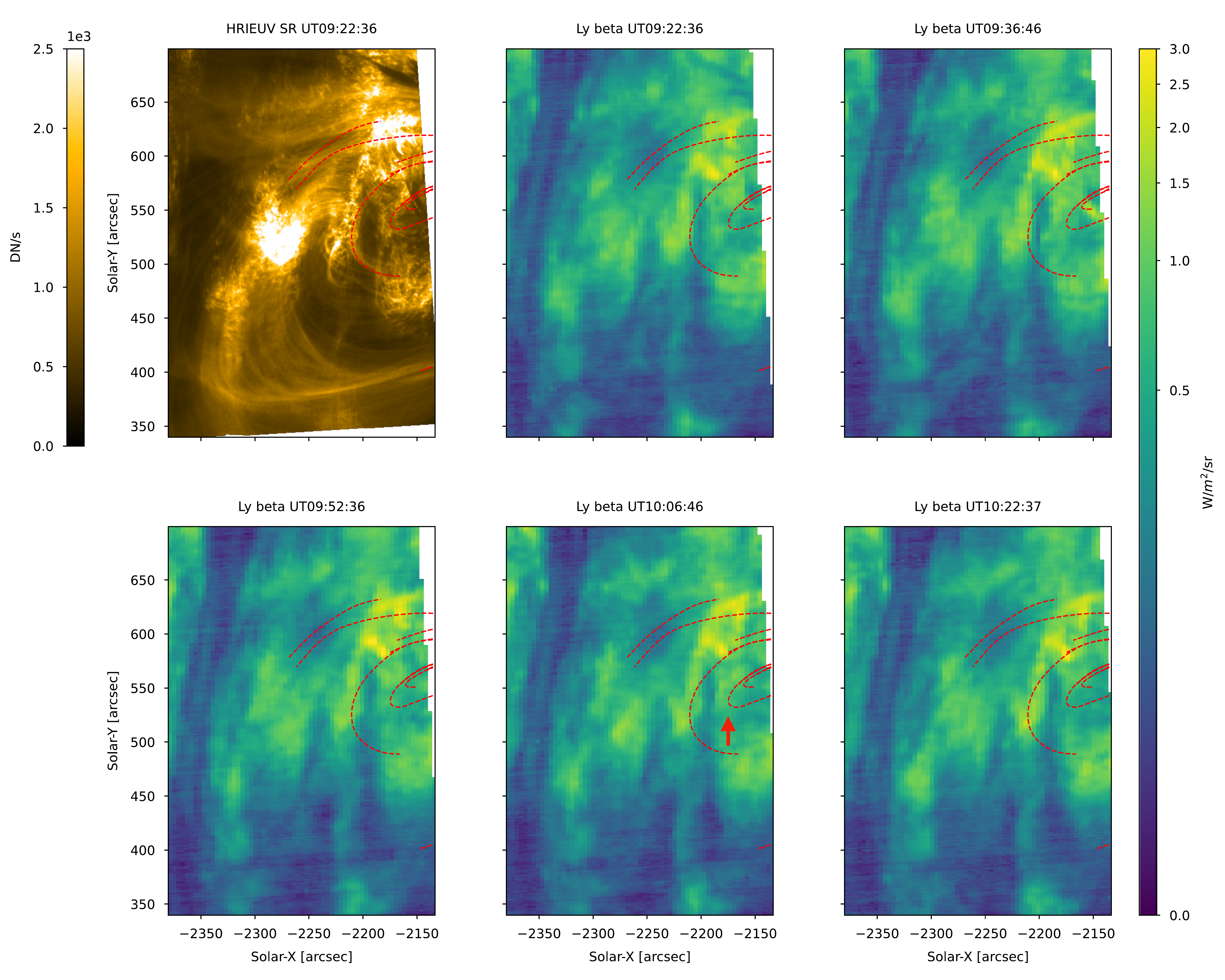}
    \caption{{Same as in Figure~\ref{fig:spice_april1_regs1-2} but for the Lyman-$\beta$ line.}}
    \label{fig:spice_april1_regs1-2_lyb}
\end{figure*}

\begin{figure}
    \centering
    \includegraphics[width=0.5\textwidth]{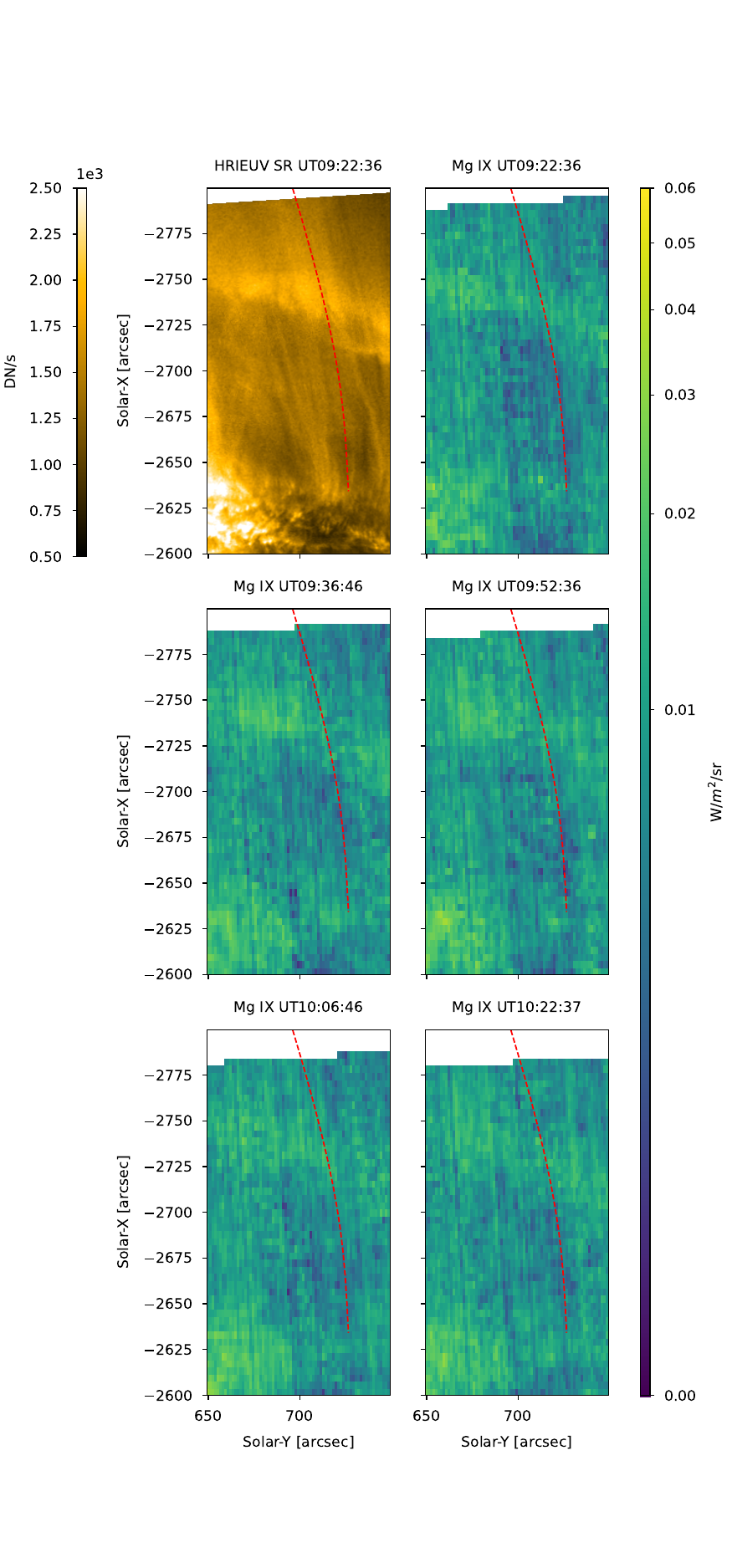}
    \caption{{Same as in Figure~\ref{fig:spice_apr1_reg4} but for the \ion{Mg}{IX} line.}}
    \label{fig:spice_apr1_reg4_mg9}
\end{figure}

\begin{figure}
    \centering
    \includegraphics[width=0.5\textwidth]{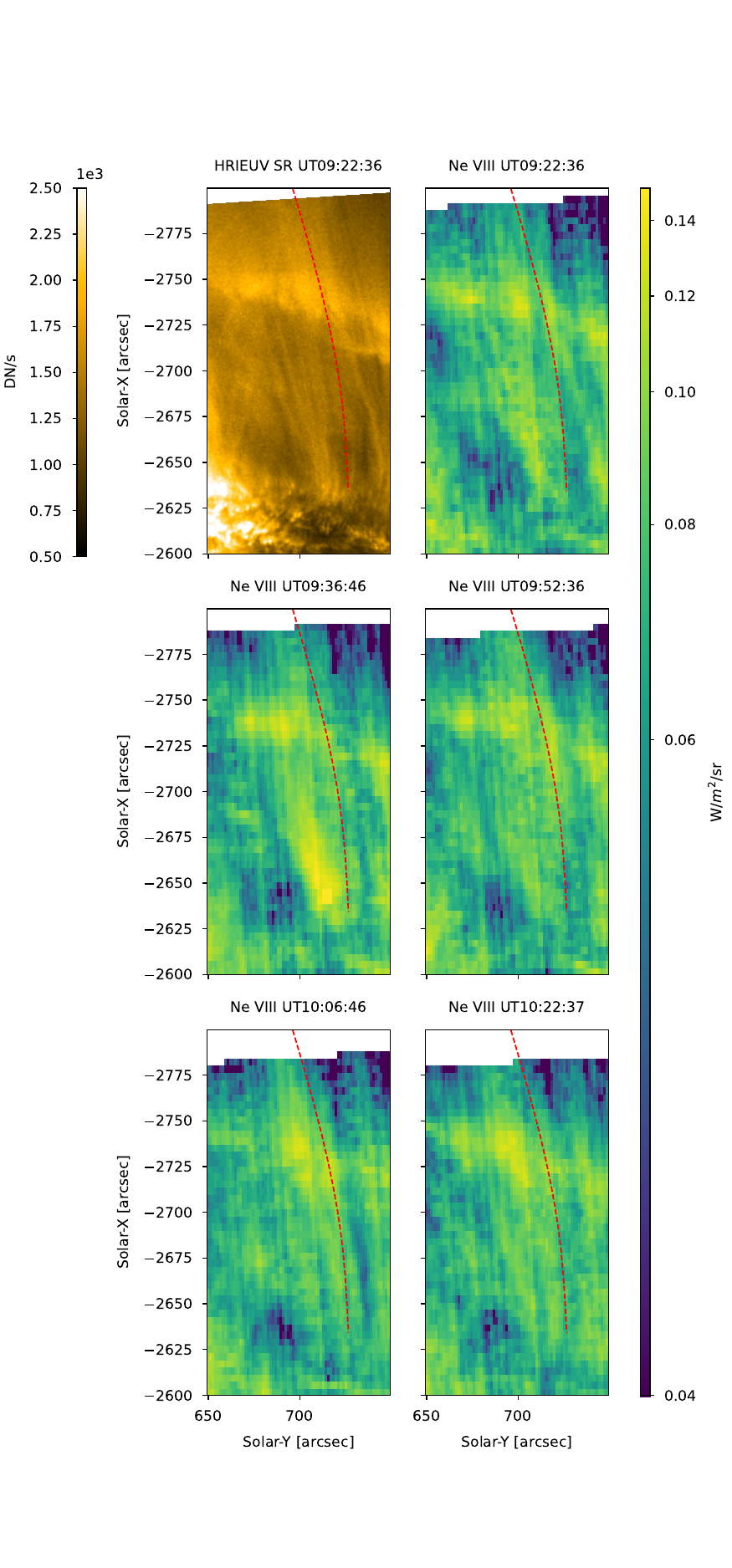}
    \caption{{Same as in Figure~\ref{fig:spice_apr1_reg4} but for the \ion{Ne}{VIII} line.}}
    \label{fig:spice_apr1_reg4_ne8}
\end{figure}

\begin{figure}
    \centering
    \includegraphics[width=0.5\textwidth]{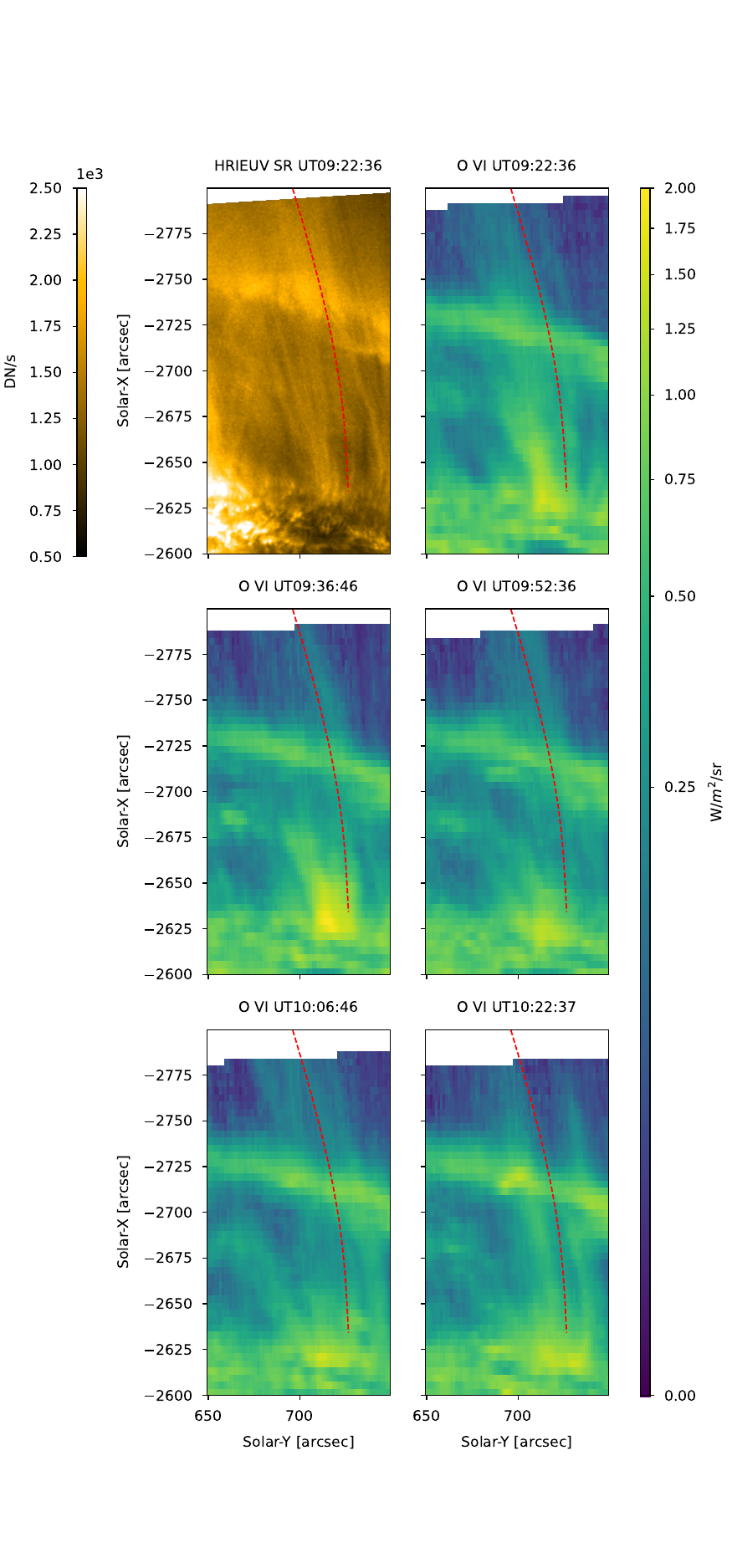}
    \caption{{Same as in Figure~\ref{fig:spice_apr1_reg4} but for the \ion{O}{VI} line.}}
    \label{fig:spice_apr1_reg4_o6}
\end{figure}

\begin{figure}
    \centering
    \includegraphics[width=0.5\textwidth]{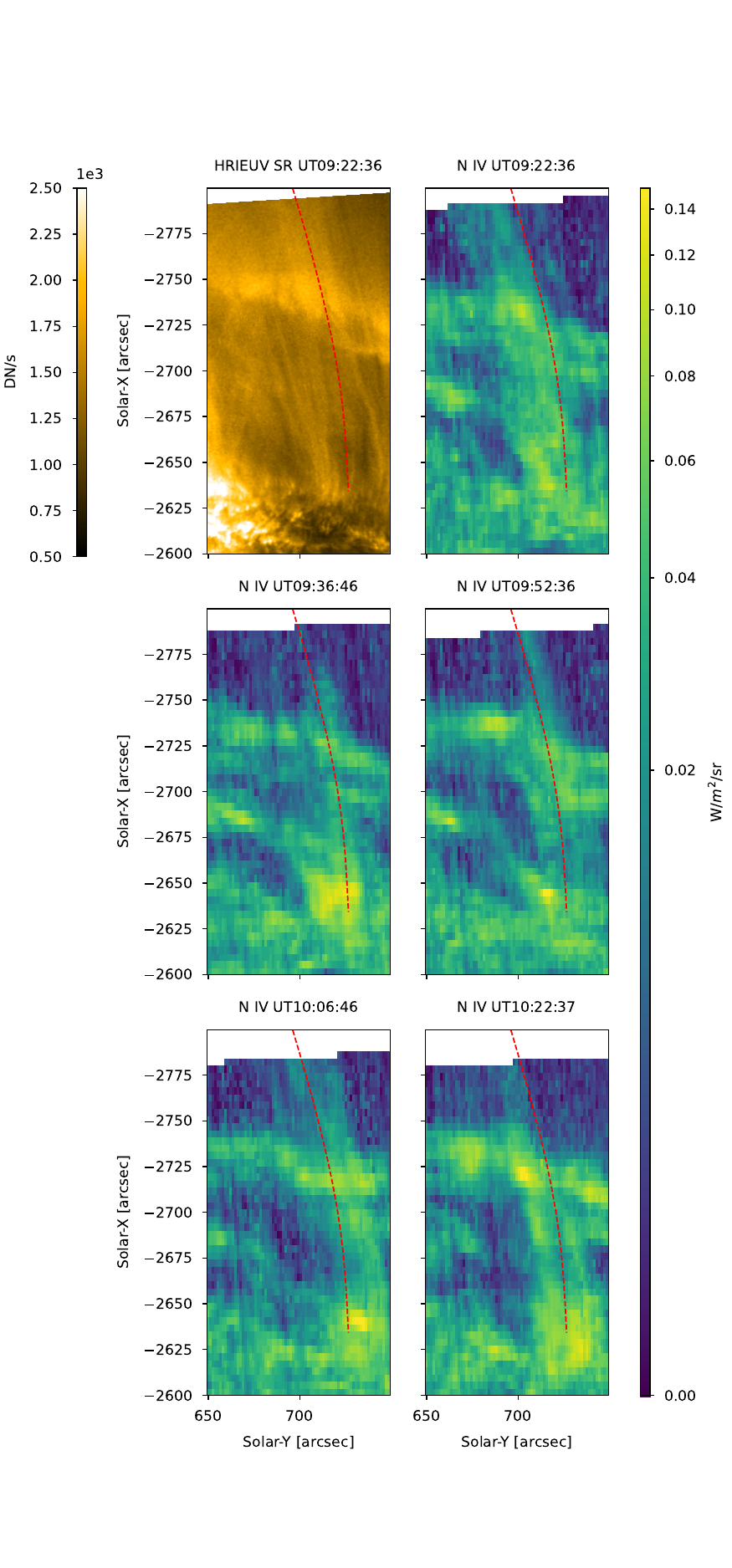}
    \caption{{Same as in Figure~\ref{fig:spice_apr1_reg4} but for the \ion{N}{IV} line.}}
    \label{fig:spice_apr1_reg4_n4}
\end{figure}

\begin{figure}
    \centering
    \includegraphics[width=0.5\textwidth]{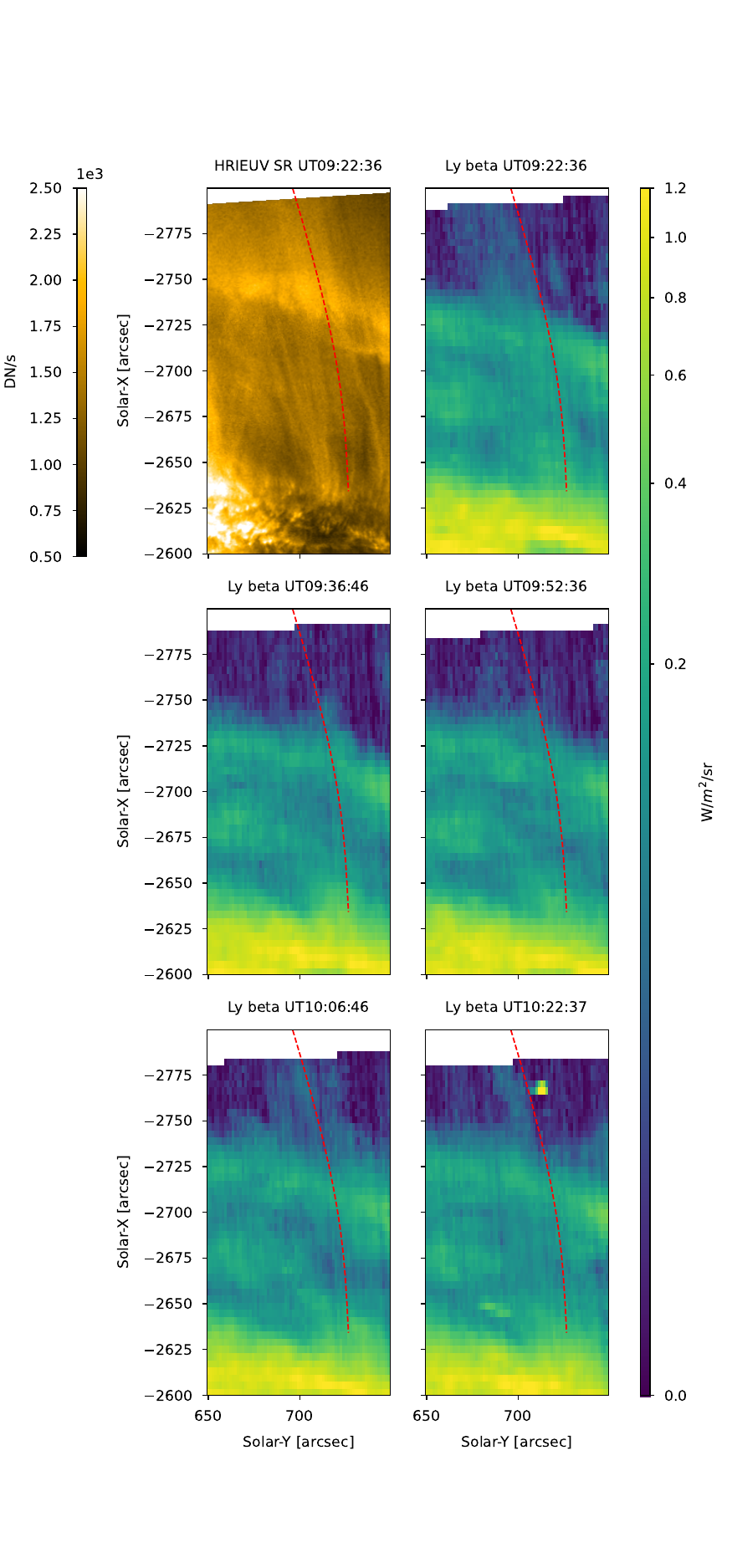}
    \caption{{Same as in Figure~\ref{fig:spice_apr1_reg4} but for the Lyman-$\beta$ line.}}
    \label{fig:spice_apr1_reg4_lyb}
\end{figure}

\section{Time-distance diagrams with \hrieuv on April~1}

\begin{figure}
    \centering
    \includegraphics[width=0.5\textwidth]{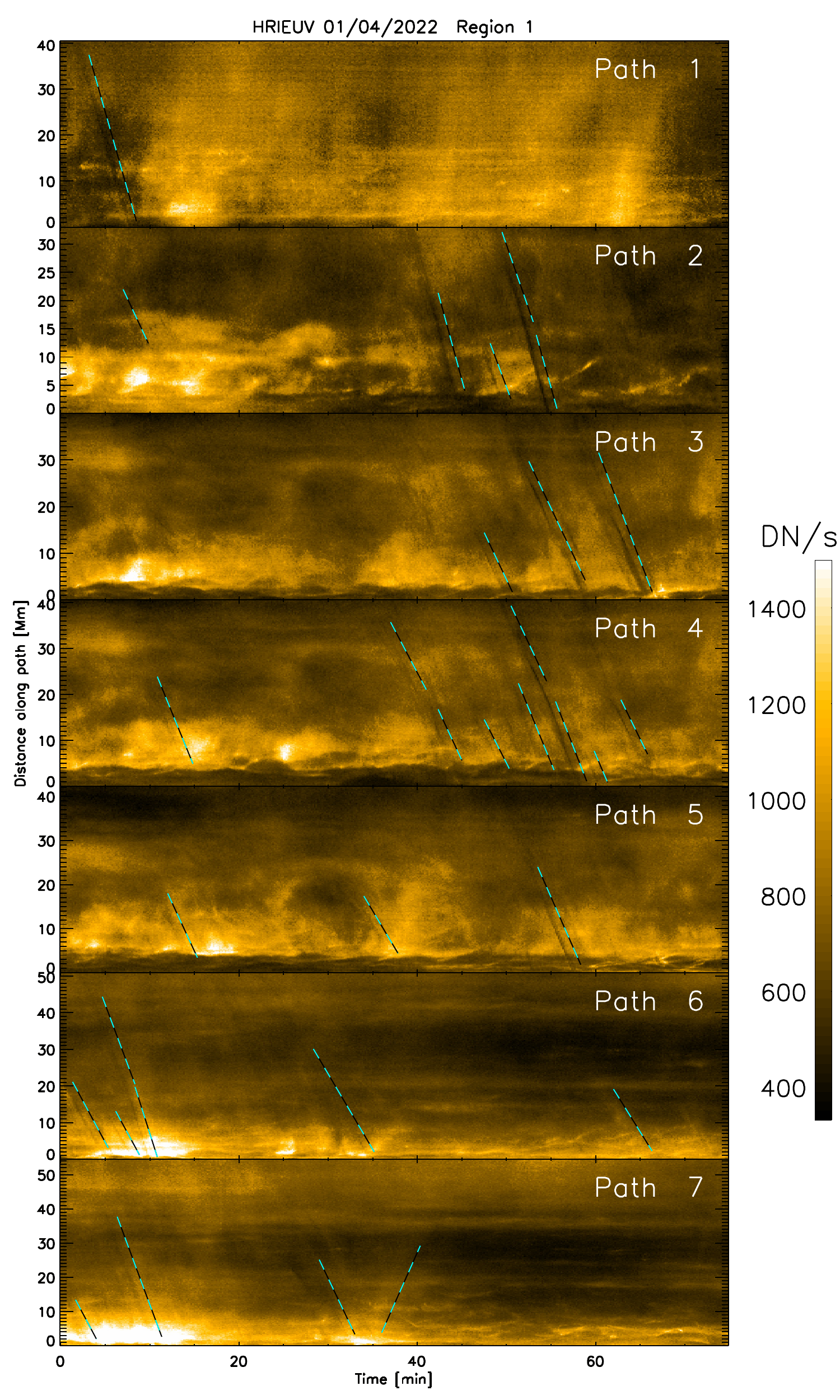}
    \caption{Time-distance diagram for the rain paths tracked in Region~1 shown in Figure~\ref{fig:april1_reg1}. The {cyan-black} dashed lines with negative slopes track the {dark/bright absorption/emission} features produced by several rain clumps falling into the chromosphere. A few bright upward propagating features can also be seen (positive slopes). Zero distance corresponds to the {footpoint of the loop} (top-right of Figure~\ref{fig:april1_reg1}).}
    \label{fig:april1st_reg1_td}
\end{figure}

\begin{figure}
    \centering
    \includegraphics[width=0.5\textwidth]{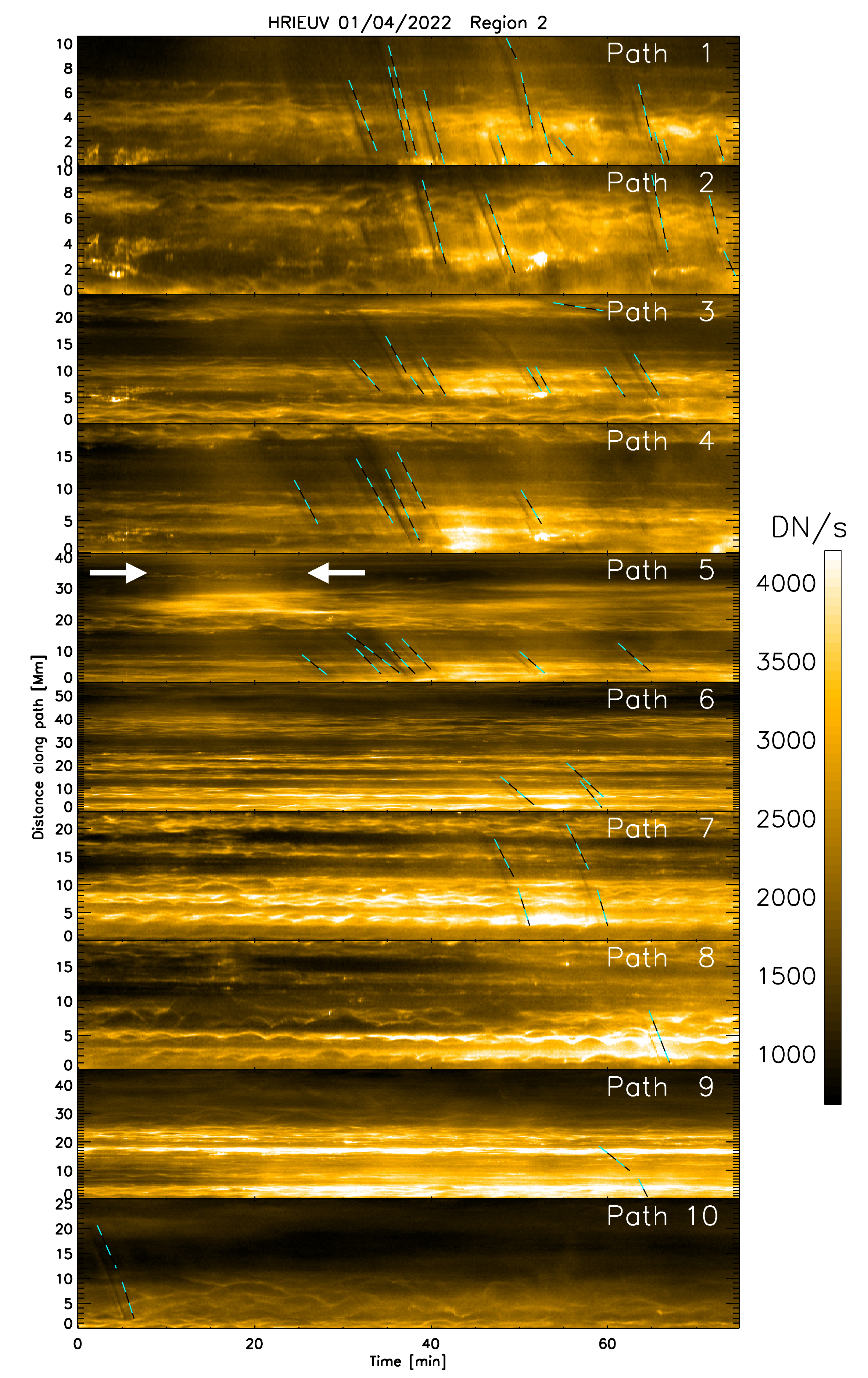}
    \caption{Time-distance diagrams for the paths of the rain clumps shown in Figure~\ref{fig:april1_reg2}. The {cyan-black} dashed lines correspond to coronal rain tracks along the paths, with zero distance corresponding to the loop {footpoints on the right side} in Figure~\ref{fig:april1_reg2}. {The lines are offset in time by 1~min to better see the rain features.}The white arrows in the time-distance diagrams for path~5 indicate the times when the loop brightens prior to the appearance of the rain. }
    \label{fig:april1_reg2_td}
\end{figure}

\begin{figure}
    \centering
    \includegraphics[width=0.5\textwidth]{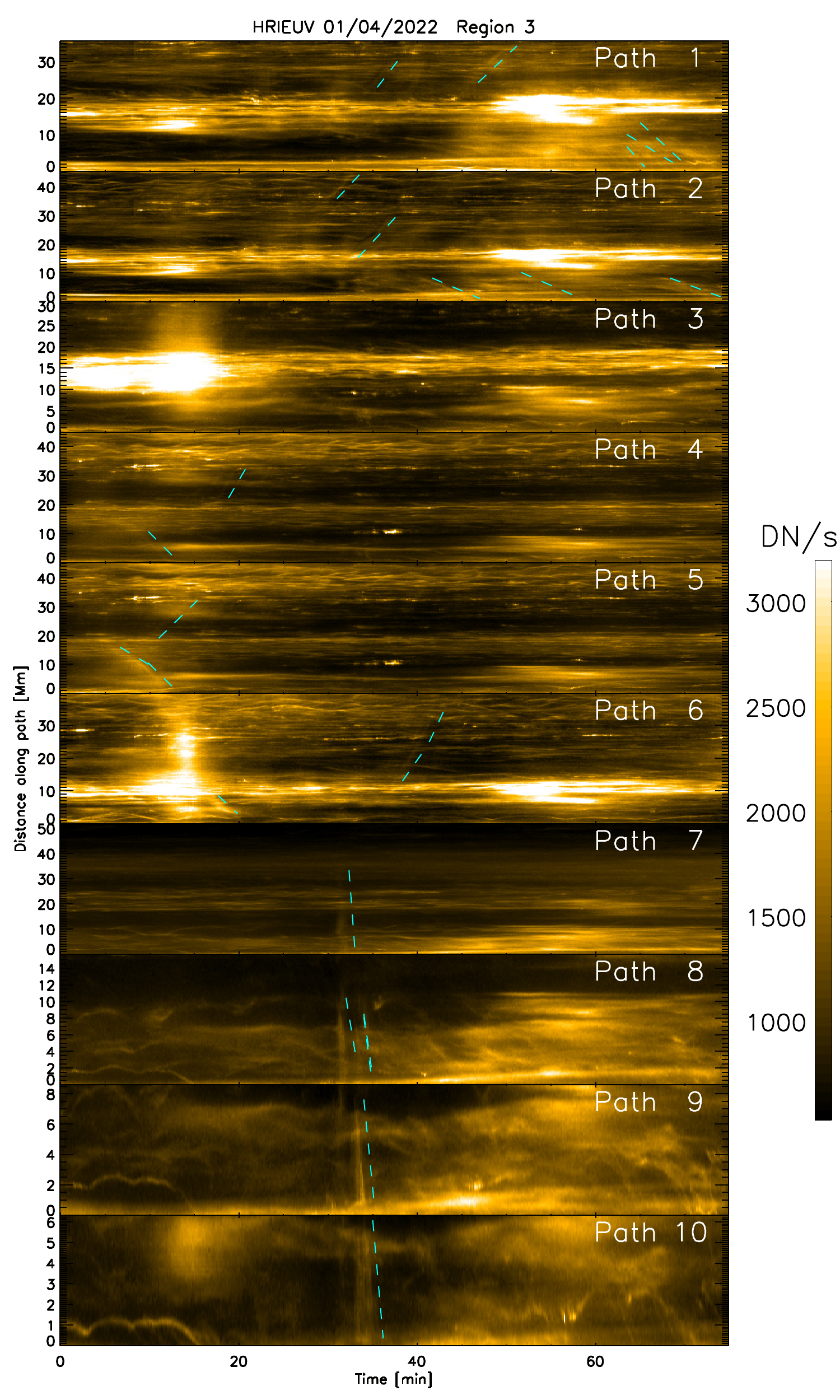}
    \caption{Time-distance diagrams for the paths of the rain clumps shown in Figure~\ref{fig:april1_reg3}. The {cyan} dashed lines correspond to coronal rain tracks along the paths, with zero distance corresponding to the loop footpoints (left in  Figure~\ref{fig:april1_reg3}).  {The lines are offset in time by 1~min to better see the rain features. Note the loop brightening seen clearly in Paths 3 and 6 at time $t=15~min$.}}
    \label{fig:april1_reg3_td}
\end{figure}

\end{document}